\documentstyle[amssymb,a4,fullpage,11pt,subfigure,epsfig]{article}

\newif\iffigure

\figuretrue

\def\agt{\gtrsim}
\def\alt{\lesssim}
\def\tsig{\hbox{\boldmath$\sigma$}}
\def\tchi{\hbox{\boldmath$\chi$}}
\def\ttau{\hbox{\boldmath$\tau$}}

\title{\bf Finite-temperature properties of doped antiferromagnets}
\author{J. JAKLI\v C$^{1,*}$ and P. PRELOV\v SEK$^{1,2}$\\
$^1$~J. Stefan Institute, 1000 Ljubljana, Slovenia \\
$^2$~Faculty of Mathematics and Physics,
University of Ljubljana, 1000 Ljubljana, \\Slovenia }

\date{ }

\begin{document}

\maketitle

\centerline{to appear in ADVANCES IN PHYSICS}

\bigskip

\begin{abstract}
We review recent results for the properties of doped antiferromagnets,
obtained by the numerical analysis of the planar $t$-$J$ model using
the novel finite-temperature Lanczos method for small correlated
systems. First we shortly summarize our present understanding of
anomalous normal-state properties of cuprates, and present the
electronic phase diagram, phenomenological scenarios and models
proposed in this connection. The numerical method is then described in
more detail. Following sections are devoted to various static and
dynamical properties of the $t$-$J$ model. Among thermodynamic
properties the chemical potential, entropy and the specific heat are
evaluated. Discussing electrical properties the emphasis is on the
optical conductivity and the d.c. resistivity. Magnetic properties
involve the static and dynamical spin structure factor, as measured
via the susceptibility measurements, the NMR relaxation and the
neutron scattering, as well as the orbital current
contribution. Follows the analysis of electron spectral functions,
being studied in photoemission experiments. Finally we discuss
density fluctuations, the electronic Raman scattering and the
thermoelectric power. Whenever feasible a comparison with experimental
data is performed. A general conclusion is that the $t$-$J$ model
captures well the main features of anomalous normal-state properties
of cuprates, for a number of quantities the agreement is even a
quantitative one. It is shown that several dynamical quantities
exhibit at intermediate doping a novel universal behaviour consistent
with a marginal Fermi-liquid concept, which seems to emerge from a
large degeneracy of states and a frustration induced by doping the
antiferromagnet.

\end{abstract}

\tableofcontents

\newpage
\setcounter{equation}{0}\setcounter{figure}{0}
\section{Introduction}

The discovery of the high-temperature superconductivity (SC) in
copper-oxide based compounds - cuprates (Bednorz and M\"uller 1986)
revived the interest in materials containing transition elements. The
main feature of these materials is the crucial role of 
electron-electron interactions.  This can result in electronic
properties, very unusual when compared to the behaviour of
conventional metals.  Although the unconventional SC is
clearly the most puzzling phenomenon in cuprates, we focus in
this review predominantly on the analysis of more or less anomalous
properties of the normal state which deviate essentially from the
standard understanding of electrons in metals and still present one of
the major theoretical challenges in the solid state physics.

\begin{figure}[ht]
\centering
\iffigure
\mbox{
\subfigure[]{
\epsfig{file=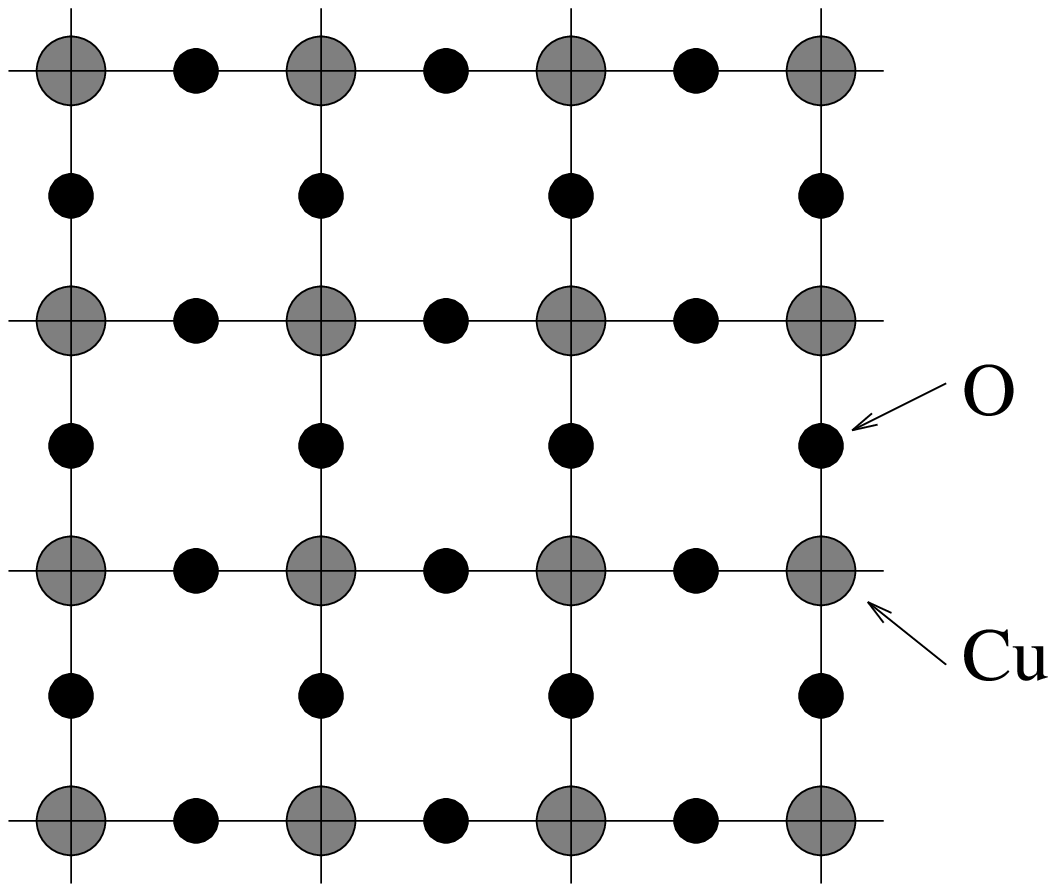,height=3.3cm}}
\quad
\subfigure[]{
\epsfig{file=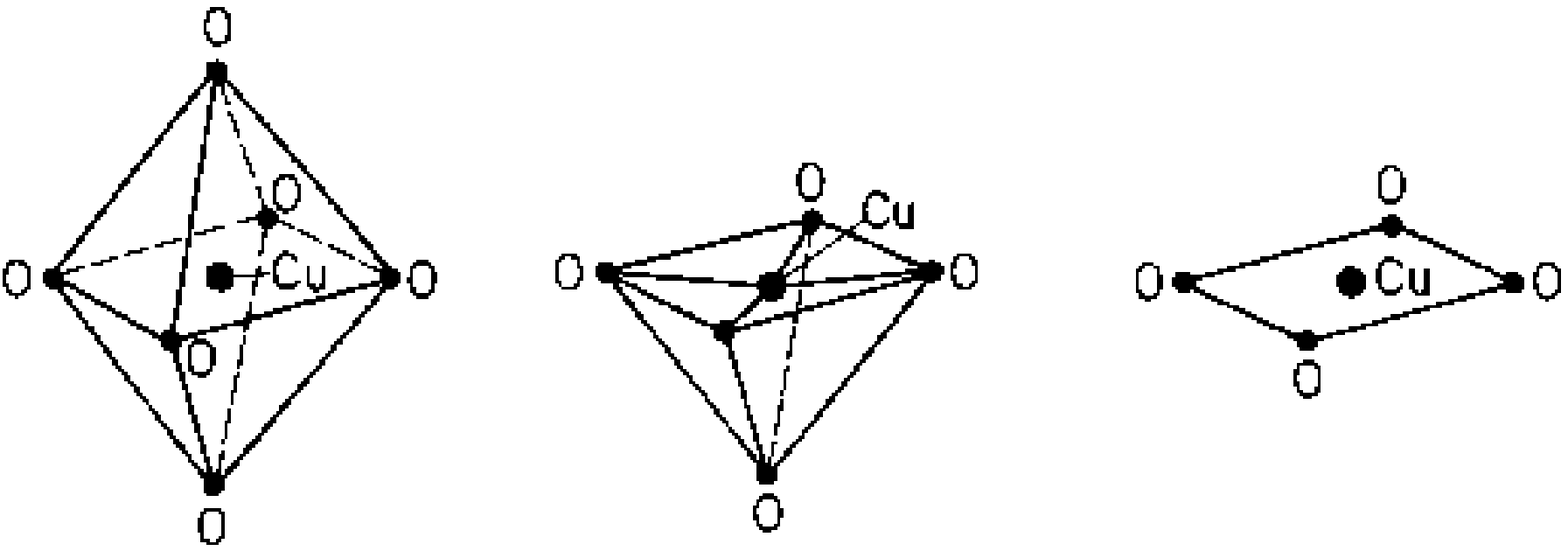,height=3.3cm}}}
\fi
\caption{(a) Schematic structure of copper-oxide planes and 
(b) three possible building blocks of the planes, after
Fulde (1991).} \label{1.1}
\end{figure}

The cuprate superconductors have a very anisotropic structure, where
the common building blocks are layers, in cuprates formed by
combining one of the three possible structural elements containing Cu
and O, as shown in Fig.~\ref{1.1}b.  The CuO layered structures are
stacked in the crystal, separated by various intercalant layers in
different cuprates. In spite of vast differences in the structure of
unit cells, electronic properties of the whole family of cuprates are
quite universal.  This can be explained by the predominant role of
generic CuO$_2$ planes, Fig.~\ref{1.1}a, where conducting electrons
reside. The electronic coupling between CuO$_2$ planes is very weak,
resulting in a huge ratio of the in-plane resistivity to the
perpendicular resistivity, in the anisotropy of the SC coherence
length etc.  Intermediate layers serve mainly as a charge reservoir
for the planes. Consequently properties of cuprates are quite
well classified according to the doping level of the reference CuO$_2$
electronic structure. 

As now well established, the reference cuprate compounds, as
La$_2$CuO$_4$ and YBa$_2$Cu$_3$O$_6$, are Mott (charge-transfer)
insulators due to strong correlations which induce in a half-filled
band a charge gap $\sim 2~$eV.  The spin degrees of planar CuO$_2$
electrons can be well mapped on the properties of a planar
antiferromagnetic (AFM) $S=1/2$ Heisenberg model. The AFM ground state
of undoped materials, emerging from strong correlations, has been
quite early recognized as a crucial starting point for theoretical
considerations (Anderson 1987) of doped materials, being strange
metals in the normal state and exhibiting high transition temperature
$T_c$ to the SC state.

There are by now numerous indications that essential features of
electronic properties of the doped AFM, as realized in cuprates, are
well represented by prototype single-band models of correlated
electrons, as the Hubbard model and the $t$-$J$ model (Rice 1995). In
spite of their apparent simplicity both models are notoriously
difficult to treat analytically, in particular in the most interesting
regime of strong correlations. The lack of analytical tools for
correlated electrons (for a general introduction see Fulde 1991)
has increased the efforts towards numerical approaches (Dagotto 1994),
which predominantly can be divided into two categories: the quantum
Monte Carlo (QMC) methods and the exact diagonalization (ED)
methods. The $t$-$J$ model, which incorporates the strong correlation
requirement explicitly, is more adapted to the ED approach. So far most
calculations were performed for the ground state (g.s.) at $T=0$,
where the standard Lanczos technique (Lanczos 1950) offers an
efficient ED analysis of small systems of reasonable sizes.  Recently,
present authors (Jakli\v c and Prelov\v sek 1994a) introduced a novel
numerical method, combining the Lanczos method with a random sampling,
which allows for an analogous treatment of statics and dynamics of
many-body quantum models at $T>0$. The latter method, further referred
to as the finite-temperature Lanczos method (FTLM), and results for
the $t$-$J$ model obtained using this method, are the main subject of
this review.

The final goal is to understand properties of doped AFM in general,
and of high-$T_c$ cuprates in particular. In the absence of reliable
analytical methods and results, numerical calculations can help to
answer several crucial questions.  Are strong correlations, as
incorporated in prototype models, enough to account for anomalous
normal-state properties of the strange metal ? Which are the relevant
energy and temperature scales in doped AFM, as represented by the
$t$-$J$ model, and in which properties do they show up ? Which is the
unifying phenomenological description of the normal state ? Is the
$t$-$J$ model sufficient, or which ingredients should be added to
account qualitatively and quantitatively for observed properties ?
Can we learn something macroscopically meaningful from the studies of
small systems, and why ?

We note that at present the SC seems to be beyond the reach of
numerical approaches including the FTLM, hence we do not investigate
here in more detail the possible existence of SC and its origin in
model systems.

\setcounter{equation}{0}\setcounter{figure}{0}
\section{Cuprates as doped antiferromagnets}

\subsection{Electronic phase diagram and properties of the normal state} 

\begin{figure}
\centering
\iffigure
\epsfig{file=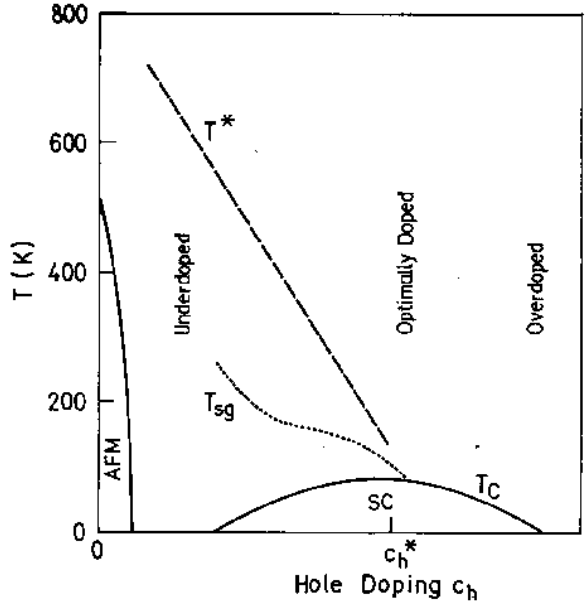,height=8cm}
\fi
\caption{
Schematic electronic phase diagram of cuprates, after Batlogg (1997).
} \label{2.1}
\end{figure}

Reference cuprate compounds are AFM insulators and are so far best
understood.  Properties of various other layered cuprates can be
interpreted in terms of doping the reference material, where (mainly)
holes are introduced into CuO$_2$ planes. One of the major conceptual
achievements, which emerged from careful experimental investigations
of high-quality materials in the last decade, has been the realization
of quite universal electronic phase diagram (Hwang {\it et al.} 1994,
Batlogg {\it et al.} 1994, Batlogg 1997), revealing characteristic
temperature scales as they develop as the function of hole doping.  It
is at present quite common to classify materials with respect to
doping as underdoped, optimally doped and overdoped, and the
corresponding phase diagram is shown in Fig.~\ref{2.1}.

\subsubsection{Optimum doping regime}

Experimentally optimum doping is chosen to correspond to materials
with the highest $T_c$ within the given class of chemically and
structurely related compounds.  It has been realized soon after the
discovery of high-$T_c$ SC that also normal state properties at
$T>T_c$ of the optimally doped materials are very anomalous, but at
the same time also most universal. The prominent feature is the
resistivity law $\rho \propto T$ (Takagi {\it et al.} 1992), valid
essentially in the whole measurable range $T>T_c$ and clearly
contradicting the normal Landau-Fermi-liquid (LFL) behaviour $\rho
\propto T^2$. Related is the observation that the dynamical
conductivity $\sigma(\omega)$ does not fall off for larger $\omega$
according to the Drude form $\sigma \propto 1/\omega^2$, but rather as
$\sigma \propto 1/\omega$ (Tanner and Timusk 1992).

The clearest evidence for an anomalous spin dynamics comes from the
NMR and NQR relaxation (Slichter 1994), where the relaxation rate
$1/T_1$ on planar $^{63}$Cu in optimally doped La$_{2-x}$Sr$_x$CuO$_4$
(LSCO) with $x_{opt} \sim 0.15$ is nearly $T$-independent (as well as
nearly doping independent) in contrast to the usual Korringa law for
metals $1/T_1 \propto T$. The qualitative difference, i.e. a large
enhancement of low-frequency spin fluctuations at low $T$, can be
related to the persistence of short range AFM fluctuations, even at
the optimum doping. The support for this comes also from neutron
scattering experiments, where a substantial AFM correlation length
$\xi$ has been measured in the same class of materials, with $\xi \sim
3.8 \AA /\sqrt{x}$ (Birgeneau {\it et al.} 1988).
   
On the other hand, several properties at the optimum doping seem to be
close to the normal LFL picture.  The angle-resolved photoemission
spectroscopy (ARPES) on cuprates (Shen and Dessau 1995), most reliable
for BiSrCaCuO (BISCCO) compounds, shows electronic excitations
consistent with a large Fermi surface (FS) and with the conserved FS
volume (Luttinger theorem), although spectral shapes are quite
distinct from a simple LFL picture.  The specific heat (Loram {\it et
al.} 1993) follows in the normal state roughly the LFL behaviour $C_v
=\gamma T$ with $\gamma$ not very far from the free-fermion value,
consistent with a nearly $T$-independent uniform susceptibility $\chi_0$,
whereby the Wilson ratio is close to the free-fermion one (Loram {\it
et al.}  1996). It is also the unifying characteristic of the optimum
doping that properties do not reveal above $T_c$ any additional
characteristic $T$ scale.

\subsubsection{Overdoped and underdoped regime}

In overdoped materials $T_c$ is decreasing and finally vanishing with
increased doping. At the same time the electronic properties are
getting closer to the usual metallic behaviour consistent with the LFL
scenario. E.g., the resistivity behaviour moves towards the normal FLF
form $\rho \propto T^2$ (Takagi {\it et al.} 1992), spectral shapes of
electronic excitations, as revealed by ARPES, become sharper near the
FS (Marshall {\it et al.} 1996) etc. These facts can be put together
with a decreasing intensity of AFM fluctuations. It is thus plausible
that we are dealing in the overdoped regime with the crossover to the
normal LFL, however this crossover is not a trivial one and so far
also not well understood either.
   
The most evident progress in the investigations of normal-state
properties has been made in last few years for the underdoped
cuprates.  In contrast to the optimum doping, experiments reveal in
this regime additional characteristic temperatures $T>T_c$ (Batlogg
{\it et al} 1994), which show up as the crossovers where particular
properties qualitatively change.  As summarized in Fig.~\ref{2.1},
there seems to be an indication for two distinct crossovers. The
existence of both as well as their distinction is still widely
debated, nevertheless we will refer to them as the AFM crossover scale
$T^*$ and the pseudogap scale $T_{sg}$ for the lower one (Batlogg
1997).

The $T^*$ scale (Batlogg {\it et al.} 1994) shows up most clearly as
the maximum of the spin susceptibility $\chi_0(T=T^*)$ (Torrance {\it
et al.} 1989).  The in-plane resistivity $\rho(T)$ is linear $\rho
\propto T$ for $T>T^*$ and decreases more steeply for
$T<T^*$. Characteristic is also the anomalous $T$-dependence of the
Hall constant $R_H(T)$ for $T<T^*$ (Ong 1990, Hwang {\it et al.} 1994). The
latter is evidently hole-like in the classical sense $R_H(T \agt T_c)
\propto 1/x$, which is also not properly understood
theoretically.  It seems plausible that the $T^*$ crossover is related to
the onset of short-range AFM correlations for $T<T^*$, since in
the undoped AFM $T^*$ corresponds just to a well understood maximum
due to a gradual transition from a disordered paramagnet to the one
with short-range AFM correlations.

The crossover $T_{sg}$ has been first identified in connection with the
decrease of the NMR relaxation $1/T_1$ for $T<T_{sg}$ (Takigawa {\it et
al.} 1991, Slichter 1994), indicating the reduction of low-energy
spin excitations interpreted as the opening of the spin pseudogap in
underdoped materials.  Most striking evidence for an additional energy
scale in underdoped cuprates is the observation of the leading-edge
shift in ARPES measurements at $T>T_c$ (Marshall {\it et al} 1996),
indicating features of the $d$-wave SC gap persisting within the
normal phase. It should be pointed out, however, that the designation
of crossover features is at present still controversial. In particular
it is not evident whether we are dealing with two or more essentially
different energy scales.
      
\subsection{Phenomenology of the normal state}

Properties of the normal LFL follow from the one-to-one correspondence
of low-energy excitations of the interacting fermion system to that of
a free-fermion gas. The prerequisite is that the volume of the FS is
conserved (Luttinger 1960). Essential are the well defined
quasiparticles (QP) with the vanishing damping at the FS, with $\Gamma
\propto (E-E_F)^2$. Consequences are the linear specific heat $C_V
=\gamma T$, nearly $T$-independent static and dynamical spin
susceptibilities, the Korringa law for NMR relaxation $1/T_1 \propto T$,
the resistivity $\rho(T) \propto T^2$ etc.  Experimental facts on cuprates
contradict the usual LFL picture.  Several more or less
elaborated scenarios have been proposed to capture main anomalous
features.

Focusing on the importance of AFM spin fluctuations, the concept of a
{\it nearly AFM Fermi liquid} (NAFL) has been elaborated (see
e.g. Monthoux and Pines 1994). Here one assumes that at
low-frequencies $\chi''(\vec{q},\omega)\propto \omega$ at all
$\vec{q}$, as expected in a LFL, but with strongly enhanced
fluctuations near the AFM wavevector $\vec{Q}=(\pi,\pi)$,
corresponding to the critical slowing down in the proximity of a
phase transition to the AFM-ordered state. The following form has been
proposed, which can be derived also via the self-consistent paramagnon
theory (Moriya {\it et al.} 1990),
\begin{equation}
\chi(\vec{q},\omega)=\frac{\chi_{\vec Q}}{ 1+
\xi^2 |\vec{q}-\vec{Q}|^2 - i \omega/\omega_{SF}}\;.
\label{cp1}
\end{equation}
In the proposal by Millis {\it et al.} (1990), originally devoted to
the interpretation of the NMR and NQR relaxation, the main
$T$-dependence is expected to arise from the AFM correlation length
$\xi$, which is assumed to show a critical behaviour as $T$ is
decreased, i.e.  $\xi^2\propto T_x/(T+T_x)$. In order to explain the
anomalous NQR relaxation $1/T_1(T) \sim const$ (Imai {\it et al.}
1993), contradicting the Korringa law, strong $T$-dependence of $\xi$
is essential with $T_x \sim 100~K$, as well as low $\omega_{SF}$ and
large $\chi_{\vec Q}$ for $T\sim T_c$ (Monthoux and Pines 1994).  The
form of spin fluctuations is the basis for further investigations of
the charge dynamics, strongly coupled to spin degrees of
freedom. The calculated response functions such as the conductivity
appear also anomalous, e.g. the resistivity is close to a linear law
$\rho \propto T$.

There are several proposals, analogous to the NAFL in the basic idea
of {\it the proximity to a critical point}, enhancing fluctuations in
the optimum doping regime. One proposal invokes the quantum critical
scaling of the spin dynamics, established in the nonlinear sigma model
(Chakravarty {\it et al.} 1989), induced by doping the AFM (Sokol and
Pines 1993). Another scenario relates the quantum critical point to
a charge-density-wave instability (Castellani {\it et al.} 1995).

An alternative interpretation of experimental facts has been provided
by the concept of a marginal Fermi liquid (MFL) (Varma {\it et al.}
1989).  The hypothesis is that there exist excitations, contributing
both to the charge and spin response, which show in a broad range of
wavevectors $\vec q$ anomalous susceptibilities of the form
\begin{equation}
\chi''(\vec{q},\omega)\sim\left\{{C~ \omega/ T ~~~{\rm for}~~~
|\omega|<T, \atop C~{\rm sgn}\,\omega ~~~{\rm for}~~~
|\omega|>T}\right. \label{cp2}
\end{equation}
Due to the scattering on bosonic excitations with the spectrum
(\ref{cp2}) the single-particle self-energy $\Sigma(\vec k,\omega)$ is
also anomalous.  Assuming its $\vec k$-independence in a broad range,
it has been postulated using phenomenological arguments (Littlewood
and Varma 1991),
\begin{equation}
\Sigma(\omega)\sim \pi \lambda \left[2 \omega\ln\frac{x}{\omega_c}
-i x\right], \qquad x=\max(|\omega|,\pi T), \label{cp3}
\end{equation}
where $\omega_c$ is a high-frequency cutoff. Hence the QP lifetime
$1/\tau \propto -\Sigma''(\omega)$ is anomalous, i.e.  $1/\tau(\omega)
\propto x$. It should be however mentioned that the Ansatz (\ref{cp3})
is not unique and also modified forms have been invoked, e.g. in the
analysis of the optical conductivity (El Azrak {\it et al.} 1994,
Baraduc {\it et al.} 1995) better fit has been obtained with
\begin{equation}
-\Sigma''(\omega)\sim \pi \lambda ( |\omega| +\pi T).
\label{cp3a}
\end{equation}
While the FS should remain well defined with the volume equal to that
of free fermions, the corresponding QP weight $\tilde Z$ at the FS,
given by
\begin{equation} 
\tilde Z^{-1} = 1- \partial \Sigma'(\omega)
/ \partial\omega |_{\omega =0}, \label{cp4}
\end{equation}
vanishes on the FS in analogy to the case of a one-dimensional
Luttinger liquid (Haldane 1981).  The MFL concept accounts for several
remarkable properties at the optimum doping, as the anomalous
resistivity $\rho(T)\propto T$, the optical conductivity
$\sigma(\omega) \propto 1/\omega$, the NMR and the NQR relaxation rate
$1/T_1(T) \sim const$.  Note that the only low-energy scale within the
MFL scenario, equations (\ref{cp2}) - (\ref{cp3a}), is given by
$T$. Although there are certain similarities to the NAFL and other
critical-point scenarios, the essential difference of the MFL concept
is a non-critical $\vec q$ dependence. Hence the critical behaviour
within the MFL is rather a local one.  The attempts to derive MFL
behaviour from a microscopic model have not been successful so far.
 
\subsection{Models of correlated electrons in cuprates}

A similarity of electronic properties in a wide class of different
cuprates serves as a strong indication that the appropriate
microscopic model should be quite universal and must in first place
describe the electrons restricted to CuO$_2$ orbitals within a single
plane.  There has been quite an extensive effort put into finding a
proper model, and at present there seems to be a wide consensus
on its main features. This should be contrasted to various other
materials with interacting electrons, e.g. heavy fermions and 1D
conductors, where microscopic models are much less known.

Since the physics of electrons in CuO$_2$ planes is governed by Cu
$3d_{x^2-y^2}$ orbitals and O $2p_{x,y}$ orbitals (see the structure
in Fig.~\ref{1.1}), quite a complete model seems to be the three-band
Hubbard model (Emery 1987), describing fermion carriers (holes) added
to closed $3d^{10}$ and $2p^6$ shells. Parameters are the Cu--O
hopping $t_{pd}$, the direct O--O hopping $t_{pp}$, the on-site
energies $\varepsilon_d$, $\varepsilon_p$, and the corresponding
Coulomb repulsions $U_d$, $U_p$ on Cu and O sites,
respectively. Parameters correspond to the charge-transfer regime with
$\Delta =\varepsilon_p-\varepsilon_d>t_{pd}$ and $U_d >
\Delta$. The reference (insulator) material contains one fermion/cell,
entering predominantly the $d$ orbitals. Due to $U_d\gg t_{pd}$, a large
charge gap $\sim 2$eV opens at half filling, while spin degrees can be
mapped on the isotropic $S=1/2$ Heisenberg model, first proposed in
connection with cuprates by Anderson (1987). Holes added by doping
enter the singlets, introduced by Zhang and Rice (1988), which
can be in a fermion model treated as empty sites (holes).  Such a
reduction, confirmed also with other analytical approaches (Zaanen and
Ole\'s 1988, Ram\v sak and Prelov\v sek 1989) and cluster methods
(Hybertsen {\it et al.} 1990), leads to a single-band $t$-$J$ model
(Rice 1995),
\begin{equation}
H=-t\sum_{\langle ij\rangle  s}(\tilde{c}^\dagger_{js}\tilde{c}_{is}+
{\rm H.c.})+J\sum_{\langle ij\rangle} (\vec{S}_i\cdot \vec{S}_j -
{1\over 4} n_i n_j), \label{cm1}
\end{equation}
describing fermions in a tight-binding band with the hopping parameter
$t \propto t_{pd}^2/\Delta$. Here
$\vec{S}_i=(1/2)\sum_{ss'}c^\dagger_{is}\vec{\sigma}_{ss'}c_{is'}$ are
the local spin operators interacting with the exchange parameter
$J\propto t^4_{pd}/U_d \Delta<t$.  Due to the strong on-site repulsion
states with doubly occupied sites are explicitly forbidden and we
are dealing with projected fermion operators
$\tilde{c}_{is}=c_{is}(1-n_{i,-s})$.

By explicitly projecting out states with doubly occupied sites, the
$t$-$J$ model only allows for charge fluctuations in terms of a hole
motion, while at half-filling it is equivalent to the $S=1/2$
Heisenberg model.  The $t$-$J$ model is expected to capture
the essential low-energy physics of doped AFM as well of cuprates
in the whole regime of dopings. The challenging regime of the model is
the one of strong correlations with $J<t$.

The $t$-$J$ model is the simplest model which describes the interplay
of magnetism and itinerant metallic properties of cuprates.  A more
rigorous reduction of the three-band model (Zaanen and Ole\'s 1988,
Ram\v sak and Prelov\v sek 1989) leads to additional terms, which
could be as well represented within a reduced space without doubly
occupied sites. Among possible generalizations most attention has been
recently devoted to the addition of the next-nearest-neighbour
(n.n.n.)  hopping ($t'$) terms, emerging from the $t_{pp}$ hopping in
the three-band model,
\begin{equation}
H_{t'}=-t'\sum_{\langle \langle ij\rangle \rangle s} (\tilde
c^\dagger_{js}\tilde c_{is} + {\rm H.c.}), \label{cm2}
\end{equation}
representing the hopping along the diagonal in
Fig.~\ref{1.1}a. Analogous is the $t''$ term for the n.n.n. hopping
along each axis. It seems necessary to include such a term to account
for the QP dispersion found in ARPES experiments on undoped material,
such as Sr$_2$CuO$_2$Cl$_2$ (Wells {\it et al.} 1995). In spite of
their apparent smallness, $t'$ and $t''$ terms could lead to relevant
corrections since they allow a free propagation of fermions even in an
AFM (N\'eel) spin background.

The $t$-$J$ model, as relevant for cuprates, should be considered on a
planar square lattice. Parameters are rather well known (Rice
1995). $J$ is measured in the undoped AFM via the inelastic neutron
scattering and the magnon dispersion, leading to $J\sim 0.13$~eV. The
hopping parameter $t$ is not accessible directly, but cluster
calculations (Hybertsen {\it el al.} 1990) and other considerations
(Rice 1995) allow only for a narrow range of $t$ values. In our
calculations we shall furtheron use (if not declared differently)
$t=0.4$~eV and $J/t =0.3$.A possible range for $t'$ (Hybertsen {\it et
al.} 1990) is more controversial, while in numerical studies (Tohyama
and Maekawa 1994, Nazarenko {\it et al.} 1995) values $-0.35 < t'/t
<-0.2$ are used.
    
Another prototype model for strongly correlated electrons is the
traditional Hubbard model (Hubbard 1963),
\begin{equation}
H=-t\sum_{\langle ij\rangle  s}(c^\dagger_{js} c_{is}+
{\rm H.c.}) +U\sum_i n_{i\uparrow}n_{i\downarrow}. \label{cm3}
\end{equation}
High-energy excitations of the Hubbard model could be different from
those of the charge-transfer regime in the three-band model, still it
is expected that the low-energy properties map well on those of the
$t$-$J$ model provided that $U\gg t$. In the following we shall mainly
consider $T>0$ properties of the $t$-$J$ model. It should be however
noted that prior to the introduction of the FTLM method most
calculations of $T>0$ properties have been performed for the Hubbard
model by applying QMC methods (Dagotto 1994).

\setcounter{equation}{0}\setcounter{figure}{0}
\section{Finite-Temperature Lanczos Method}

This chapter is devoted to the description of the FTLM which we
developed (Jakli\v c and Prelov\v sek 1994a) for studying correlated
systems at $T>0$ and which is used to obtain results described
furtheron.  The goal was to calculate $T>0$ properties in small model
systems and to find a method, comparable in efficiency to
g.s. calculations employing ED methods, used in the past decade
extensively in the study of correlated systems (Dagotto 1994).

Here we should stress  that the advantage of $T>0$ calculations
is twofold. It is evident that we are interested in static and
dynamical properties at nonzero $T$, in particular in their
$T$-variation.  On the other hand, the use of finite but small $T>0$
represents the proper approach to more reliable g.s. calculations in
small systems. Namely, it is well known that g.s. ED results, in
particular for dynamical quantities, are strongly influenced by finite
size artifacts. At $T>0$ the latter effects can to large extent
average out, leading to more macroscopic-like results. Still the
understanding of remaining finite-size restrictions is important for
the proper application of the method, as will be described in Sec.~3.7.
In Sec.~3.8. we put our approach in perspective with other methods
yielding $T>0$ results for models of correlated electrons. These
includes mainly various QMC methods and the high-$T$ expansion (HTE)
technique.

\subsection{Lanczos algorithm and matrix elements}

The scarcity of well-controlled analytical approaches to models of
strongly correlated electrons has stimulated the development of
computational methods. Conceptually the simplest is the ED method of
small systems.  In models of correlated electrons, however, one is
dealing with the dimension of the basis (number of basis states) which
grows exponentially with the size of the system. In the Hubbard model
there are 4 basis states for each lattice site, therefore the number
of basis states in the $N$-site system is $N_{st} \propto 4^N$. In the
$t-J$ model $N_{st}$ still grows as $\propto 3^N$.  In the ED of such
systems one is therefore representing operators with matrices
$N_{st}\times N_{st}$, which become large already for very modest
values of $N$. The helpful circumstance is that for most interesting
operators and lattice models only a small proportion of matrix
elements is nonzero within the local basis.  Then, the operators can
be represented by sparse matrices with $N_{st}$ rows and at most
$f(N)$ nonzero elements in each row. In this way memory requirements
are relaxed and matrices up to $N_{st} \sim 10^7$ are considered in
recent applications.  Finding eigenvalues and eigenvectors of such
large matrices is not possible with standard algorithms performing the
full diagonalization. One must instead resort to the power algorithms
(see Parlett 1980), among which the Lanczos algorithm (Lanczos 1950)
is one of the most widely known.

The Lanczos algorithm starts with a normalized random vector
$|\phi_0\rangle$ in the vector space in which the Hamiltonian operator
$H$ is defined. $H$ is applied to $|\phi_0\rangle$ and the resulting
vector is split in components parallel to $|\phi_0\rangle$, and
$|\phi_1\rangle$ orthogonal to it, respectively,
\begin{equation}
H|\phi_0\rangle=a_0 |\phi_0\rangle + b_1|\phi_1\rangle.
\label{fl1}
\end{equation}
Since $H$ is Hermitian, $a_0=\langle\phi_0|H|\phi_0\rangle$ is real,
while the phase of $|\phi_1\rangle$ can be chosen so that $b_1$ is
also real.  In the next step $H$ is applied to $|\phi_1\rangle$,
\begin{equation}
H|\phi_1\rangle=b_1'|\phi_0\rangle +a_1 |\phi_1\rangle + b_2|\phi_2\rangle,
\label{fl2}
\end{equation}
where $|\phi_2\rangle$ is orthogonal to $|\phi_0\rangle$ and
$|\phi_1\rangle$. It follows also $b_1'=\langle\phi_0|H|\phi_1\rangle
= b_1$. Proceeding with the iteration one gets in $i$ steps
\begin{equation}
H|\phi_i\rangle=b_i|\phi_{i-1}\rangle +a_i |\phi_i\rangle + 
b_{i+1}|\phi_{i+1}\rangle,\qquad 1\leq i \leq M. \label{fl3}
\end{equation}
By stopping the iteration at $i=M$ and putting the last coefficient
$b_{M+1}=0$, the Hamiltonian can be represented in the basis of
orthogonal Lanczos functions $|\phi_i\rangle$ as the tridiagonal
matrix $H_M$ with diagonal elements $a_i$ with $i=0\ldots M$, and
offdiagonal ones $b_i$ with $i=1\ldots M$.  Such a matrix is easily
diagonalized using standard numerical routines to obtain approximate
eigenvalues $\epsilon_j$ and the corresponding orthonormal
eigenvectors $|\psi_j\rangle$,
\begin{equation}
|\psi_j\rangle=\sum_{i=0}^M v_{ji} |\phi_i\rangle,\;\;\;j=0\ldots M.
\label{fl5}
\end{equation}
It is important to realize that $|\psi_j\rangle$ are (in general) not
exact eigenfunctions of $H$, but show a remainder
\begin{equation}
H|\psi_j\rangle-\epsilon_j|\psi_j\rangle= b_{M+1}v_{jM}|\phi_{M+1}\rangle.
\label{fl6}
\end{equation}
On the other hand it is evident from the diagonalization of $H_M$ that
matrix elements
\begin{equation}
\langle\psi_i|H|\psi_j\rangle=\epsilon_j\delta_{ij},\;\;\;i,j=0\ldots M
\label{fl7}
\end{equation}
are exact, without restriction to the subspace $L_M$.

If in the equation (\ref{fl3}) $b_{M+1}=0$, we have found a
$(M+1)$-dimensional eigenspace where $H_M$ is already an exact
representation of $H$.  This inevitably happens when $M=N_{st}-1$, but
for $M<N_{st}-1$ it can only occur if the starting vector is
orthogonal to some invariant subspace of $H$. This should not be the
case if the input vector $|\phi_0\rangle$ is random, without any
hidden symmetries.

The number of operations needed to perform $M$ Lanczos iterations
scales as $MN_{st}$. Numerically the Lanczos procedure is subject to
roundoff errors, introduced by the finite-precision arithmetics.  This
problem usually only becomes severe at larger $M>100$ (more than
needed to get accurate g.s. $|\psi_0\rangle$) and is seen in the loss
of the orthogonality of vectors $|\phi_i\rangle$.  It can be remedied
by successive reorthogonalization (and normalization) of new states
$|\phi'_i\rangle$, plagued with errors, with respect to previous
ones. However this procedure requires $\sim M^2N_{st}$ operations, and
can become computationally more demanding than Lanczos iterations
alone. This effect prevents one to use the Lanczos method e.g. to
tridiagonalize large matrices.

The identity (\ref{fl7}) already shows the usefulness of the Lanczos
method for the calculation of particular matrix elements. As an aid in a
further discussion of the Lanczos method we consider the
calculation of a matrix element
\begin{equation}
W_{kl}=\langle n|H^k B H^l A|n\rangle, \label{fe1}
\end{equation}
where $|n\rangle$ is an arbitrary normalized vector, and $A, B$ are
general operators. One can calculate this expression exactly by
performing two Lanczos procedures with $M=\max(k,l)$ steps. The first
one, starting with the vector $|\phi_0\rangle=|n\rangle$, produces the
subspace $L_M=\{|\phi_j\rangle,\;j=0\ldots M\}$ along with approximate
eigenvectors $|\psi_j\rangle$ and eigenvalues $\epsilon_j$. The second
Lanczos procedure is started with the normalized vector
\begin{equation}
|\tilde\phi_0\rangle=A|\phi_0\rangle/\sqrt{\langle\phi_0| A^\dagger
A|\phi_0\rangle}, \label{fe2}
\end{equation}
and results in the subspace $\tilde
L_M=\{|\tilde\phi_j\rangle,\;j=0\ldots M\}$ with approximate
$|\tilde\psi_j\rangle$ and $\tilde \epsilon_j$. We can now define projectors
\begin{equation}
P_m=\sum_{i=0}^m|\phi_i\rangle\langle\phi_i|,\;\;
\tilde P_m=\sum_{i=0}^m|\tilde\phi_i\rangle\langle\tilde\phi_i|, \label{fe3}
\end{equation}
which for $m=M$ can also be expressed as
\begin{equation}
P_M=\sum_{i=0}^M|\psi_i\rangle\langle\psi_i|,\;\;
\tilde P_M=\sum_{i=0}^M|\tilde\psi_i\rangle\langle\tilde\psi_i|.
\label{fe4}
\end{equation}
By taking into account definitions (\ref{fe3}), (\ref{fe4}) we show
that
\begin{equation}
H P_m=P_{m+1}HP_m=P_M H P_m, \qquad m<M. \label{fe5}
\end{equation}
Since in addition $|n\rangle=|\phi_0\rangle=P_0|\phi_0\rangle$
and $A|n\rangle\propto|\tilde\phi_0\rangle=P_0|\tilde\phi_0\rangle$,
by successive use of the first equality in (\ref{fe5}) we 
arrive at 
\begin{equation}
W_{kl}= \langle \phi_0|P_0HP_1H\ldots HP_kB\tilde P_l H\ldots\tilde P_1
H\tilde P_0 A|\phi_0\rangle. \label{fe6}
\end{equation}
Using the second equality in the equation (\ref{fe5}) and identities
$P_0|\phi_0\rangle=P_M|\phi_0\rangle$, 
$\tilde P_0 A|\phi_0\rangle=\tilde P_M A|\phi_0\rangle$
we can rewrite $W_{kl}$ as
\begin{equation}
W_{kl}=
\langle \phi_0|P_MHP_MH\ldots HP_MB\tilde P_MH\ldots\tilde P_M
H\tilde P_M A|\phi_0\rangle. \label{fe7}
\end{equation}
We note that the necessary condition for the equation (\ref{fe7}) is
$M\ge k,l$. We finally expand the projectors according to expressions
(\ref{fe4}) and take into account the orthonormality relation
(\ref{fl7}) for matrix elements, and get
\begin{eqnarray}
W_{kl}&=&
\sum_{i_0=0}^M\ldots\sum_{i_k=0}^M\sum_{j_0=0}^M\ldots\sum_{j_l=0}^M
\langle\phi_0|\psi_{i_0}\rangle\langle\psi_{i_0}|
H|\psi_{i_1}\rangle\ldots \langle\psi_{i_{k-1}}|H|\psi_{i_k}\rangle 
\nonumber \\
& &\times \langle\psi_{i_k}|B|\tilde\psi_{j_l}\rangle
\langle\tilde\psi_{j_l}|
H|\tilde\psi_{j_{l-1}}\rangle\ldots \langle \tilde\psi_{j_1}|H
|\tilde\psi_{j_0}\rangle
\langle\tilde\psi_{j_0}|A|\phi_0\rangle =\nonumber \\
&=&
\sum_{i=0}^M\sum_{j=0}^M\langle\phi_0|\psi_{i}\rangle\langle\psi_{i}|
B|\tilde\psi_{j}\rangle\langle\tilde\psi_{j}|A|\phi_0\rangle
(\epsilon_i)^k (\tilde \epsilon_j)^l. \label{fe8}
\end{eqnarray}
We have thus expressed the desired quantity in terms of the
Lanczos (approximate) eigenvectors, and eigenvalues alone.

\subsection{Dynamical response in the ground state}

Within the Lanczos algorithm the extreme (smallest and largest)
eigenvalues $\epsilon_i$, along with their corresponding
$|\psi_i\rangle$, are rapidly converging to exact eigenvalues $E_i$
and eigenvectors $|\Psi_i\rangle$. It is quite characteristic that
usually (for nondegenerate states) $M=30-60 \ll N_{st}$ is sufficient
to achieve the convergence to the machine precision of the g.s. energy
$E_0$ and the wavefunction $|\Psi_0\rangle$, from which various
static and dynamical correlation functions at $T=0$ can be evaluated.

After $|\Psi_0\rangle$ is obtained, the g.s. dynamic correlation
functions can be calculated within the same framework (Haydock {\it et
al.} 1972). Let us consider the autocorrelation function
\begin{equation}
C(t)=-i\langle\Psi_0|A^\dagger(t)A|\Psi_0\rangle
=-i\langle\Psi_0|A^\dagger e^{i(E_0-H)t}A|\Psi_0\rangle
\label{fd1}
\end{equation}
with the transform, 
\begin{equation}
\tilde C(\omega)=\int_0^\infty dt e^{i\omega^+t}C(t) =\langle\Psi_0|
A^\dagger \frac{1}{\omega^+ +E_0-H}A|\Psi_0\rangle. \label{fd2}
\end{equation}
where $\omega^+=\omega+i\epsilon,~\epsilon>0$.  To calculate $\tilde
C(\omega)$, one has to run the second Lanczos procedure starting with
the normalized function $|\tilde\phi_0\rangle$, equation (\ref{fe2}).
The matrix for $H$ in the new basis $\tilde L_M$, with elements
$\langle\tilde\phi_i|H|\tilde\phi_j\rangle=[\tilde H_M]_{ij}$, is
again a tridiagonal one with $\tilde a_i$ and $\tilde b_i$ elements,
respectively.  Terminating the Lanczos procedure at given $M$, one can
evaluate the $\tilde C(\omega)$ as a resolvent of the $\tilde H_M$
matrix which can be expressed in the continued-fraction form (Haydock
{\it et al.}  1972),
\begin{equation}
\tilde C(\omega)=\frac{\langle\Psi_0|A^\dagger A|\Psi_0\rangle}
{\omega^+ +E_0-\tilde a_0-{\displaystyle\frac{\tilde b_1^2}
{\omega^+ +E_0-\tilde a_1-{\displaystyle\frac{\tilde b_2^2}
{\omega^+ +E_0-\tilde a_2-\ldots}}}}}\;, \label{fd3}
\end{equation}
terminating with $\tilde b_{M+1}=0$, although other termination functions
have also been employed.

The spectral function $C(\omega)=-(1/\pi) {\rm Im} \tilde C(\omega)$
is characterized by frequency moments,
\begin{equation}
\mu_l=\int_{-\infty}^\infty \omega^l C(\omega) d\omega
=\langle\Psi_0|A^\dagger(H-E_0)^lA|\Psi_0\rangle,
\label{fd4}
\end{equation}
which are particular cases of the expression (\ref{fe1}) for
$B=A^\dagger$, $k=0$, and $|n\rangle=|\Psi_0\rangle$.  Using the
equation (\ref{fe8}) we can express $\mu_l$ for $l\le M$ in terms of
Lanczos quantities alone
\begin{equation}
\mu_l=\sum_{j=0}^M 
\langle\Psi_0|A^\dagger|\tilde\psi_j\rangle
\langle\tilde\psi_j|A|\Psi_0\rangle
(\tilde \epsilon_j-E_0)^l. \label{fd5}
\end{equation}
Hence moments are exact for given $|\Psi_0\rangle$ (as would be also
for any other starting $|n\rangle$) provided $l\leq M$.  The
corresponding approximation for $C(t)$, equation (\ref{fd1}), within
the restricted set of eigenfunctions $|\tilde\psi_j\rangle, j=0\ldots
M$, can be written at given $M$ (assuming $\tilde b_{M+1}=0$) as
\begin{equation}
C(t) = -i\sum_{j=0}^M 
\langle\Psi_0|A^\dagger|\tilde\psi_j\rangle
\langle\tilde\psi_j|A|\Psi_0\rangle e^{-i(\tilde \epsilon_j-E_0)t}. \label{fd6}
\end{equation}
Note that such $C(t)$ expanded as a series in $t$ (short-time
expansion) has exact $M$ terms, since the coefficients are just
moments $\mu_l$, equation (\ref{fd5}).

As a practical matter we note that
$|\langle\tilde\psi_j|A|\Psi_0\rangle|^2=\tilde v_{j0}^2$, hence no
matrix elements need to be evaluated within this approach.  In
contrast to the continued fraction (\ref{fd3}), the expression
(\ref{fd6}) allows also the treatment of more general correlation
functions $\langle B(t)A\rangle$, with $B\ne A^\dagger$.  In this case
the matrix elements $\langle\Psi_0|B|\tilde\psi_j\rangle$ have to be
evaluated explicitly.  From the above one sees that the Lanczos method
is very convenient to calculate the frequency moments as well as the
dynamical $C(\omega)$.  Certain ideas presented above will be used to
construct the algorithm for $T>0$, discussed in next subsections.

\subsection{High-temperature expansion}

The novel method for $T>0$ is based on the application
of the Lanczos iteration, reproducing correctly high-$T$ and
large-$\omega$ series. The method is then combined with the
reduction of the full thermodynamic trace to the random sampling. We
present these ingredients in the following.

We first consider the expectation value of the operator $A$ in the
canonical ensemble
\begin{equation}
\langle A\rangle=\sum_{n=1}^{N_{st}}\langle n|e^{-\beta H}A|n\rangle
\biggm/
\sum_{n=1}^{N_{st}}\langle n|e^{-\beta H}|n\rangle,
\label{fh1}
\end{equation}
where $\beta=1/k_BT$. A straightforward calculation of $\langle
A\rangle$ requires the knowledge of all eigenstates $|\Psi_n\rangle$
and corresponding energies $E_n$, obtained by the full diagonalization
of $H$,
\begin{equation}
\langle A\rangle=\sum_{n=1}^{N_{st}}e^{-\beta E_n}\langle \Psi_n|A|
\Psi_n\rangle \biggm/
\sum_{n=1}^{N_{st}} e^{-\beta E_n}, \label{fh1a}
\end{equation}
computationally accessible only for $N_{st} \sim 5000$. Instead let us
perform the HTE of the exponential $e^{-\beta H}$,
\begin{eqnarray}
\langle A\rangle &=& Z^{-1}
\sum_{n=1}^{N_{st}}\sum_{k=0}^\infty
\frac{(-\beta)^k}{k!}\langle n|H^kA|n\rangle, \nonumber\\
Z&=&\sum_{n=1}^{N_{st}}\sum_{k=0}^\infty
\frac{(-\beta)^k}{k!}\langle n|H^k|n\rangle. \label{fh2}
\end{eqnarray}
Terms in the expansion $\langle n|H^k A|n\rangle$ can be calculated
exactly using the Lanczos procedure with $M \geq k$ steps and with
$|\phi^n_0\rangle=|n\rangle$ as a starting function, since this is a
special case of the expression (\ref{fe1}). Using the relation
(\ref{fe8}) with $l=0$ and $B=1$, we get
\begin{equation}
\langle n|H^k A|n\rangle=
\sum_{i=0}^M\langle n|\psi^n_{i}\rangle\langle\psi^n_{i}|
A|n\rangle (\epsilon^n_i)^k. \label{fh3}
\end{equation}
Working in a restricted basis $k\leq M$, we can insert the expression
(\ref{fh3}) into sums (\ref{fh2}), extending them to $k >M$.
The final result can be expressed in analogy to the equation
(\ref{fd6}) as
\begin{eqnarray}
\langle A \rangle &\approx& Z^{-1}\sum_{n=1}^{N_{st}}\sum_{i=0}^M
e^{-\beta \epsilon^n_i}\langle n|\psi^n_i\rangle\langle\psi^n_i|A|n
\rangle, \nonumber \\
Z &\approx& \sum_{n=1}^{N_{st}}\sum_{i=0}^M e^{-\beta
\epsilon^n_i}\langle n|\psi^n_i\rangle\langle\psi^n_i|n
\rangle, \label{fh4}
\end{eqnarray}
and the error of the approximation is of the order of 
$\beta^{M+1}$.

Evidently, within a finite system the expression (\ref{fh4}), expanded
as a series in $\beta$, reproduces exactly the HTE series to the order
$M$. In addition, in contrast to the usual HTE, it becomes (remains)
exact also for $T\to 0$. Let us assume for simplicity that the
g.s. $|\Psi_0\rangle$ is nondegenerate.  For initial states
$|n\rangle$ not orthogonal to $|\Psi_0\rangle$, already at modest
$M\sim 50$ the lowest function $|\psi^n_0\rangle$ converges to
$|\Psi_0\rangle$. We thus have for $\beta \to \infty$,
\begin{eqnarray}
\langle A\rangle
&=&\sum_{n=1}^{N_{st}}
\langle n|\Psi_0\rangle\langle\Psi_0|A|n\rangle\bigg/
\sum_{n=1}^{N_{st}}\langle n|\Psi_0\rangle\langle\Psi_0|n\rangle =\nonumber \\
&=&\langle\Psi_0|A|\Psi_0\rangle/\langle\Psi_0|\Psi_0\rangle,
\label{fh5}
\end{eqnarray}
where we have taken into account the completeness of the set $|n\rangle$.
Obtained result is just the usual g.s. expectation value of an operator.

\subsection{Large-frequency expansion at $T>0$} 

In order to calculate dynamical quantities, the HTE must be supplemented
by the high-frequency (short-time) expansion analogous to the one used
at $T=0$ in deriving the equation (\ref{fd6}) from (\ref{fd5}).  The goal
is to calculate the dynamical correlation function at $T>0$,
\begin{equation}
\langle B(t)A\rangle={\rm Tr}\left[e^{-\beta H}e^{iHt}Be^{-iHt}A\right]/
{\rm Tr}~e^{-\beta H}. \label{ff1}
\end{equation}
Expressing the trace explicitly and expanding the exponentials, we get
\begin{equation}
\langle B(t)A\rangle = Z^{-1}
\sum_{n=1}^{N_{st}}\sum_{k=0}^\infty\sum_{l=0}^\infty
\frac{(-\beta+it)^k}{k!}\frac{(-it)^l}{l!}
\langle n|H^kBH^lA|n\rangle. \label{ff2}
\end{equation}
Expansion coefficients in equation (\ref{ff2}) can be again obtained
via the Lanczos method, as discussed in Sec.~3.1.  Performing two
Lanczos iterations with $M$ steps, started with normalized
$|\phi^n_0\rangle=|n\rangle$ and $|\tilde\phi^n_0\rangle \propto
A|n\rangle$, respectively, we calculate coefficients $W_{kl}$
following the equation (\ref{fe8}), while $Z$ is approximated by the
static expression (\ref{fh4}). Extending and resumming series in $k$
and $l$ into exponentials, we get
\begin{equation}
\langle B(t)A\rangle \approx Z^{-1}
\sum_{n=1}^{N_{st}}
\sum_{i=0}^M\sum_{j=0}^M e^{-\beta \epsilon^n_i}
e^{it(\epsilon^n_i-\tilde \epsilon^n_j)}
\langle n|\psi^n_{i}\rangle\langle
\psi^n_{i}|B|\tilde\psi^n_{j}\rangle\langle\tilde\psi^n_{j}|A
|n\rangle.
\label{ff3}
\end{equation}

We check again a nontrivial $T=0$ limit of the above expression.  If
$|n\rangle$ are not orthogonal to the g.s. $|\Psi_0\rangle$, then for
large enough $M$ the lowest-lying state converges to $\epsilon^n_0\sim
E_0$ and $|\psi^n_0\rangle\sim|\Psi_0\rangle$, respectively.  In this
case we have in analogy to the equation (\ref{fh5})
\begin{equation}
\langle B(t)A\rangle\approx\sum_{n=1}^{N_{st}}\sum_{j=0}^M
e^{it(E_0- \tilde \epsilon^n_j)}
\langle\Psi_0|B|\tilde\psi^n_{j}\rangle
\langle\tilde\psi^n_{j}|A|\phi^n_0\rangle\langle\phi^n_0|
\Psi_0\rangle\bigg/\langle\Psi_0|\Psi_0\rangle. \label{ff4}
\end{equation}
Generally larger $M\gg 100$ are needed in order that relevant
higher-lying states $|\tilde\psi^n_j\rangle$ and $\tilde \epsilon^n_j$
become independent of $|n\rangle$.  Only in such a limit we recover
strictly the g.s. result, corresponding for $A^{\dagger} \to B$ to the
equation (\ref{fd2}).  Note however that similar restrictions apply to
the continued fraction expansion (\ref{fd3}) which reproduces
correctly moments $\mu_l$ (\ref{fd4}) up to $l<M$, but not necessarily
the details (e.g. positions and weights of peaks) of the $C(\omega)$
spectrum.

\subsection{Random sampling}

The computation of static quantities (\ref{fh4}) and dynamical ones
(\ref{ff3}) still involves the summation over the complete set of
$N_{st}$ states $|n\rangle$, which is not feasible in practice.  To
obtain a useful method, one further approximation must be made which
replaces the full summation by a partial one over a much smaller set
of random states (Imada and Takahashi 1986). Such an approximation
analogous to Monte Carlo methods is of course hard to justify
rigorously, nevertheless we can estimate the errors involved.

We consider the expectation value $\langle A \rangle$ at $T>0$, as
defined by the expression (\ref{fh1}).  Instead of the whole sum in
equation (\ref{fh1}) we first evaluate only one element with respect
to a random state $|r\rangle$, which is a linear combination of basis
states
\begin{equation}
|r\rangle=\sum_{n=1}^{N_{st}}\beta_{rn}|n\rangle, \label{fr1}
\end{equation}
i.e. $\beta_{rn}$ are assumed to be distributed  randomly.
Let us discuss then the random quantity
\begin{eqnarray}
\tilde A_r&=&\langle r|e^{-\beta H}A|r\rangle/
\langle r|e^{-\beta H}|r\rangle =\nonumber \\
&=&
\sum_{n,m=1}^{N_{st}}\beta^*_{rn}\beta_{rm}\langle n|e^{-\beta H}A|m\rangle
\biggm/
\sum_{n,m=1}^{N_{st}}\beta^*_{rn}\beta_{rm}\langle n|e^{-\beta H}|m\rangle.
\label{fr2}
\end{eqnarray}
We choose for convenience that here basis  states $|n\rangle$
correspond to the eigenstates of $H$.  We first assume in addition
that $[H,A]=0$, to diagonalize simultaneously both $A$ and
$H$. Then we have
\begin{equation}
\tilde A_r=
\sum_{n=1}^{N_{st}}|\beta_{rn}|^2\langle n|e^{-\beta H}A|n\rangle
\biggm/
\sum_{n=1}^{N_{st}}|\beta_{rn}|^2\langle n|e^{-\beta H}|n\rangle.\label{fr3}
\end{equation}
We can express $|\beta_{rn}|^2=1/N_{st}+\delta_{rn}$, where random
deviations $\delta_{rn}$ are not correlated with matrix elements
$\langle n|e^{-\beta H}|n\rangle=Z_n$ and $\langle n|e^{-\beta
H}A|n\rangle=Z_n A_n$.  It is then easy to see that $\tilde A_r$ is
close to $\langle A\rangle$, and the statistical deviation is 
related to the effective number of terms $\tilde Z$ in the
thermodynamic sum, i.e.
\begin{equation}
\tilde A_r = \langle A\rangle +{\cal O}(1/\sqrt{\bar Z}),\qquad
\bar Z=e^{\beta E_0}\sum_n Z_n.
\label{fr4}
\end{equation}
Note that for $T\to \infty$ we have $\bar Z\to N_{st}$ and therefore
at large $N_{st}$ a close estimate of the average (\ref{fr4}) can be
obtained from a single random state (Imada and Takahashi 1988, Silver
and R\"oder 1994). On the other hand, at finite $T<\infty$ the
statistical error of $\tilde A_r$ increases with decreasing $\bar
Z$. Still, strictly at $T=0$ and for a nondegenerate g.s. we obtain
from equation (\ref{fr3}) again the correct result.

In the FTLM we replace the full summation in the expression
(\ref{fh1}) with a restricted one over several random vectors
$|r\rangle$, $r=1\ldots R$,
\begin{equation}
\tilde A=
\sum_{r=1}^R\langle r|e^{-\beta H}A|r\rangle\biggm/
\sum_{r=1}^R\langle r|e^{-\beta H}|r\rangle \nonumber. \label{fr5}
\end{equation}
From equations (\ref{fr3}) and (\ref{fr4}) it follows that the
statistical error is even reduced,
\begin{equation}
\tilde A = \langle A\rangle +{\cal O}(1/\sqrt{R\bar Z}).
\label{fr6}
\end{equation}

For a general $A$, not commuting with $H$, we have to consider also the
contribution of offdiagonal terms in the equation (\ref{fr2}).  Since
the phases of random coefficients $\beta_{rn}$ are randomly distributed,
we can expect that vectors $\beta_{rn}$ are approximately orthogonal,
\begin{equation}
\sum_{r=1}^R\beta_{rn}^*\beta_{rm}\sim {R\over N_{st}}(\delta_{nm} + 
\zeta_{nm}/\sqrt{R}), \label{fr7}
\end{equation}
where $|\zeta_{nm}| = {\cal O}(1)$. The relative contribution of
offdiagonal terms is then given by
\begin{equation}
w(R,N_{st})=\frac{1}{\sqrt{R}}\left|
\sum_{n,m=1\atop n\ne m}^{N_{st}}\langle n|e^{-\beta H}A|m\rangle
\zeta_{nm}\right|\biggm/
\sum_{n=1}^{N_{st}}\langle n|e^{-\beta H}A|n\rangle.
\end{equation}
It is not easy to estimate the ratio $w$ in general. First we note
that at $\beta \to 0$ we could choose $|n\rangle$ to diagonalize $A$,
so offdiagonal terms could be avoided anyhow (for a static operator
$A$). This is however not the case for low $T$.  If we assume that the
sign of $\zeta_{nm}$ is random and uncorrelated with matrix elements
$\langle n|e^{-\beta H}A|m\rangle$, we can put an upper bound
$w(R,N_{st}) < 1/\sqrt{R}$.  There are however several arguments,
e.g. operators $A$ are usually local leading to a sparse-matrix
representation, which seem to indicate a much smaller contribution of
offdiagonal terms.

To conclude, taking into account all assumptions mentioned, the
approximation $\tilde A$ (\ref{fr5}) should therefore yield a good
estimate of the thermodynamic average $\langle A\rangle$ at all
$T$. For low $T$ the error is expected to be of the order of ${\cal
O}(1/\sqrt{R})$, while for high $T$ the error is expected to scale
even as ${\cal O}(1/\sqrt{N_{st}R})$. Since arguments leading to these
estimates rely on several assumptions which are not easy to verify, it
is essential to test the method for particular cases.

\subsection{Implementation and tests}

We comment now on the practical implementation of the FTLM and present
few tests in order to get a quantitative estimate of approximations
mentioned above.  First we consider the calculation of static
quantities. Joining the HTE and the random sampling we approximate the
average of the operator $A$ as
\begin{eqnarray}
\langle A \rangle &\approx&
\frac{N_{st}}{ZR}\sum_{r=1}^{R}\sum_{j=0}^M
e^{-\beta \epsilon^r_j}\langle r|\psi^r_j\rangle\langle\psi^r_j|A|
r \rangle, \nonumber \\
Z &\approx& \frac{N_{st}}{R}\sum_{r=1}^{R}\sum_{j=0}^M e^{-\beta
\epsilon^r_j}|\langle r|\psi^r_j\rangle|^2. \label{fi1}
\end{eqnarray}
The sampling is over $R$ random states $|r\rangle=|\phi^r_0\rangle$,
which serve as initial functions for the $M$-step Lanczos procedure,
resulting in $M$ approximate eigenvalues $\epsilon^r_j$ with
corresponding eigenvectors $|\psi^r_j\rangle$.

For a general operator $A$ the calculation of eigenfunctions
$|\psi_j^r\rangle$ and corresponding matrix elements
$\langle\psi^r_j|A| r \rangle$ is needed. On the other hand, the
calculation effort is significantly reduced if $[H,A]=0$ and $A$
can be diagonalized simultaneously.  Then
\begin{equation}
\langle A \rangle \approx
\frac{N_{st}}{ZR}\sum_{r=1}^{R}\sum_{j=0}^M
e^{-\beta \epsilon^r_j}|\langle r|\psi^r_j\rangle|^2 A^r_j. \label{fi2} 
\end{equation}
In this case the evaluation of eigenfunctions is not necessary since
the element $\langle r|\psi^r_j\rangle=v_{j0}^r$, equation
(\ref{fl5}), is obtained directly from eigenvectors of the
tridiagonal matrix $H_M^r$.

Few remarks on the implementation are in order here. Already in 
usual g.s. Lanczos calculations the use of symmetries of the model
Hamiltonian is crucial in order to reduce computational and storage
requirements. This is even more important when using the FTLM where
the computational burden is increased due to the sampling and due to
the calculation of matrix elements. At $T>0$ in general all symmetry
sectors must be taken into account and these can differ significantly
regarding the number of basis states they contain. Formulas
(\ref{fi1}), (\ref{fi2}) must then be generalized to allow for the
varying number of samples in each sector, so that sectors containing
more states are more thoroughly sampled. If in the symmetry sector $s$
containing $N_{st}^s$ basis states $R_s$ samples are evaluated, then
the random sampling summation is modified as
\begin{equation}
\frac{N_{st}}{R}\sum_{r=1}^R\longrightarrow
\sum_s \frac{N^s_{st}}{R_s}\sum_{r=1}^{R_s}. \label{fi3}
\end{equation}
Usually we choose $R_s\propto N_{st}^s$. The number of Lanczos steps can
also be taken as sector dependent, $M_{s}\leq N^s_{st}$. This is
important in sectors with small dimensions $N^s_{st}$.

Calculations on finite systems can be carried out on different
lattices with various boundary conditions. For planar problems it is
convenient to use tilted square lattices of sizes $N=n^2+m^2$ (Oitmaa
and Betts 1978) with periodic boundary conditions (p.b.c.). The
translational invariance of lattice Hamiltonians is preserved on such
systems, which makes crystal momenta $\vec{k}$ good quantum numbers
and enables to reduce the basis of states and their dimension
$N^s_{st}$.

\begin{figure}
\centering
\iffigure
\mbox{
\subfigure[]{
\epsfig{file=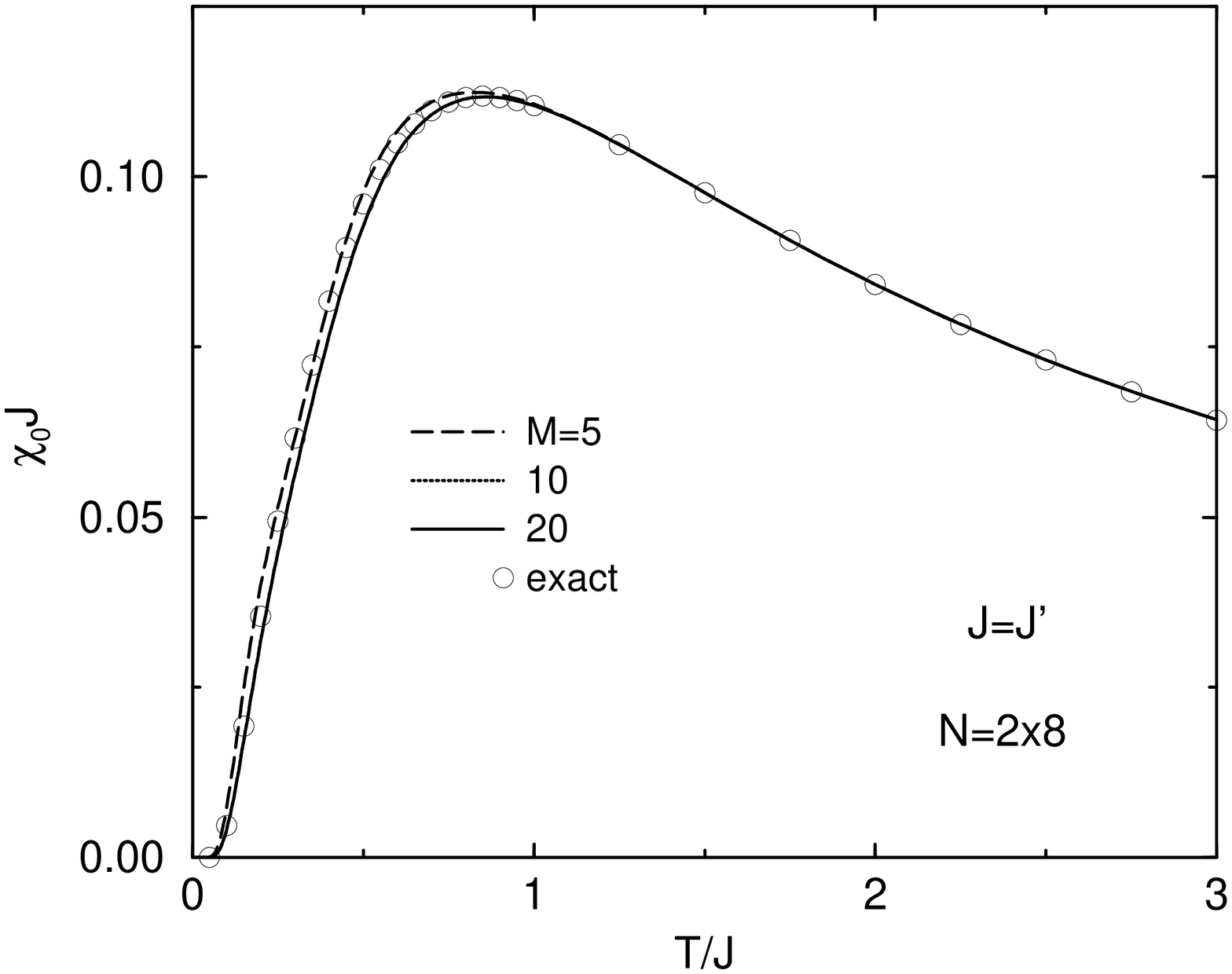,height=8cm,angle=-90}}
\quad
\subfigure[]{
\epsfig{file=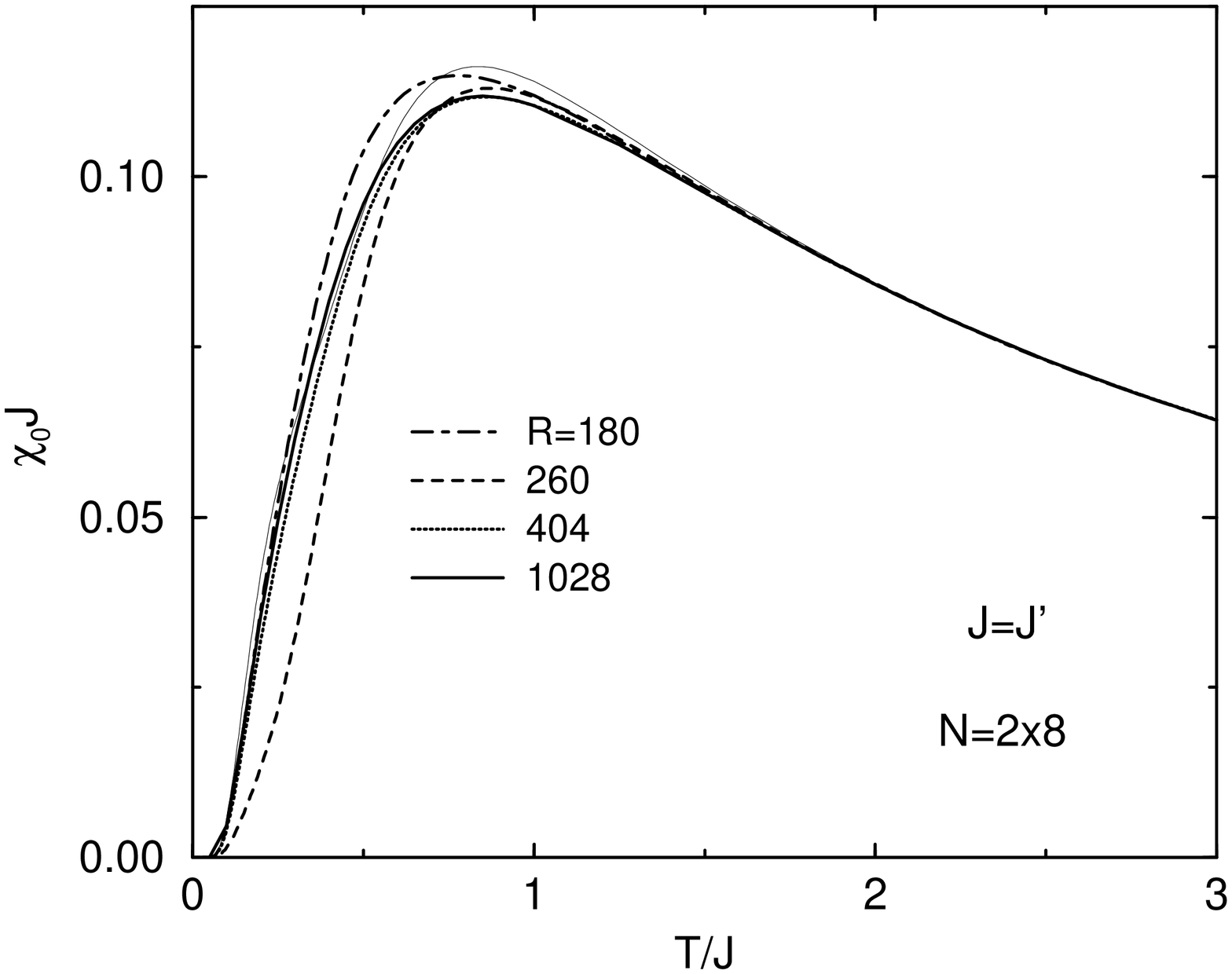,height=8cm,angle=-90}}}
\fi
\caption{
Uniform susceptibility $\chi_0$ vs. $T$ for the $2\times 8$-site
Heisenberg spin ladder, as obtained using the FTLM with a) fixed
sampling $R=1028$ and various number of Lanczos steps $M$, and b)
fixed $M=30$ and different number of random samples $R$. Exact results
are taken from Barnes and Riera (1994).  }
\label{3.1}
\end{figure}

Let us test the method for static quantities on the problem of a
Heisenberg model on a two-leg ladder. We discuss the case $J=J'$
(exchange equal along and perpendicular the ladder) which exhibits the
spin gap. One of the most interesting quantities in this system is the
uniform susceptibility $\chi_0=\langle (S_z^{tot})^2\rangle/T N$
(defined and discussed in more detail for doped AFM in Sec.~6.2) and
its $T$ dependence. As a test we choose the model on a $N=2\times 8$
ladder, for which the exact results obtained by full diagonalization
are available (Barnes and Riera 1994).  In Fig.~\ref{3.1}a we study the
influence of the number of Lanczos steps $M$ on the accuracy of
results. These are shown for fixed sampling $R=1028$, while $M$ is
varied from 5 to 20. It is rather surprising that even with the
smallest $M=5$ in the largest symmetry sector we obtain very good
agreement with the exact result not only at high $T>J$, but as well as
at low $T<J$.  In this case this is likely to be due to the gap in the
energy spectrum, as expected for ladders with even number of legs.

In Fig.~\ref{3.1}b we fix $M= 30$, while the number of random samples
$R$ varies. Note that using the translational symmetry and the
conservation of total $S_z^{tot}=S_z$ the maximum number of states in
a symmetry sector is $N_{st}^s=820$, while the total number of states
is $N_{st}=2^{16}=65536$. The number of samples within each symmetry
sector $R_s$ is chosen to be proportional to the number of basis
states in the sector. In the largest sector $N_{st}^s=820$ this
amounts to $R_s=2 - 13$ for the cases with total $R=180 - 1028$,
respectively. This corresponds to the sampling in the range
$R/N_{st}=0.003-0.016$.  We first observe that almost regardless of
the sampling results agree closely with the exact one at higher
$T>J$, as expected from the discussion of the statistical error in the
equation (\ref{fr6}).  At $T<J$ results start to disagree, still $R_s
\gg 1$ leads also to an improved accuracy in this regime.

Let us turn now to the calculation of dynamical quantities.  By joining
equation (\ref{ff3}) with the random sampling (\ref{fr5}), we get the
frequency-dependent correlation function 
\begin{equation}
\langle B(t)A\rangle_{\omega} \approx \frac{\pi N_{st}}{ZR}\sum_{r=1}^R
\sum_{i=0}^M\sum_{j=0}^M e^{-\beta \epsilon^r_i} 
\delta(\omega - \epsilon^r_i+\tilde \epsilon^r_j)
\langle r|\psi^r_{i}\rangle\langle
\psi^r_{i}|B|\tilde\psi^r_{j}\rangle\langle\tilde\psi^r_{j}|A
|r\rangle. \label{fi4}
\end{equation}
The sampling is over $R$ random states $|r\rangle$, resulting
in $M$ approximate eigenfunctions $|\psi^r_i\rangle, |\tilde
\psi^r_j\rangle$, and corresponding $\epsilon_i^r, \tilde
\epsilon_j^r$, respectively.

At the full sampling $R=N_{st}$ and within the chosen system the
number of Lanczos steps $M$ determines the number of exact frequency
moments $\mu_l$, analogous to g.s. moments (\ref{fd5}). This is
evident from the expansion (\ref{ff2}) at least for $\beta \to 0$,
while for lower $T$ we are dealing with a double expansion,
i.e. $\beta$ and $t$ series, and combined moments are exact. It
follows that at least at high $T$ the frequency resolution in spectra
is $\Delta \omega \sim \Delta E/M$, $\Delta E$ representing
typically the energy span of the model.  Since the information content
in higher moments is limited, in particular due to finite-size
effects, there is no point in using very large $M$, hence we restrict
our calculation in most cases to $M<200$.  The effects of using a
reduced sampling $R\ll N_{st}$ are expected to be most pronounced at
low $T$ where moreover only a minority of symmetry sectors with the
lowest energies contributes.

\begin{figure}[ht]
\centering
\iffigure
\epsfig{file=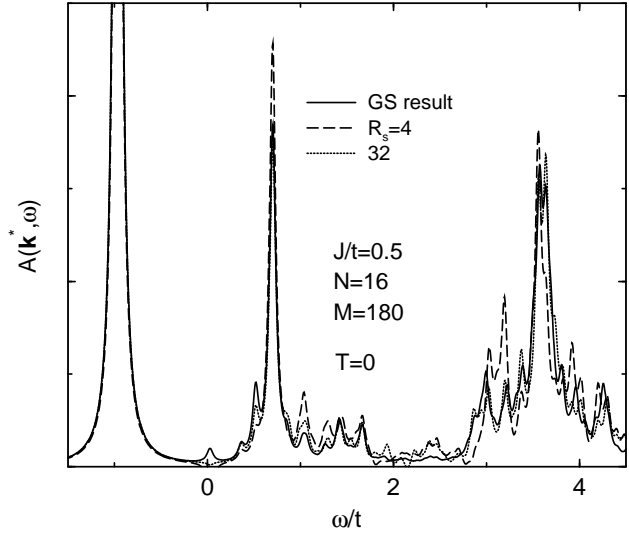,height=10cm,angle=-90}
\fi
\caption{
Spectral function $A(\vec k^*,\omega)$ of a single hole in an AFM at
$T=0$. The g.s. ED result and the FTLM results for different
random sampling $R_s$ are shown, at fixed $M=180$.  }
\label{3.2}
\end{figure}

Several tests for dynamical quantities within the $t$-$J$ model
(\ref{cm1}) have been already presented by Jakli\v c and Prelov\v
sek (1994a, 1995c). Here we consider in addition the spectral function
of a single hole injected in the undoped AFM,
$A(\vec{k},\omega)\propto{\rm Re}
\langle c^\dagger_{\vec{k}s}(t)c_{\vec{k}s}(0)\rangle_\omega$, defined and
discussed in more detail in Sec.~7.2.  We choose the system of
$4\times 4$ sites with $J/t=0.5$ and $\vec{k}^*= (\pi/2,\pi/2)$
corresponding to the g.s. wavevector. We compare $T=0$ results, being
the most stringent test for the FTLM, with the g.s. ED results
(Stephan and Horsch 1990, Eder {\it et al.} 1994).  In Fig.~\ref{3.2}
we first show the convergence of the spectral function at $T=0$ with
the sampling $R_s$ (note that for $T=0$ only g.s. symmetry sectors
have to be considered) for a fixed number of Lanczos steps
$M=180$. Note that $M$ determines also the number of correct $T=0$
frequency moments.  We observe that the position of peaks in the
spectrum is mainly unaffected by the sampling. Low-$\omega$ peaks are
known to be quite accurate within the $T=0$ ED method, hence also
their positions within the FTLM. Their intensities are less reliable
at smaller sampling and also frequency moments are expected to have
larger errors.  However, by increasing $R_s$ the accuracy of peak
intensities is improved.

\begin{figure}
\centering
\iffigure
\epsfig{file=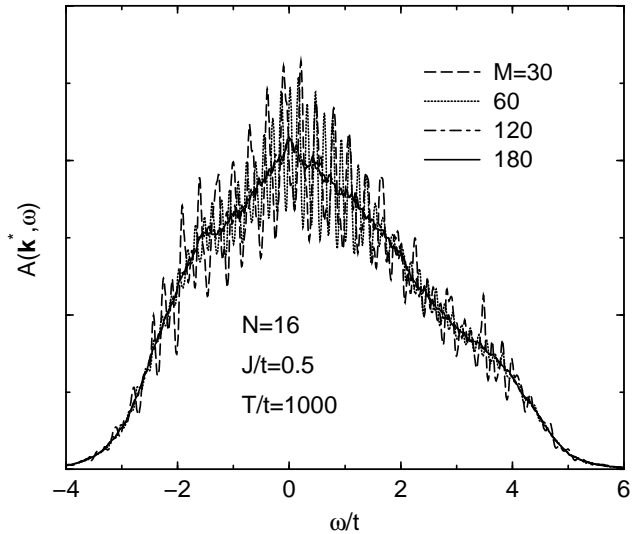,height=10cm,angle=-90}
\fi
\caption{
$A(\vec k^*,\omega)$ at large $T \gg t$. The dependence on $M$ is shown at
fixed $R=32$.  } \label{3.3}
\end{figure}

We investigate further the effect of reducing the number of Lanczos
steps $M$. In Fig.~\ref{3.3} the spectral function at $T \to\infty$ is
calculated with $R=32$ and varying $M$. We observe regular
oscillations for lowest $M =30, 60$, appearing in frequency
intervals $\Delta\omega \sim\Delta E/M$, where $\Delta E$ is the
maximum energy span in the model. We have only a partial explanation
of this phenomenon, typical for high $T$. While the Lanczos algorithm
obtains correct lowest and highest eigenvalues, Lanczos
eigenvalues in the middle of the spectrum do not have any
correspondence with true ones (as evident already from the discrepancy
in their number $M \ll N^s_{st}$) and appear almost equidistant.  At
$T \to \infty$ they all contribute and yield observed
oscillations. Since oscillations do not contain any relevant
information, they can be easily smoothened out by a suitable
filtering. They also become much less pronounced at lower $T$, where
the predominant contribution is given by transitions from the states
in the lower part of the spectrum to excited states.

\subsection{Finite size effects}

\begin{figure}
\centering
\iffigure
\epsfig{file=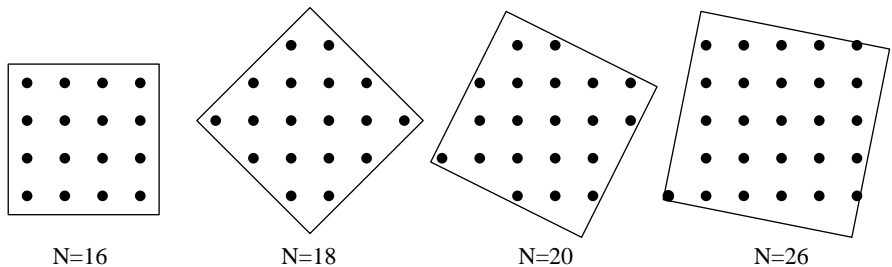,height=15cm,
bbllx=230,bblly=60,bburx=410,bbury=688,
angle=-90,clip=}
\fi
\caption{
Tilted square lattices with p.b.c. corresponding to different sizes.
} \label{3.4}
\end{figure}

We introduce and justify the FTLM as a method to calculate $T>0$
properties on small systems. We also argue that choosing
appropriate $M$, and the sampling $R$ one can reproduce exact results
to prescribed precision on a given system. In this sense the method is
very effective, in its computational effort comparable (although more
time and memory consuming) to g.s. ED calculations on the same system.
Still the well known deficiency of the ED method is the smallness of
available lattices. Hence it is important to understand the finite
size effects and their role at $T>0$. In the following we
predominantly study planar systems corresponding to the tilted
square lattice with p.b.c. (Oitmaa and Betts 1978) where $N=n^2+m^2$.
Mostly, we employ $N=16, 18, 20, 26$, as presented in Fig.~\ref{3.4}.

We claim that generally $T>0$ reduces finite size effects.  This is
related to the fact that at $T=0$ both static and dynamical quantities
are calculated only from one wavefunction $|\Psi_0\rangle$, which can
be quite dependent on the size and on the shape of the system. In
particular, g.s. spectra of dynamical quantities (see Dagotto 1994),
e.g. the optical conductivity (Sega and Prelov\v sek 1990) and the
single-particle spectral function (Stephan and Horsch 1991), quite
generally appear as a restricted number of delta functions. While
lowest frequency moments, in the sense of equations (\ref{fd4}), can
be quite representative of a large system, the peak-like structure
and details of spectra are mostly not.

$T>0$ introduces the thermodynamic averaging over a larger number of
eigenstates. This reduces directly finite-size effects for static
quantities, whereas for dynamical quantities spectra become
denser. From the equation (\ref{fi4}) it follows that we get in spectra at
elevated $T>0$ generally $RM^2$ different peaks leading to nearly
continuous spectra. This is also evident from high-$T$ result in
Fig.~\ref{3.3}, as compared to the $T=0$ result in Fig.~\ref{3.2}.
 
The effect of $T>0$ can be expressed also in another way.  There are
several characteristic length scales in the system of correlated
electrons, e.g. the AFM correlation length $\xi$, the transport mean
free path $l_s$, etc.  These lengths decrease with increasing $T$ and
results for related quantities have a macroscopic relevance provided
that the lengths become shorter than the system size, e.g. $l_s < L$
where $L$ is the linear size of the system. This happens for
particular $T> T_s$, where clearly $T_s$ depends also on the quantity
considered.  For certain quantities one can monitor such conditions
directly, e.g. for $l_s$ as discussed in more detail in Sec.~5.

As a criterion for finite size effects we use the characteristic
finite-size temperature $T_{fs}$. It is chosen so that in a given
system the thermodynamic sum
\begin{equation}
\bar Z(T)= {\rm Tr}e^{-\beta(H-E_0)} \label{fs1}
\end{equation}
is appreciable, i.e. $\bar Z(T_{fs})=Z^* \gg 1$.

\begin{figure}
\centering
\iffigure
\mbox{
\subfigure[]{
\epsfig{file=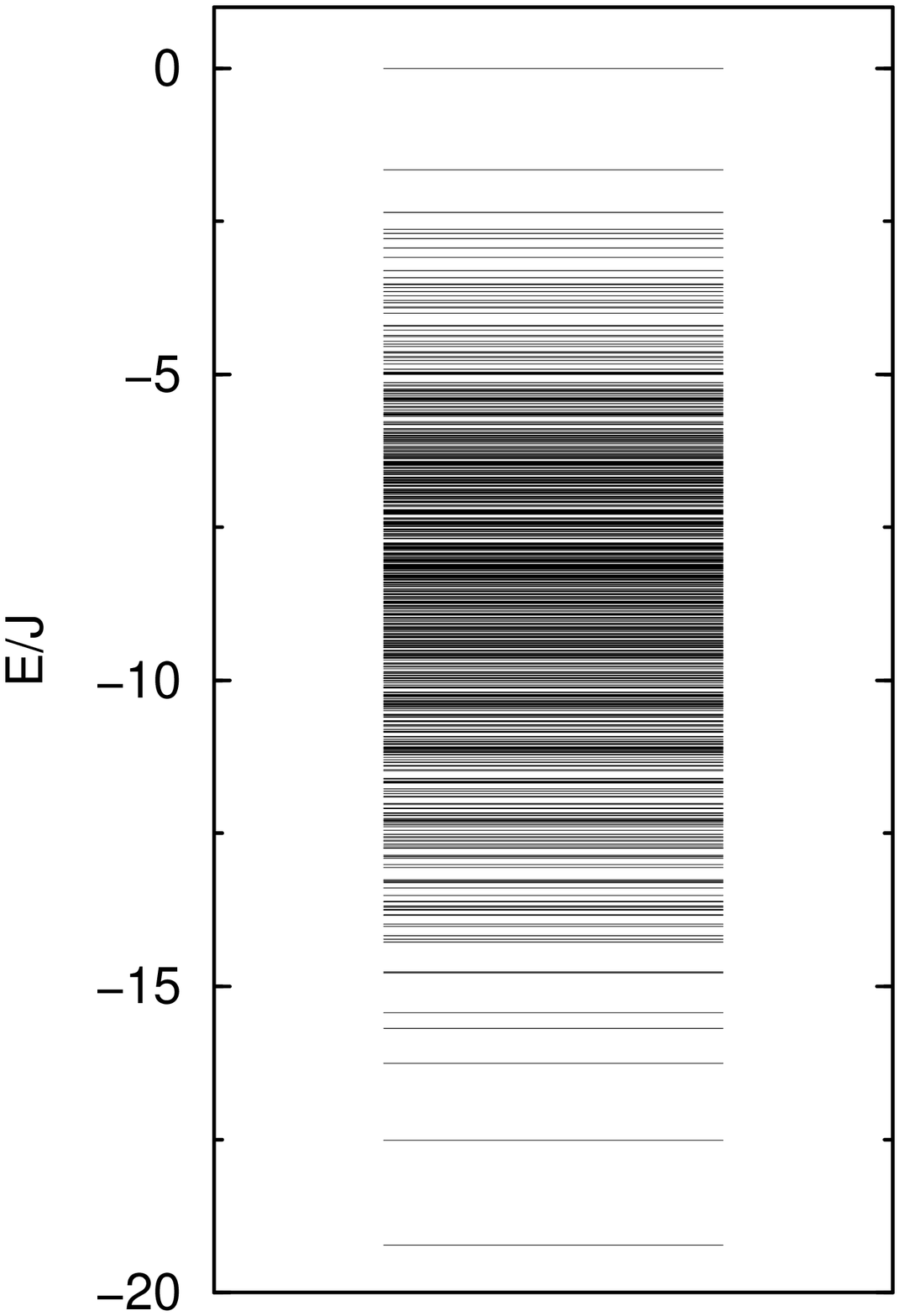,height=8cm}}
\quad
\subfigure[]{
\epsfig{file=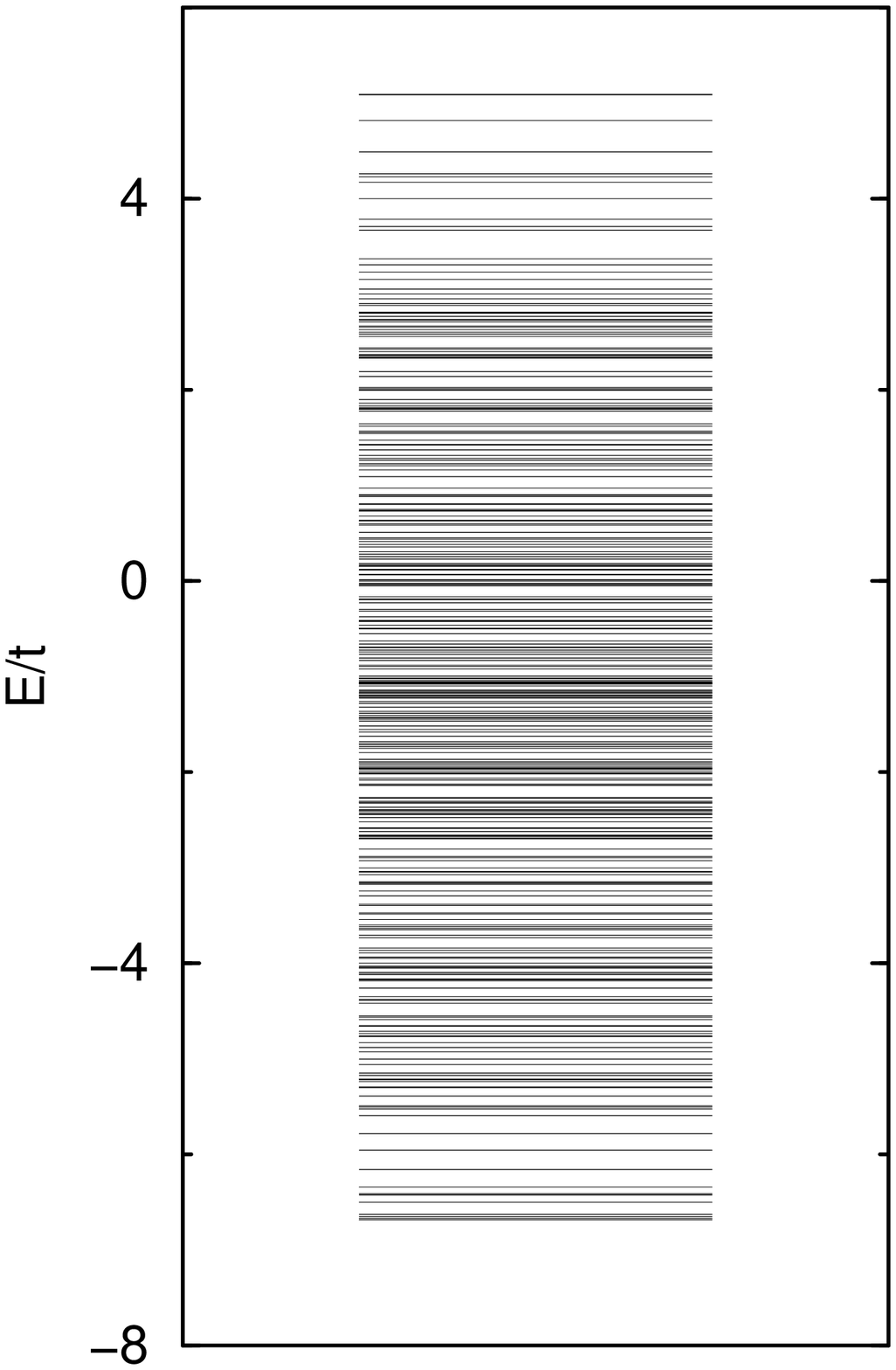,height=8cm}}}
\fi
\caption{
Many-body levels for: (a) the 2D Heisenberg model on $N=16$ sites, (b)
the 2D $t$-$J$ model with $N_h=2$ on $N=10$ sites. In both cases only
the $S_z=0, k=0$ sector is presented.  } \label{3.5}
\end{figure}

To get the size dependence of $T_{fs}$ it is important to understand
general features of many body spectra. In Fig.~\ref{3.5}a we present
the levels of the Heisenberg model on a square lattice with $N=16$
sites for the $S_z=0, k=0$ sector.  Note that energy span is $\Delta
E\propto NJ$ while the number of states scales as $N^s_{st} \propto
2^N$. This means that the density of states far from spectral edges
scales exponentially with $N$.  On the other hand, near the edges
spectra become sparse and the spacing between lowest levels decreases
rather slowly with the size, i.e. $\Delta \epsilon \propto N^{-p}$. We
expect that also $T_{fs}$ scales as $\Delta
\epsilon$. Still the character and the density of low-lying states can
change qualitatively from one regime to another. In Fig.~\ref{3.5}b we
show for comparison the levels within the $t$-$J$ model with $N_h=2$
holes on the tilted square lattice with $N=10$ sites, again only for
the $S_z=0, k=0$ sector.  While both cases have similar $N^s_{st}$, it
is evident that the low-energy regime shows much higher density of
states in doped $t$-$J$ model, Fig.~\ref{3.5}b, indicating a large
degeneracy of states at $E\agt E_0$.

\begin{figure}
\centering
\iffigure
\epsfig{file=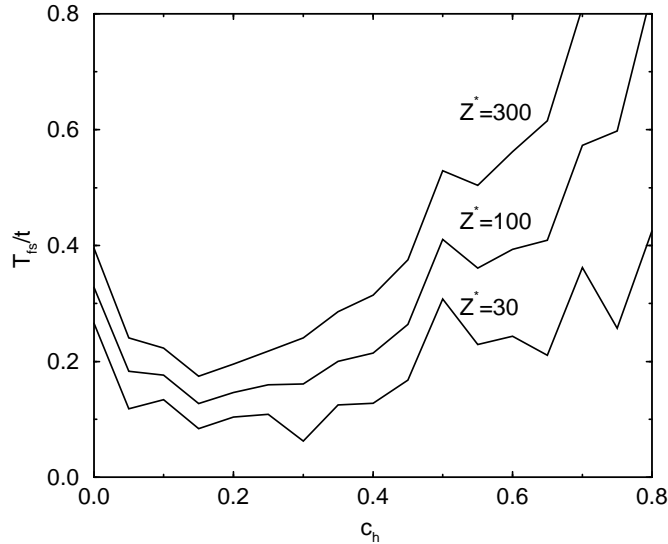,height=10cm,angle=-90}
\fi
\caption{
The variation of $T_{fs}$ with doping $c_h$ in the $t$-$J$ on $N=18$
sites with $J/t=~0.3$. Curves correspond to different choices of
$Z^*=\bar Z(T_{fs})$.  } \label{3.6}
\end{figure}

The FTLM is best suited just for quantum many-body systems with a
large degeneracy of states, i.e. large $\bar Z$ at low $T$.  This is
the case with doped AFM and the $t$-$J$ model in the strong
correlation regime $J<t$. To be concrete we present in Fig.~\ref{3.6}
the variation of $T_{fs}$ with the doping $c_h=N_h/N$, as calculated
from the system of $N=18$ sites and $J/t=0.3$.  For convenience we fix
$T_{fs}$ with criteria $Z^*=30, 100, 300$, respectively. It is
indicative that $T_{fs}$ reaches the minimum for intermediate
(optimum) doping $c_h =c^*_h \sim 0.15$, where we are able to reach
$T_{fs}/J \alt 0.3$. Away from such optimum case $T_{fs}$ is larger.
In the undoped (and underdoped) AFM this happens due to rather large
finite-size gaps in magnon excitations, while in the overdoped system
electrons behave closer to free electrons with a nearly unrenormalized
bandwidth and hence large gaps between single-electron excited
states. We claim that small $T_{fs}$ and related large degeneracy of
low-lying states are the essential features of strongly correlated
system in their most challenging regime, being a sign of a novel
quantum frustration. On the other hand this gives an advantage to the
FTLM which performs best where several other methods, like QMC, fail
due to the same frustration (sign) problems.
         
\subsection{Relation to other numerical methods}

It is not our intention to give an exhaustive overview of other
numerical (or partly analytical) methods which are used in the
analysis of the problem of strongly correlated electrons, and for
doped AFM in particular.  We mainly list below methods which are
alternative to the FTLM, together with their limitations and
advantages.  Since $T>0$ calculations are often used as a proper
approach to $T=0$ results, as is also the case for several quantities
evaluated within the FTLM furtheron, we comment also on some
g.s. calculations.

Clearly the closest relation is with the ED studies of small systems.
Apart from the FTLM only few studies of $T>0$ properties of the
$t$-$J$ model and of the Heisenberg model have been performed (Tohyama
{\it et al.}  1993, Sokol {\it et al.}  1993, Tsunetsugu and Imada
1997), restricted to small $N$ and in particular small $N^s_{st} \alt
1000$ due to the full diagonalization employed in these calculations.
We note that such restricted $N_{st}$ can influence in particular
dynamical spectra, which can show up finite size effects even at
higher $T$. We should also note that quite an analogous method to the
FTLM, based instead on the Chebyshev iteration, has been introduced
recently (Silver and R\"oder 1994), but has not been exploited much so
far.

Much more extensive ED calculations have been performed for g.s.
properties (see Dagotto 1994). While systems considered could be
possibly somewhat larger than those reachable within the FTLM, we
claim that in particular dynamical spectra as evaluated within the
g.s. should be interpreted with care due to their sparse
structure. This seems to be the case with the g.s. optical
conductivity, the spin structure factor, spectral functions etc., as
discussed in detail furtheron.

Closely related to the FTLM (see Sec.~3.3) is the HTE approach, which
in principle deals with a large (infinite) system and is one of few
straightforward (at least partly) analytical methods for correlated
systems.  It uses $\beta$ as a small parameter for the series
expansion, while appropriate extrapolations (e.g. using Pade
approximants) are needed to obtain results for $T$ in the physically
interesting regime. So far it has been used in several studies of the
$t$-$J$ model, e.g. to address the question of the phase separation
(Putikka {\it et al.} 1992), magnetic correlations (Singh and
Glenister 1992a), the momentum-distribution function (Singh and Glenister
1992b), the charge-spin separation (Putikka {\it et al.}  1994), the
Hall effect (Shastry {\it et al.} 1993) etc. The advantage of the
method is the absence of any finite cluster bounds. On the other hand
the method requires a careful and a nonunique extrapolation procedure
which is reliable only for static quantities.

Widely used QMC technique yields in general results for $T>0$.  There
are various methods, which have been covered in several recent reviews
(von der Linden 1992, Suzuki 1993).  We mention here only approaches
which are relevant for studies of planar undoped and doped AFM.  The
world-line QMC has been very successful in the evaluation of static
properties of the Heisenberg model (see Manousakis 1991). Analogous
results have been obtained via the QMC for the insulating Hubbard
model at half-filling (Hirsch 1985). Away from the half-filling the
sign problem becomes the major difficulty for the QMC studies of
fermionic models. It is particularly severe at low $T$ within the
intermediate-doping regime, being thus connected with the large
degeneracy of fermionic states.  Still various static quantities have
been evaluated within the Hubbard model as a function od doping both
for $T=0$ and $T>0$ (see Dagotto 1994).

The calculation of dynamical quantities within the QMC is possible via
the deconvolution of the imaginary-time dynamics into a real-frequency
one using the maximum entropy analysis (Jarrell {\it et al.}
1991). The latter appears to be quite delicate due a large influence
of statistical errors.  Still there has been in recent years several
studies of dynamic properties of spin systems (Makivi\'c and Jarrell
1992) and of the planar Hubbard model, in particular of spectral functions
(Bulut {\it et al.} 1994, Preuss {\it et al.} 1995, 1996).  In spite
of much larger systems reachable within QMC studies, it is well
conceivable that due to inherent difficulties QMC results for dynamics
are less reliable than those obtained within the FTLM. 
 
There are other powerful numerical methods which are only partly
relevant to studies of doped AFM. Particularly successful and
promising is the Density Matrix Renormalization Group (DMRG) approach,
as developed by White (1992) and extensively applied to problems of
correlated electrons. While designed mainly for 1D systems, it has
been extended to ladder systems (White and Scalapino 1997a) as well as
to planar models (White and Scalapino 1997b).  So far appropriate
generalizations in order to study the $T>0$ and dynamical properties
have only been attempted (Pang {\it et al.} 1996).

\setcounter{equation}{0}\setcounter{figure}{0}
\section{Thermodynamic properties}

We first consider thermodynamic properties of the $t$-$J$ model
(Jakli\v c and Prelov\v sek 1996).  These include quantities directly
derivable from the grand-canonical sum $\Omega$: free energy density
${\cal F}$, chemical potential $\mu$, charge compressibility
$\kappa$, entropy density $s$, specific heat $C_V$ etc.  Some
of these have been already studied using other methods.  Within the
HTE some thermodynamical quantities have been calculated within the
$t$-$J$ model. Results indicate the ferromagnetic phase at $J \ll
t$ and at low doping (Putikka {\it et al.}  1992), but also a large
enhancement of the entropy in a doped AFM (Putikka,
unpublished). Within the Hubbard model the projector ($T=0$) QMC
method has been employed to study the phase diagram and to calculate
the charge compressibility (Furukawa and Imada 1992). Several
calculations of the chemical potential vs. doping within the Green's
function QMC have been presented in order to establish the regime of
the phase separation (Kohno 1997, Hellberg and Manousakis 1997),
although with contradictory conclusions for the most interesting
regime $J<t$.
   
In order to study continuously varying particle densities, we
perform the averaging within the grand-canonical ensemble, involving
all possible numbers of electrons $N_e$,
\begin{equation}
\langle A\rangle = \sum_{N_e}{\rm Tr}_{N_e}
[e^{-\beta(H-\mu N_e)}A]/\sum_{N_e}{\rm Tr}_{N_e}e^{-\beta(H-\mu
N_e)},
\label{t1}
\end{equation}
where $\mu$ is the chemical potential.  For each $N_e$ the problem
thus reduces to the evaluation of the canonical thermal average, which
we achieve with the FTLM as described in Sec.~3.

The implementation of the FTLM can be further simplified for operators
$A$ which are conserved quantities, i.e. commute with $H$, as shown in
Sec.~3.6.  Examples include $H$ itself, the particle number $N_e$, the
total spin $S_z$ etc. By choosing random functions $|r\rangle$
to have good quantum numbers $N_e$ and $S_z$, we can evaluate the
expectation value of an arbitrary function $f(N_e,S_z,H)$
\begin{eqnarray}
\langle f\rangle &\approx& \frac{N_{st}}{R\Omega }\sum_{r=1}^{R}
\sum_{j=0}^{M}|\langle r|\psi_j^r \rangle|^2 f(N_e^r,S_z^r,\epsilon_j^r) 
e^{-\beta(\epsilon_j^r-\mu N_e^r)}, \nonumber \\
\Omega &\approx& \frac{N_{st}}{R}\sum_{r=1}^{R}
\sum_{j=0}^{M}|\langle r|\psi_j^r \rangle|^2
e^{-\beta(\epsilon_j^r-\mu N_e^r)}. \label{t2}
\end{eqnarray}
As noted in the equation (\ref{fi2}), in this case $|\langle
r|\psi_j^r\rangle|^2$ can be evaluated from the tridiagonal matrix
directly. Since also $M\alt 100$ is enough, the reorthogonalization of
Lanczos functions can be avoided. This eliminates the need to store
wavefunctions $|\phi_j^r\rangle$ and systems with considerably larger
$N_{st}$ can be studied.  Consequently the computational effort in
this case is equal to that of a g.s. Lanczos procedure, repeated $R$
times. We employ in the following typically $R\sim 200-1000$ in each
$N_e$ sector.  Calculations are performed on systems with $N=16$, $18$,
and $20$ sites for arbitrary filling $c_h$, while for the undoped
system we reach $N=26$ sites. Note also that we fix $J/t = 0.3$.

\subsection{Chemical potential}

\begin{figure}
\centering
\iffigure
\epsfig{file=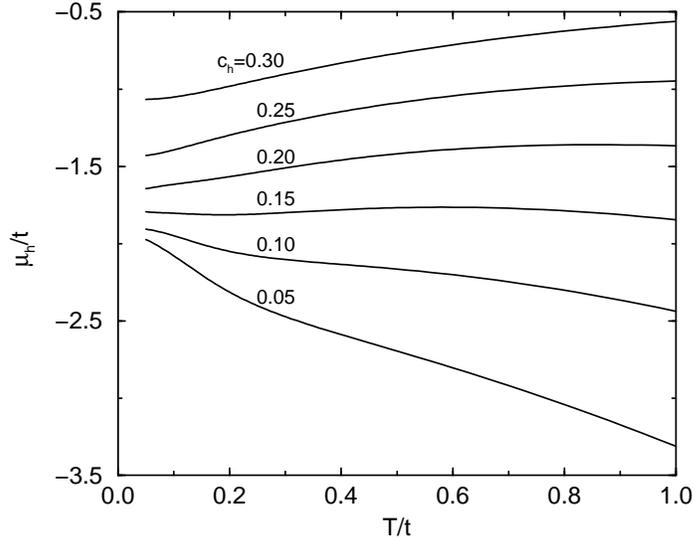,height=10cm,angle=-90}
\fi
\caption{
Hole chemical potential $\mu_h$ vs. $T$ at several dopings $c_h$.  }
\label{4.1}
\end{figure}

We first analyze the hole chemical potential $\mu_h=-\mu$ as a
function of $T$ and of the hole density $c_h=1-\langle
N_e\rangle/N$. Results are obtained by first calculating $c_h$ at
fixed $\mu, T$ from equations (\ref{t2}) with $f=N_e$, and then
inverting the dependence $c_h(\mu,T)$.  In Fig.~\ref{4.1} we present
curves $\mu_h(T)$ for several $c_h$. Note that for thermodynamic
quantities results seem somewhat less sensitive to finite-size
effects, hence we follow them in Fig.~\ref{4.1} to $T \sim 0.05~t$
(for $c_h\leq 0.1$ we would still estimate higher $T_{fs} \sim
0.1~t$). In order to interpret $\mu_h(T)$ at low $c_h \ll 1$ it is
essential to note that at $T=0$ the system contains no holes in the
equilibrium provided $\mu_h<\mu_h^0$. For chosen $J/t=0.3$ it has been
established $\mu_h^0 \sim -1.99~t$ (Dagotto 1994), related to the
minimum energy of a single hole added to the undoped AFM.

Analyzing Fig.~\ref{4.1} we mostly do not find a $T^2$ dependence of
$\mu_h$ at low $T$, as expected for a normal LFL, except within the
extremely overdoped regime $c_h\geq 0.3$. In particular, in a broad
range $0.05 <c_h <0.3$ we find a very unusual roughly linear
variation,
\begin{equation}
\mu_h(T)= \mu_h(T=0) + \alpha T, \qquad T_{fs}<T<J, \label{t2a}
\end{equation}
whereby the slope $\alpha$ changes the sign at $c_h = c_h^*
\agt 0.15$. It is remarkable that the marginal doping $c_h^*$ appears to
be quite system independent, as checked quantitatively for different
system sizes $N=16 - 20$.

\begin{figure}
\centering
\iffigure
\epsfig{file=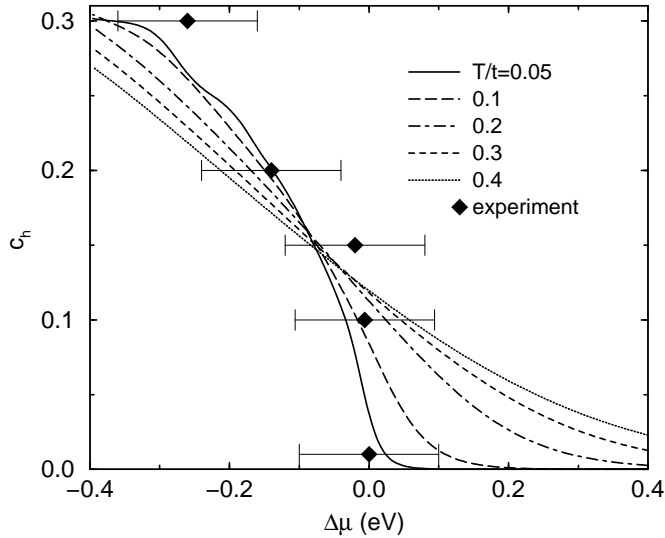,height=10cm,angle=-90}
\fi
\caption{
Hole concentration $c_h$ vs. $\Delta \mu$ at various $T$.  For
comparison we show experimental results for LSCO at low $T$, obtained
from the shift of photoemission spectra by Ino {\it et al.} (1997a).
}
\label{4.2}
\end{figure}

The marginal $c_h^*$ shows up again in Fig.~\ref{4.2}, displaying the
variation $c_h(\Delta \mu)$ at various temperatures $T$.  Here $\Delta
\mu= \mu+\mu_h^0$ is the difference to the undoped AFM case and is
displayed in eV to allow the comparison with experiments, using the
usual correspondence $t=0.4~$eV.  $c_h^*$ represents in this case the
crossing of curves at different $T$, whereby $\Delta\mu(T)$ is essentially
pinned at the value $\Delta\mu\sim -0.17~t$. This pinning is active
in a wide range of $T$.  Analyzing the regime $c_h<c^*_h$ in
Fig.~\ref{4.2}, we note again that $c_h(T\to 0)$ remains finite only
for $\mu_h<\mu_h^0$.  One can evaluate from these results the
compressibility of the hole fluid $\kappa
\propto -dc_h/d\mu$.  We find that $\kappa <\infty $ for all $T>0$,
indicating the absence of the phase separation in the system at chosen
$J/t$. This is in contrast with some recent QMC studies (Hellberg and
Manousakis 1997) claiming the phase separation in the $t$-$J$ model at
all $J/t$, as first put forward by Emery {\it et al.}  (1990).  Our
results in Fig.~\ref{4.2} do not support such a behaviour, at least
not in the range $T/t>0.05$.  Still Fig.~\ref{4.2} reveals that
$\kappa(T\to 0)$ is increasing and becoming larger on approaching
$\Delta \mu \sim 0$.

A proper interpretation of the regime $c_h \to 0$ at low $T$ is
clearly one of the major challenges. One frequently used picture is
that holes doped in an AFM could be described as degenerate fermions
with small FS - hole pockets (Trugman 1990, Eder and Becker 1991).  In
this case one would expect in a 2D lattice $\kappa = {\cal N}(\mu) =
1/2\pi t^*$, where ${\cal N}(\epsilon)$ is the single-particle (hole)
density of states (DOS) and $t^*$ an effective hopping
parameter. Since the hole effective mass can be quite enhanced,
i.e. $t^*\ll t$, $\kappa$ can become quite large.  Results in
Fig.~\ref{4.2} put a lower bound to the possible enhancement,
i.e. $t/t^*>5$.

Analyzing $T=0$ QMC results for the Hubbard model near half filling a
variation $c_h(\mu_h)$, similar to ours, has been found (Furukawa and
Imada 1992, Assaad and Imada 1996). Results were interpreted in terms
of a singular behaviour $\kappa \propto (\mu_h-\mu_h^0)^{-1/2}$. In
contrast to the hole-pocket picture, the latter form does not allow
for a regime with a degenerate gas of holes, at least not with holes
having a doping-independent mass. From our results it is hard to
exclude any of these scenarios, whereby the regime of hole pockets
should be in any case restricted to very low doping $c_h<0.1$.

It is tempting to interpret the existence of the marginal
concentration $c_h^*$ as a change of the character of the FS. To
establish the relation, we have to rely on arguments which apply
to the gas of noninteracting fermions. A simple Sommerfeld expansion
yields that the electron density at fixed $\mu$ is given by
\begin{equation}
c_e(T)=c_e(T=0)+\frac{(\pi k_BT)^2}{6} {\cal N}'(\mu). \label{t3}
\end{equation}
Indirectly this gives an information on the FS, since one would
plausibly associate ${\cal N}'(\mu)>0$ for $c_e\alt 1$ with a large
electron FS, and oppositely ${\cal N}'(\mu)<0$ with a hole-like FS or
small hole pockets vanishing for $c_e \to 1$.  At least for free
electrons it is easy to establish the connection of ${\cal N}'(\mu)$
with the curvature $K^{-1}$ of the FS (Jakli\v c and Prelov\v sek
1996), analogous to the relation for the Hall resistivity (Tsuji
1958). At least in the region of the $\vec{k}$ space, where the
effective-mass tensor is positive-definite, ${\cal N}'(\mu)>0$ implies
also that the average FS curvature $K^{-1}$ is positive.

The observed nonquadratic $T$ dependence in Fig.~\ref{4.1} questions
the interpretation in terms of the free-electron DOS (\ref{t3}).
Still we may interpret $dc_h/dT<0$ for $c_h>c_h^*$, as deduced from
Fig.~\ref{4.2}, as an indication for ${\cal N}'(\mu)>0$, i.e.
positive average curvature of the FS. This in turn implies a
transition at $c_h \sim c_h^*$ from a hole-pocket picture at low
doping (Trugman 1990, Eder and Becker 1991), to an electron-like large
FS (Stephan and Horsch 1991, Singh and Glenister 1992b).

Recently the variation of $\mu$ with the hole doping in LSCO has been
deduced experimentally from the shift of photoemission spectra by Ino
{\it et al.} (1997a).  For comparison we plot also these results in
Fig.~\ref{4.2}, noting that they apply to low $T$ in terms of our
model parameters. The overall agreement is quite reasonable, taking
into account the uncertainty of PES results. Again the flatness of
$\mu(c_h<c_h^*)$ is quite remarkable, indicating on the possibility of
divergent $\kappa \to \infty$ for $c_h \to 0$. As noted already by the
authors the variation $\mu(c_h)$ is highly nontrivial and cannot be
accounted for by a simple LFL results as e.g. obtained in band
structure calculations.

\subsection{Entropy}

Let us consider the entropy density (per unit cell)
\begin{equation}
s={1\over N} \left(k_B\ln\Omega+ {\langle H\rangle-\mu\langle
N_e\rangle \over T}\right), \label{t4}
\end{equation}
where averages and $\Omega$ are calculated using the FTLM (Jakli\v c
and Prelov\v sek 1995b, 1996) with equations (\ref{t2}).  Within the
$t$-$J$ model $s$ has been studied also via the HTE (Putikka, unpublished),
while the QMC method has been recently used to calculate the entropy
within the Hubbard model (Duffy and Moreo 1997).

\begin{figure}
\centering
\iffigure
\mbox{
\subfigure[]{
\epsfig{file=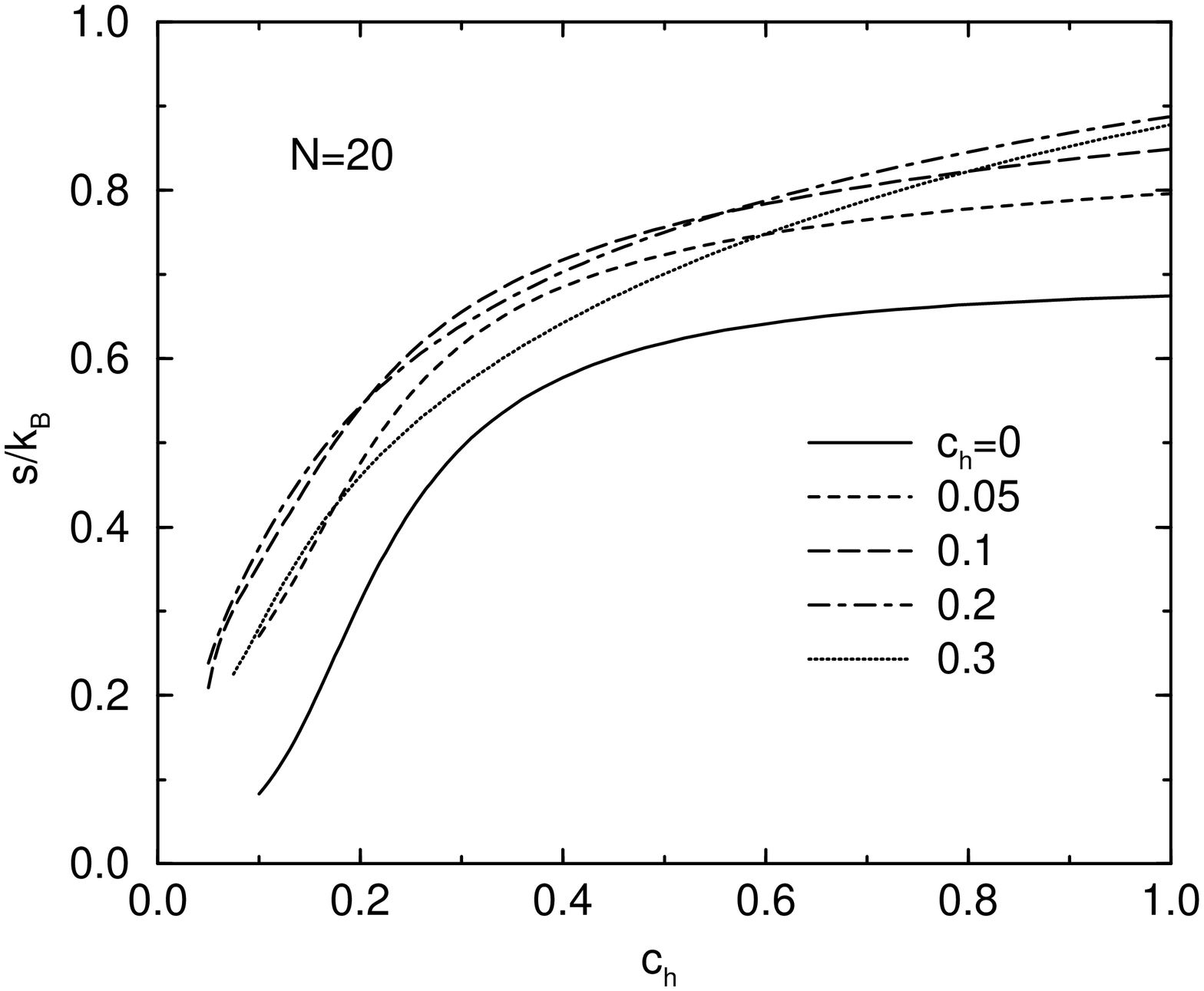,height=8cm,angle=-90}}
\quad
\subfigure[]{
\epsfig{file=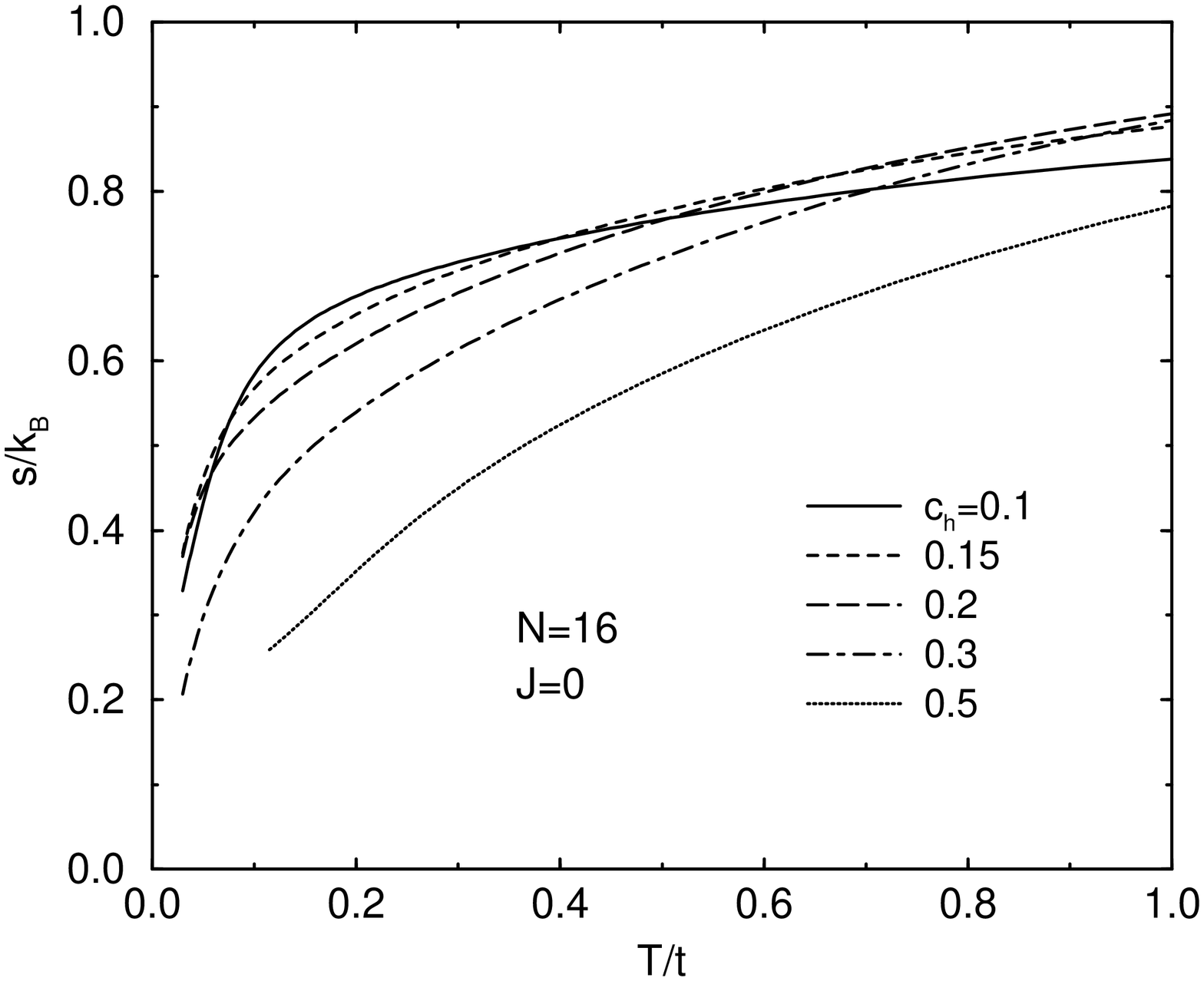,height=8cm,angle=-90}}}
\fi
\caption{
Entropy density $s$ vs. $T$ for different dopings $c_h$ for a)
$J/t=0.3$, and b) $J=0$.  }
\label{4.3}
\end{figure}

The $T$ variation of $s$ at various $c_h$ is presented in
Fig.~\ref{4.3}a. Note again that within the grand-canonical calculation
$c_h$ can be followed continuously. Results shown for $N=20$ are quite
close to those for lattices with $N=16, 18$ sites provided that
$T>0.1~t$. It is evident that in the undoped AFM $s(T)$ at low $T<J/2$
is consistent with the magnon contribution $s \propto T^2$. This
dependence changes however already for smallest finite doping $c_h =
0.05$ to $s \propto T^{\alpha}$ with $\alpha \alt 1$.

In order to understand the role of AFM correlations induced by $J>0$
on the entropy $s$, we present in Fig.~\ref{4.3}b also results
obtained for $J=0$. It is clear that here the undoped case $c_h=0$ is
singular, since $s(T>0)=\ln 2$. It is also plausible that for $c_h
\agt 0.3$ we find $s(T)$ essentially independent of $J$, i.e. the spin
exchange becomes irrelevant in the overdoped regime. Comparison again
confirms that the role of $J$ is crucial in the underdoped regime and
at optimum doping.

\begin{figure}
\centering
\iffigure
\epsfig{file=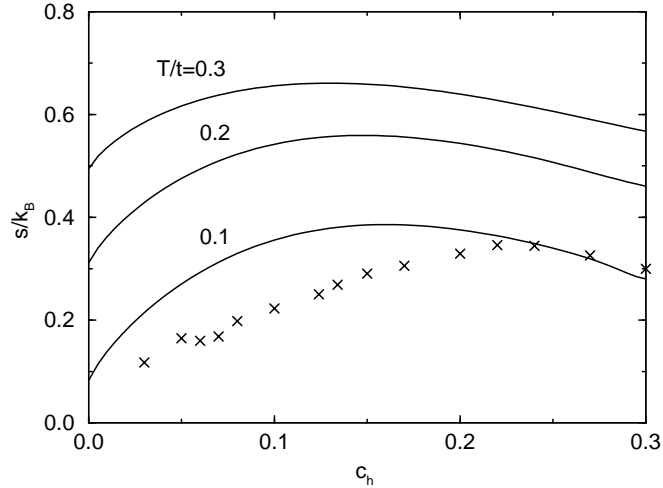,height=10cm,angle=-90}
\fi
\caption{
$s$ vs. $c_h$ at several $T$, calculated in a system with $N=20$
sites. We show for comparison also experimental results for LSCO by
Loram {\it et al.} (1996) at highest $T=320~K \sim 0.07~t$.  }
\label{4.4}
\end{figure}

Alternatively we can discuss the doping dependence of $s$, as shown in
Fig.~\ref{4.4} at different $T\le J$.  As realized already from the
discussion of the thermodynamic sum $Z(T)$ and related $T_{fs}(c_h)$
in Fig.~\ref{3.6}, the entropy displays a broad maximum at $c_h\sim
0.15$, indicating the highest density of many-body states in the
optimum-doping regime. The appearance of the maximum in $s(c_h)$ is
intimately related to $\mu(T)$ discussed in Sec.~4.1. Namely from
general thermodynamic relations (equality of mixed derivatives) for
the free energy density ${\cal F}(c_h,T)$ it follows
\begin{equation}
\left. {\partial s\over \partial c_h}\right|_T=\left. -{\partial \mu_h\over
\partial T}\right|_{c_h}, \label{t5}
\end{equation}
taking into account that $s=-\partial{\cal F}/\partial T$ and
$\mu_h=\partial{\cal F}/\partial c_h$. The relation (\ref{t5})
connects $s(c_h)=max$ with the pinning of $\mu_h(T)$ seen in
Figs.~\ref{4.1},~\ref{4.2} at the optimum doping $c_h \sim c_h^*$.

Besides the enhancement of $s$ with doping, the surprising fact is
also its magnitude at $T<J$, i.e. at small $T$ in terms of model
parameters.  $s$ in the optimum regime appears very large, e.g. at
$T=0.1~t=J/3$ the entropy per site is $s\sim 0.39~k_B$, which is
almost 40\% of $s(T=\infty)$ for the same $c_h$, although $T < J$ and
the energy span of excitations extends well beyond the scale $\Delta E
> t$. This should be contrasted with the situation in an undoped AFM,
where $s$ becomes relatively significant only for $T>J/2$, and
saturates for $T \agt J$. Moreover, in the case of noninteracting
fermions one gets $s\sim k_B$ only at the Fermi temperature $T^0_F
\sim W$, where the bandwidth is $W=8~t$. On the other hand, by
introducing the degeneracy temperature within the $t$-$J$ model as
$s(T_{{deg}})=s(T=\infty)/2$, we get for $c_h\sim c_h^*$ only
$T_{{deg}}\sim 0.17~t$, being small in comparison with any reasonable
effective QP bandwidth.

It is indicative that the entropy of such a magnitude has been deduced
from the electronic specific heat measurements in oxygen deficient
YBa$_2$Cu$_3$O$_{7-\delta}$ (YBCO) materials (Loram {\it et al.}
1993). E.g., for the optimally doped material with $\delta=0.03$ at
$T=300~{K}$ the experimental result is $\Delta s = 0.35~k_B$ per planar
copper site ($\Delta s = 0.7~k_B$ per formula unit), relative to the
undoped $\delta=1$ sample. We find the corresponding value $\Delta
s=s(c_h=0.15)-s(c_h=0)\sim 0.30~k_B$ at $T=0.1~t \sim
450~{K}$. Recently, $s$ has been measured also for LSCO in a large
doping range $0<x<0.4$ (Loram {\it et al.} 1996). We plot results at
fixed $T=320 K \sim 0.07~t$ as a function of doping in Fig.~\ref{4.4},
for comparison with our model results.  The qualitative and
quantitative agreement is quite promising, also in view of possible
uncertainties in the experimental determination of $s$. When comparing
results we note that our curves start at higher $T$ and so the main
difference is in the location of the entropy maximum, which in LSCO
appears at somewhat higher $c_h\sim 0.22$.

\subsection{Specific heat}

\begin{figure}
\centering
\iffigure
\epsfig{file=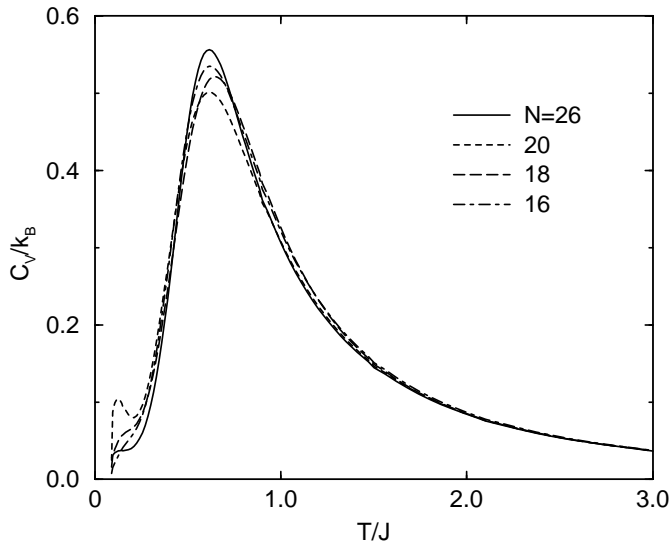,height=10cm,angle=-90}
\fi
\caption{
Specific heat $C_V$ vs. $T$ for the undoped AFM for several
system sizes. } \label{4.5}
\end{figure}

The same results can be discussed in terms of the specific heat (per
unit cell)
\begin{equation}
C_V= T\left(\frac{\partial s}{\partial T}\right)_{c_h},
\label{t6}
\end{equation}
which can be as well represented directly with expectation values at
given $T$, analogous to the equation (\ref{t4}) and a differentiation
with respect to $T$ is not needed.  Let us first show as a test $C_V$
for the undoped AFM (Heisenberg model) for several system sizes. In
Fig.~\ref{4.5} results are shown for systems ranging in size from
16 to 26 sites. Except at the lowest $T\alt J/3$, results do not vary
appreciably with the system size, particularly regarding the position
and the height of the maximum. Our results seem to be even superior to
those obtained by the QMC method (Gomez-Santoz {\it et al.} 1989).
Calculated $C_V$ is strongly $T$ dependent in the whole $T$ range,
with a maximum at $T\sim 2J/3$, and as expected $C_V \propto T^2$ at
low $T$ consistent with the magnon excitations dominating this regime
(Manousakis 1991).

\begin{figure}
\centering
\iffigure
\epsfig{file=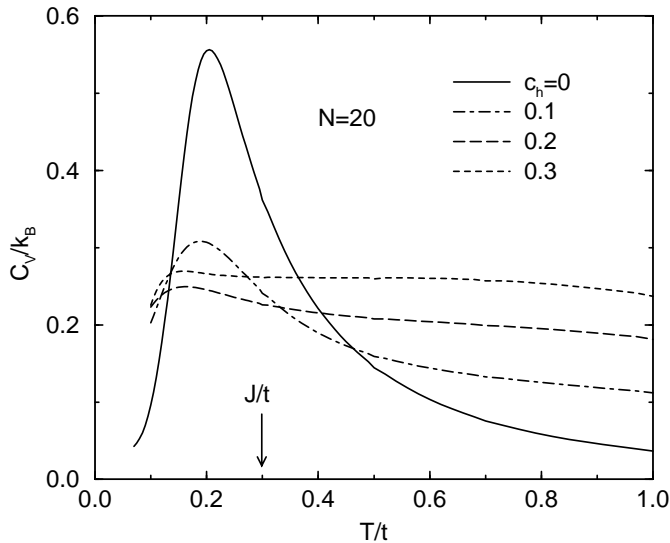,height=10cm,angle=-90}
\fi
\caption{ $C_V$ vs. $T$ for different hole dopings $c_h$.  } \label{4.6}
\end{figure}

In Fig.~\ref{4.6} we present $C_V(T)$ at different $c_h$. As the AFM
is doped, $C_V(T)$ still exhibits a maximum, which is however strongly
suppressed and gradually moves to lower $T$ with increasing $c_h$. The
peak can be attributed to the thermal activation of spin degrees of
freedom. The latter are still characterized by the exchange scale $J$
which persists in the doped system, as observed also in dynamical spin
correlations (Jakli\v c and Prelov\v sek 1995a) discussed in
Sec.~6. The exchange energy scale however disappears in the overdoped
regime $c_h\ge 0.3$. Results indicate a possible LFL behaviour with
$C_V \sim \gamma T$ only for $T<0.1~t$.  It is characteristic (and
consistent with the vanishing role of $J$) that in the optimally doped
regime $c_h\sim 0.2$ we find $C_V(T)\sim const$ for $0.15<T/t<1$,
being far from a FL behaviour.

Results in Fig.~\ref{4.6} confirm the recent conjecture (Vollhardt
1997) that in correlated systems the specific heat $C_V(T,X)$ can show
universal crossings as a function of a thermodynamic variable $X$. In
our case we consider $X=c_h$, and realize that $C_V(T)$ cross for
different $c_h$ at two $T$, whereby the lower crossing at $T \sim
0.13~t$ seems to be nearly independent of $c_h$.

\setcounter{equation}{0}\setcounter{figure}{0}
\section{Electrical Properties}

The anomalous normal-state character of electrical transport
properties has been realized since the discovery of high-$T_c$
cuprates and remains the challenge for theoreticians ever since.
Among these properties the prominent example is nearly linear in-plane
resistivity $\rho \propto T$ in the normal state (for a review see Iye
1992, Batlogg {\it et al.} 1994). It is however an experimental fact
that such a behaviour is restricted to the optimum doping regime,
while deviations from linearity appear both in the underdoped and
overdoped regimes, being still universal for a number of materials
with a similar doping (Takagi {\it et al.} 1992).  The
d.c. resistivity is intimately related to the optical conductivity
$\sigma(\omega)$, which has been also extensively studied (for a
review see Tanner and Timusk 1992) and shows in the normal state the
unusual non-Drude behaviour (Schlesinger {\it et al.} 1990, Romero
{\it et al.} 1992, Cooper {\it et al.} 1993, El Azrak {\it et al.}
1994, Puchkov {\it et al.}  1996, Startseva {\it et al.} 1997).
Another challenging set of experimental findings concern the
d.c. resistivity $\rho_c(T)$ perpendicular to the CuO$_2$ planes and
the corresponding $\sigma_c(\omega)$ (see Uchida 1997), which we will
not consider here.

The central question is whether these anomalous static and dynamical
transport properties can be accounted for by strong correlations
alone, or possible other mechanisms such as the electron-phonon
coupling have to be invoked (Zeyher 1991).  In spite of considerable
efforts so far there are very few microscopical theories of electron
transport dealing with planar or higher-dimensional strongly
correlated systems.

Brinkman and Rice (1970) solved the problem of a single mobile hole in
the extreme case $J=0$ within the retraceable-path approximation (RPA)
and evaluated the d.c. mobility $\mu_0(T)$. Analogous results have
been obtained via the HTE by Ohata and Kubo (1970).  Within the RPA also
$\sigma(\omega)$ has been evaluated (Rice and Zhang 1989). It is not
easy to find the range of validity and relevance of these results, in
particular for $J>0$ where one would expect possibly a crossover to a
different relaxation mechanism for $T<J$.  Nevertheless it is clear
that the analysis, treating holes as independent, is more appropriate
for the weak-doping regime $c_h\ll 1$, at least when the behaviour at
low $\omega<t$ and $T<t$ is concerned.  It appears much more difficult
to approach analytically the electron transport at $c_h>0$. An
attractive proposal remains that of spinons and holons as basic
low-energy excitations (Anderson and Zou 1988), as well as related
gauge theories (Nagaosa and Lee 1990) and slave-boson approaches,
which have been applied also to the calculation of optical
conductivity (Bang and Kotliar 1993). Very fruitful have been recent
studies of infinite-dimensional models, in particular for the Hubbard
model (for a review see Pruschke {\it et al.} 1995, Georges {\it et
al.} 1996), which allow also for the evaluation of $\sigma(\omega)$
and $\rho(T)$.  Since these numerical results are also hard to
interpret, it is still under debate to what extent they contain
features relevant to lower dimensions, e.g. to planar systems
discussed here.

Several conclusions on the charge dynamics have been reached using the
ED dealing with the g.s.  behaviour.  For a single hole in the AFM the
planar optical conductivity $\sigma(\omega)$ (Sega and Prelov\v sek
1990, Poilblanc {\it et al.} 1993) and the charge stiffness (Zotos
{\it et al.}  1990) have been interpreted in terms of partially
coherent hole motion with a substantially enhanced effective mass (at
$J/t < 1$) and with a mid-infrared peak at $\omega \sim 2J$. Analogous
results were presented for larger doping (Dagotto 1994), however with
considerable finite-size effects, so that their interpretation does
not appear well settled.  We discuss in the following results for the
charge transport in the $t-J$ model as obtained with the FTLM, in
part already presented elsewhere (Jakli\v c and Prelov\v sek 1994b,
1995c).

\subsection{Current response}

Let us consider the real optical conductivity $\tsig(\omega)$, which
is a tensor in general.  We are dealing here with a system without 
any external magnetic field. On a square lattice the tensor is then
diagonal, so that $\sigma_{\alpha\alpha}(\omega)=\sigma(\omega)$.
Within the linear-response theory (see e.g. Mahan 1990) the regular
part of $\sigma(\omega>0)$ is given by
\begin{equation}
\sigma_{reg}(\omega)=e_0^2{1-{\rm e}^{-\beta\omega}\over \omega} 
C(\omega), \qquad C(\omega)={1\over N}{\rm Re}\int_0^\infty dt 
{\rm e}^{i\omega t} \langle j_\alpha(t) j_\alpha (0)\rangle, \label{ec1}
\end{equation}   
where $\vec j$ is the (total) particle current operator. In a finite
system one can write $C(\omega)$ in terms of exact eigenstates
$|\Psi_n\rangle$ with corresponding energies $E_n$,
\begin{equation}
C(\omega)= {1\over NZ} \sum_{n\ne m}{\rm e}^{-\beta E_n}
|\langle \Psi_m|j_{\alpha}|\Psi_n \rangle|^2 \delta(\omega - E_m+ E_n).
\label{ec2}
\end{equation} 
It is however well known that in general one has to take into account
also the singular  contribution to the charge dynamical response, i.e.
\begin{equation}
\sigma(\omega)= 2\pi e_0^2 D_c \delta(\omega) +\sigma_{reg}(\omega),
\label{ec3}
\end{equation}
where $D_c$ represents the charge stiffness.  We study in the following
$\sigma(\omega)$ in more general tight-binding models, e.g. including
also the n.n.n. hopping. Since the analysis of this case, together
with the derivation of a proper $D_c$ and the optical sum rule, is not
usual in the literature we present it shortly below.

We follow the approach by Kohn (1964) introducing a (fictitious) flux
$\phi$ through a torus representing the square lattice with p.b.c. Such
a flux induces a vector potential $\vec A$, being equal on all lattice
sites.  In lattice models with a discrete basis for electron
wavefunctions $\vec A$ can be introduced with a gauge transformation
(Peierls construction) $ c^\dagger_{js}\to c^\dagger_{js} {\rm
exp}(-ie_0\vec{A}\cdot\vec{R}_j)$, which effectively modifies hopping
matrix elements.  Taking $\vec A$ as small we can express the modified
tight-binding Hamiltonian allowing also for more general hopping
elements $t_{ij}$
\begin{eqnarray}
H(\vec{A})&=&-\sum_{i,j,s}
t_{ij}e^{-ie_0\vec{A}\cdot\vec{R}_{ij}}c^\dagger_{js}c_{is}+H_{int} 
\approx \nonumber \\
&\approx&H(0)+e_0\vec{A}\cdot\vec{j}+{e_0^2\over 2}
\vec{A}\cdot \ttau \vec{A}, \label{ec5}
\end{eqnarray}
where $\vec{R}_{ij}=\vec{R}_j-\vec{R}_i$, $\ttau$ is the kinetic
stress tensor, and
\begin{eqnarray}
\vec{j}&=&i\sum_{i,j,s}
t_{ij}\vec{R}_{ij} c^\dagger_{js}c_{is}, \nonumber \\
\ttau&=&\sum_{i,j,s}
t_{ij}\vec{R}_{ij}\otimes\vec{R}_{ij} c^\dagger_{js}c_{is}.
\label{ec6}
\end{eqnarray}
Note that in usual n.n. tight-binding models $\ttau$ is directly
related to the kinetic energy operator, $\tau_{\alpha\alpha} =
(H_{kin})_{\alpha\alpha}$.

The electrical current $\vec{j}_e$ is from the equation (\ref{ec5})
expressed as a sum of the particle-current and the diamagnetic
contribution,
\begin{equation}
\vec{j}_e=-\partial H/\partial\vec{A}=-e_0\vec{j}-e_0^2\ttau\vec{A}.
\label{ec7}
\end{equation}
The above analysis applies also to an oscillating  $\vec A(t) =
\vec{A}(\omega){\rm exp}(-i\omega^+ t)$. This
induces an electric field in the system $\vec E(t)= -
\partial\vec{A}(t)/\partial t$.  We are interested in the response of
$\langle \vec j_e\rangle(\omega)$. Evaluating $\langle\vec{j}\rangle$
within the standard linear response (Mahan 1990), and with $\vec
A(\omega)= \vec E(\omega)/i\omega^+$, we arrive at the complex optical
conductivity
\begin{eqnarray}
\tilde {\tsig}(\omega)&=&\frac{i e_0^2}{\omega^+ N}
(\langle \ttau\rangle - \tchi(\omega)),\nonumber \\
\tchi(\omega)&=& i\int_0^{\infty} dt e^{i\omega^+t}\langle[\vec j(t), 
\vec j(0)]\rangle. \label{ec8}
\end{eqnarray}
Complex $\tilde {\tsig}(\omega)= {\tsig}(\omega)+i\tilde
{\tsig}^{\prime\prime}(\omega)$ satisfies the Kramers-Kronig
relation. Since $\tchi(\omega \to \infty) \to 0$, we get from the
equation (\ref{ec8}) a condition for $\tilde {\tsig}^{\prime\prime}
(\omega\to \infty)$,
\begin{equation}
\int_{-\infty}^\infty \tsig(\omega)d\omega=
\frac{\pi e_0^2}{N}\langle \ttau \rangle,  \label{ec9}
\end{equation}
which corresponds to the optical sum rule.  It reduces to the well
known one for continuum electronic systems, as well as for
n.n. hopping models where $\langle \tau_{\alpha\alpha} \rangle=
-\langle H_{kin}\rangle/d$ (Maldague 1977).  We can now make contact
with the definition (\ref{ec3}). From the expression (\ref{ec8}) it
follows
\begin{eqnarray}
\sigma_{reg}(\omega)&=&\frac{e_0^2}{N\omega}\chi''_{\alpha\alpha}(\omega), 
\nonumber \\
D_c =\frac{1}{2e_0^2}\lim_{\omega\to 0}\omega \tilde
\sigma_{\alpha\alpha}^{\prime\prime}(\omega)&=&
\frac{1}{2N}[\langle \tau_{\alpha\alpha} \rangle-\chi_{\alpha\alpha}'(0)]. 
\label{ec10}
\end{eqnarray}

\subsection{Charge stiffness}

Nonzero charge stiffness $D_c^0=D_c(T=0)>0$ is a characteristic
signature of a metallic state (Kohn 1964, Scalapino {\it et al.}
1993), in contrast to an insulator with $D_c^0=0$. The evaluation of
$D_c^0$ has been recently applied to a number of correlated fermionic
systems, both analytically for 1D systems (Shastry and Sutherland
1990) and numerically for planar Hubbard and $t$-$J$ models (see
Dagotto 1994).  At $T>0$ one would expect for normal resistors
$D_c=0$. Since we are working with small systems with p.b.c., where
ballistic response of carriers can persist at $T>0$, we find in
general $D_c(T)\ne 0$. Note however that recently a nontrivial
possibility of a nonergodic behaviour in a macroscopic limit has been
realized, i.e. with $D_c(T>0)>0$, being related to the integrability
of the fermionic model (Castella {\it et al.} 1995, Zotos and Prelov\v
sek 1996).

The $t$-$J$ model at half-filling has $D_c(T)=0$ at any $T$, since
charge fluctuations are projected out by construction of the model
(\ref{cm1}).  For a doped system one expects $D_c^0
\propto c_h$ at low $c_h$.  Studying the charge transport at
$T>0$ within the planar $t-J$ model we adopt the view that there exist
scattering processes which in a macroscopically large systems cause
$D_c(T>0)=0$, i.e. the model is ergodic.  However, in our numerical
calculations we are dealing with small systems and a small portion of
the total current can propagate through the system unscattered,
thereby establishing in the system a persistent current. Hence we
characterize observed $D_c(T>0) \ne 0$ as a finite-size artifact.
Nevertheless, a variation of $D_c(T)$ brings a valuable information.
At $T>0$ the scattering processes result in a finite mean-free path of
charge carriers $l_s(T)$.  When the linear system size $L$ exceeds
$l_s$, it is reasonable to expect that $D_c(T)\sim 0$. On the other
hand for $L< l_s(T)$ we obtain $D_c(T) \sim D_c^0$.  By following
$D_c(T)$ we can thus independently monitor the transport mean free
path. Since $l_s$ is strongly $T$-dependent, we can approximately
locate the crossover temperature as $L \sim l_s(T_{fs})$.

We can express $D_c$ in the $t$-$J$ model on a square lattice from
the equation (\ref{ec10}) as
\begin{equation}
D_c= -{1\over 4N} \langle H_{kin}\rangle - {1\over \pi e_0^2}
\int_{0^+}^\infty \sigma(\omega) d \omega. \label{es1}
\end{equation}
Within the FTLM $\langle H_{kin}\rangle$ and $\sigma(\omega)$ are
calculated separately. It should be however mentioned that due to
finite $M$ there could be some ambiguity in the cutoff employed at low
$\omega$, the problem which is however restricted to otherwise
unproblematic $T\gg t$ regime.

\begin{figure}
\centering
\iffigure
\epsfig{file=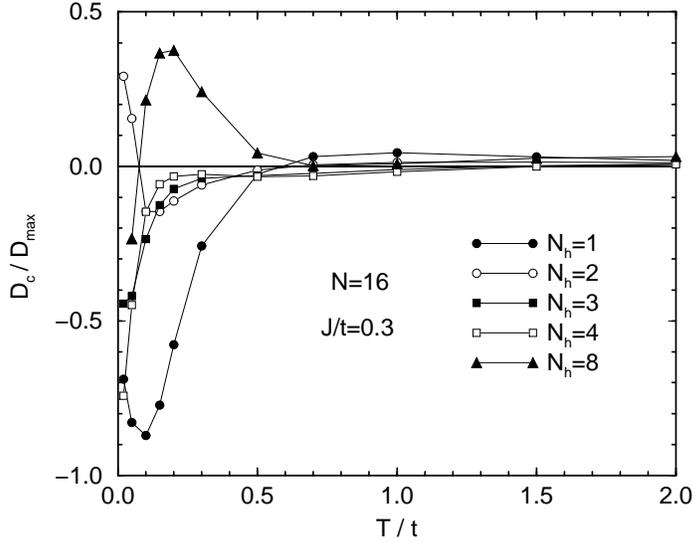,height=10cm,angle=-90}
\fi
\caption{
Normalized charge stiffness $D_c/D_{max}$ for various number on holes
$N_h$ in a system with $N=16$ sites.  } \label{5.1}
\end{figure}

Results for a single hole in the $t-$($J=0$) model, as obtained by
Jakli\v c and Prelov\v sek (1995c) show that $D_c(T)$ interpolates
quite smoothly between $D_c=0$ at $T=\infty$ and $D_c^0$, the
crossover becoming sharper in larger systems.  For the AFM case
results for $D_c(T)$ are less regular, as presented in Fig.~\ref{5.1}
for various $N_h\geq 1$ on a system with $N=16$ sites.  When judging
the extent of deviations of $D_c$ from zero at $T>0$, it is useful to
compare values with the maximum possible ones, i.e. with the value of
the sum rule at $T=0$, $D_{max}=|\langle H_{kin}\rangle(T=0)|/4N$, as
follows from the equation (\ref{es1}). We notice from Fig.~\ref{5.1}
that $D_c$ typically shows a rather abrupt transition from $D_c\sim 0$
to $D_c \neq 0$ at the crossover temperature $T_{fs}$, depending
mainly on $c_h$. For $T<T_{fs}$ the variation $D_c(T)$ can become
quite unphysical, i.e. we get in some cases even $D_c<0$. These
phenomena are influenced by particular p.b.c. and more sensible
results can be obtained by the introduction of twisted boundary
conditions or fluxes (Poilblanc 1991). Nevertheless we are here
interested only in the regime $T>T_{fs}$. It follows again that
$T_{fs}$ is minimum, i.e. $T_{fs} \sim 0.1~t$, for the intermediate
doping $0.1<c_h<0.3$.  We should stress the striking message that at
intermediate doping even at such low $T$, representing for cuprates
$T\sim 450K$, the mean free path does not exceed $l_s \sim 4$ sites
and this entirely due to correlation effects.

\subsection{Single-hole mobility}

It is expected that at low doping the conductivity scales linearly
with doping, hence it is meaningful to introduce the dynamical mobility
$\mu(\omega)$ which is a single-hole property
\begin{equation}
\sigma(\omega) = e_0 c_h \mu(\omega),  \label{em1}
\end{equation}
but still highly nontrivial for a Mott or an AFM insulator.
 
The most conclusive theoretical results (in 2D or higher D systems)
have been so far obtained for a problem of a single mobile hole
introduced into a reference insulator. Brinkman and Rice (1970) solved
the problem for $J=0$ within RPA. They pointed out of an essentially
incoherent hole motion and evaluated the d.c. mobility $\mu_0(T)$,
exhibiting $\mu_0 \propto T$ for $T>t$. $\mu(\omega)$ within RPA
(Rice and Zhang 1989) shows an incoherent motion, resulting in a
slow non-Drude fall-off $\mu(\omega) \propto 1/\omega$ for larger
$\omega$. The RPA has been recently justified and applied more
rigorously for infinite-D lattices (Metzner {\it et al.} 1992). An
analogous approach is the evaluation of frequency moments of
$\mu(\omega)$, starting at $T\gg t$, as applied to the $J=0$
problem by Ohata and Kubo (1970). On the other hand, $\mu(\omega)$
at $T=0$ has been in recent years well established by numerical
studies of small systems via the ED method (Sega and Prelov\v sek
1990, Dagotto 1994).  Nevertheless there are important unsolved
questions even for the single-hole problem. Is $\mu(\omega)$ on a
planar lattice qualitatively and quantitatively well described within the
RPA, at least for $T>t$ ?  Which are new qualitative dynamical features
at $T<t$, both for the $J=0$ and the $J>0$ case ?

Results for $\mu(\omega)$ at $J=0$, obtained via the FTLM by Jakli\v c
and Prelov\v sek (1995c), show an overall agreement with the
RPA. However, in contrast to the smooth RPA curve the actual
$\mu(\omega)$, evaluated at $T\gg t$, seems to be nonanalytical,
i.e. it shows a cusp at $\omega=0$. The phenomenon seems to be
characteristic for $J=0$, but not for $J>0$.  $\mu(\omega)$ for $J=0$
retains its high-$T$ form for all $T > t$, but changes qualitatively
for $T<t$.  Here the central peak due to the formation of the FM
polaron (Nagaoka 1966) starts to emerge at low $\omega$, and the
incoherent broad background steadily vanishes on approaching $T
\rightarrow 0$.

\begin{figure}[ht]
\centering
\iffigure
\epsfig{file=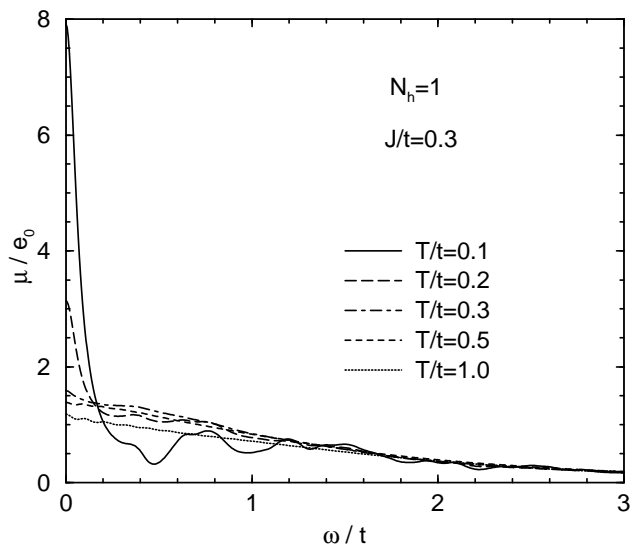,height=10cm,angle=-90}
\fi
\caption{
Dynamical mobility $\mu(\omega)$ for a single hole in an AFM
at various $T$.
} \label{5.2}
\end{figure}

We are more interested in the AFM case, as shown in Fig.~\ref{5.2}.
For $T>t$ the spin background is disordered and the mobility retains
the high-$T$ form $\mu(\omega) \propto \beta$, leading to $\mu_0
\propto 1/T$.  As seen from Fig.~\ref{5.2}, there is a qualitative
change already in the regime $J<T<t$.  Namely, it appears that here
$\mu(\omega)$ has a weaker $T$-dependence. For $T\rightarrow 0$ we are
however approaching a nontrivial response of an AFM polaron, which has
been analyzed numerically by several authors. At $T=0$ one expects
in $\mu(\omega)$ for $\omega \to 0$ a coherent response of an AFM
polaron with an enhanced mass, but also a nonvanishing incoherent part
at $\omega>J$ (Sega and Prelov\v sek 1990, Dagotto 1994).  The latter
component seems to have a nontrivial internal structure being related
to the mid-infrared peak, as realized also in Fig.~\ref{5.2} for the
lowest $T=0.1~t$.

\begin{figure}[ht]
\centering
\iffigure
\epsfig{file=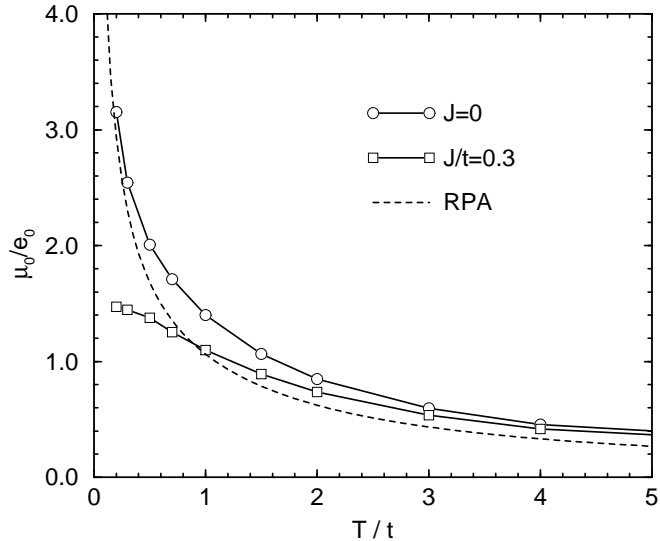,height=10cm,angle=-90}
\fi
\caption{
D.c. mobility $\mu_0$ vs. $T$ for $J=0$ and $J/t=0.3$. The RPA result
is shown for comparison. } \label{5.3}
\end{figure}

From $\mu(\omega)$ one can extract the d.c. mobility $\mu_0$.  Results
for $J=0.3~t$ and $J=0$ are presented in Fig.~\ref{5.3}, with the RPA
($J=0$) result for comparison.  While for $T>2t$ we see the high-$T$
result $\mu_0 \propto 1/T$ for both $J$, we get for the AFM case a
nontrivial $\mu_0 \sim const$ within the regime $J<T<t$.  It is
plausible that $J>0$ reduces $\mu_0$ as follows also from the
frequency-moment analysis at $T \gg t$ (Jakli\v c and Prelov\v sek
1995c), indicating that in an AFM the hole motion is more frustrated
leading to stronger scattering.  As one can realize from
Fig.~\ref{5.2}, it is hard to get meaningful results for $T< J \sim
T_{fs} $, since in this regime the central QP peak is essentially
undamped. This is an indication that we are already dealing with
finite-size effects, as discussed in Sec.~5.2, and the mean free path
is beyond our system size $l_s>L$. Naively one would expect from
Fig.~\ref{5.3} that in the regime $T<J$ $\mu_0$ would increase with
decreasing $T$, whereby the scattering of the QP (AFM polaron) should
be on thermally excited magnons. There are however unsolved problems
with such a description.  In an AFM the QP dispersion (Kane {\it at
al.} 1989, Martinez and Horsch 1991) is strongly renormalized and is
effectively narrower than the magnon one, hence it seems hard to find
appropriate allowed scattering processes.  These questions become
important when discussing the relation of calculated $\mu_0$ to the
resistivity of cuprates at low doping, as discussed in Sec.~5.5.
 
\subsection{Optical conductivity at finite doping}

Even more challenging questions appear at finite hole doping.  Whereas
at low doping one could treat the transport within the
semiconductor-like model of independent QP - AFM polarons, this
concept fails even for a moderate $c_h\agt 0.1$ due to the overlap
between extended spin deformations around holes. Alternative
indications for such a phenomenon have been already discussed in Sec.~4
in connection with $c_h(\mu_h)$.

Since in a planar system we are evaluating the sheet conductivity
$\sigma(\omega)$ it is convenient to discuss at finite doping the
dimensionless quantity $\bar \sigma= \sigma \rho_0$, where
$\rho_0=\hbar / e_0^2 = 4.1 k\Omega$ is the universal 2D sheet
resistance. Such a quantity has an additional meaning since it makes a
direct contact with the theory of localization, where $\bar
\sigma_0 \sim \bar \sigma_{min}$ is a characteristic marginal value
associated with the 2D minimum metallic conductivity (Mott and Davis
1979). The value of $\bar \sigma_{min}$ and its relevance is however
controversial, and it ranges from $\bar \sigma_{min}=0.1$ (Mott and
Davis 1979) to $\bar \sigma_{min} \sim 0.5$ found in experiments
(Mandrus {\it et al.}  1991) as a borderline between insulating and
conducting cuprates at low $T$.

\subsubsection{Intermediate doping}

\begin{figure}
\centering
\iffigure
\epsfig{file=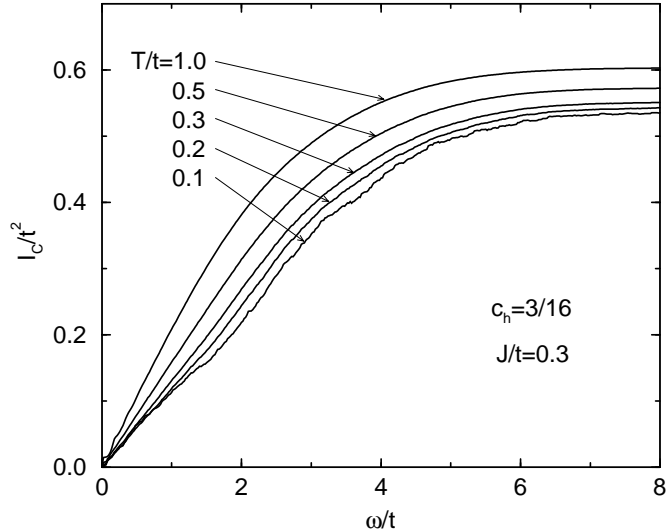,height=10cm,angle=-90}
\fi
\caption{
Integrated current correlation spectra $I_C(\omega)$ at $c_h=3/16$ for
different $T\le t$.  } \label{5.4}
\end{figure}

Let us go straight to results at the intermediate (optimum) doping,
where we can reach lowest $T_{fs}$.  Instead of $\sigma(\omega)$ it is
more instructive to present the current correlation function
$C(\omega)$, equation (\ref{ec1}). To avoid the ambiguities with an
additional smoothing, we plot the corresponding integrated spectra
\begin{equation}
I_C(\omega)=\int_0^{\omega}
C(\omega') d\omega'. \label{eo1}
\end{equation} 
In Fig.~\ref{5.4} we present $I_C(\omega)$ for $c_h=3/16$, for various
$T\le t$.  Spectra reveal several remarkable features (Jakli\v c and
Prelov\v sek 1995b):

\noindent (i) For $T \le J$ spectra $I_C(\omega)$ are essentially
independent of $T$, at least for available $T>T_{fs}$.

\noindent (ii) Simultaneously the slope of $I_C(\omega < 2~t)$
is nearly constant, i.e.  we find $C(\omega) \sim C_0$ in a wide
$\omega$ range. At the same time $C_0$ is only weakly dependent on $J$
as tested for $J/t=0.2 - 0.6$.

\noindent (iii) Even for higher $T>J$ the
differences in the slope $C_0$, as also in $I_C(\infty)$,
appear as less essential. Note that for $T\gg t$ we know exactly
$I_C(\infty)=\pi t^2 c_h(1-c_h)$ (Jakli\v c and Prelov\v sek 1995c).

We conclude that $C(\omega<2~t) \sim C_0$ implies
a simple universal form (Jakli\v c and Prelov\v sek 1995b),
\begin{equation}
\sigma(\omega)=C_0 e_0^2{1-e^{-\beta\omega} \over \omega}. \label{eo2}
\end{equation}
Such a $\sigma(\omega)$ shows a nonanalytic behaviour at $\omega
\rightarrow 0$, starting with a finite slope.  This is already an
indication that $\sigma(\omega)$ is not consistent with the usual
Drude form, but rather with a marginal concept (Varma {\it et al.}
1989) where the only $\omega$ scale is given by $T$.  It is also
remarkable that the form (\ref{eo2}) trivially reproduces the linear
law $\rho \propto T$ as well as the non-Drude fall-off at $\omega
>T$. It is evident that the expression (\ref{eo2}) is universal
containing the only parameter $C_0$ as a prefactor.

\begin{figure}
\centering
\iffigure
\mbox{
\subfigure[]{
\epsfig{file=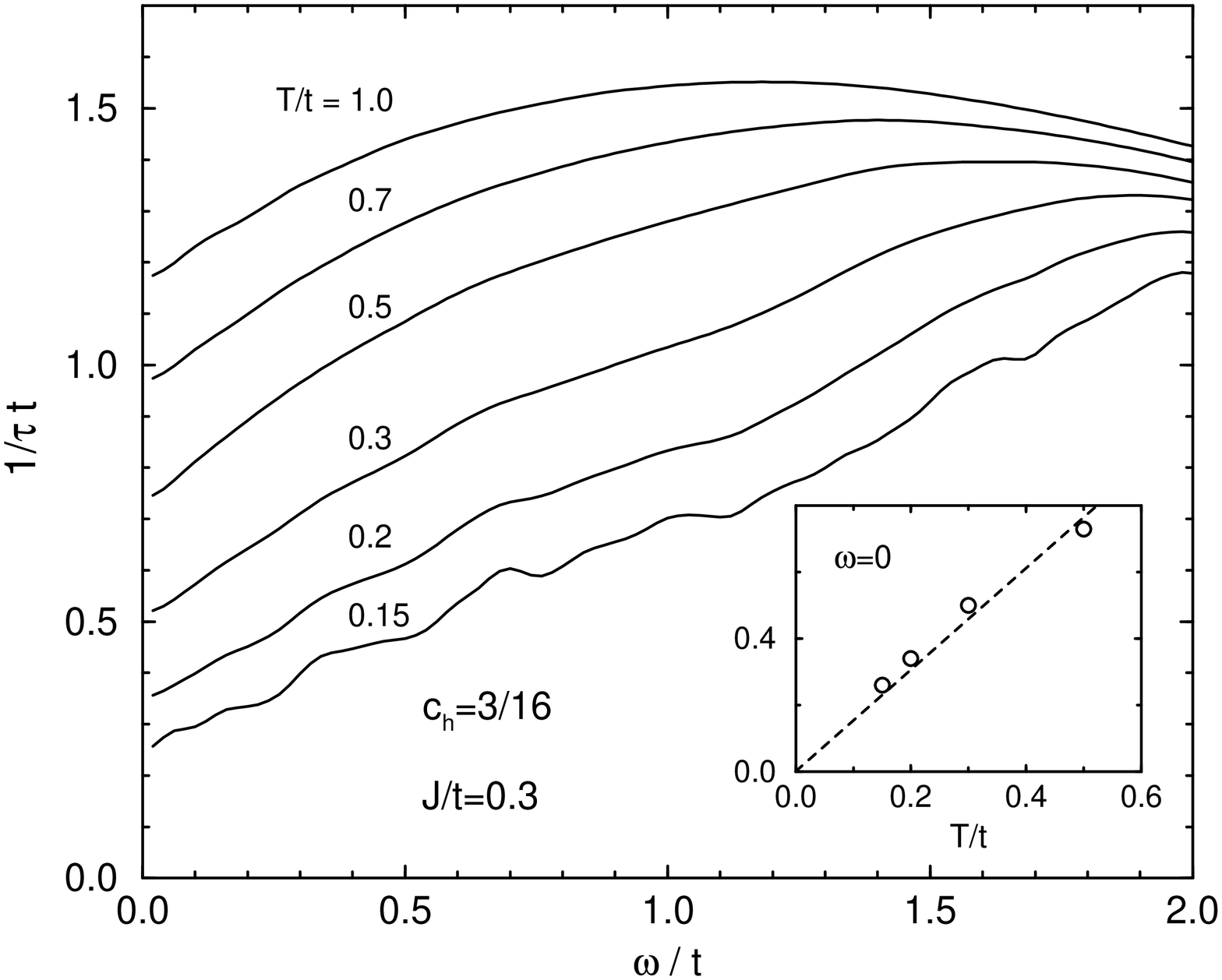,height=7.5cm,angle=-90}}
\quad
\subfigure[]{
\epsfig{file=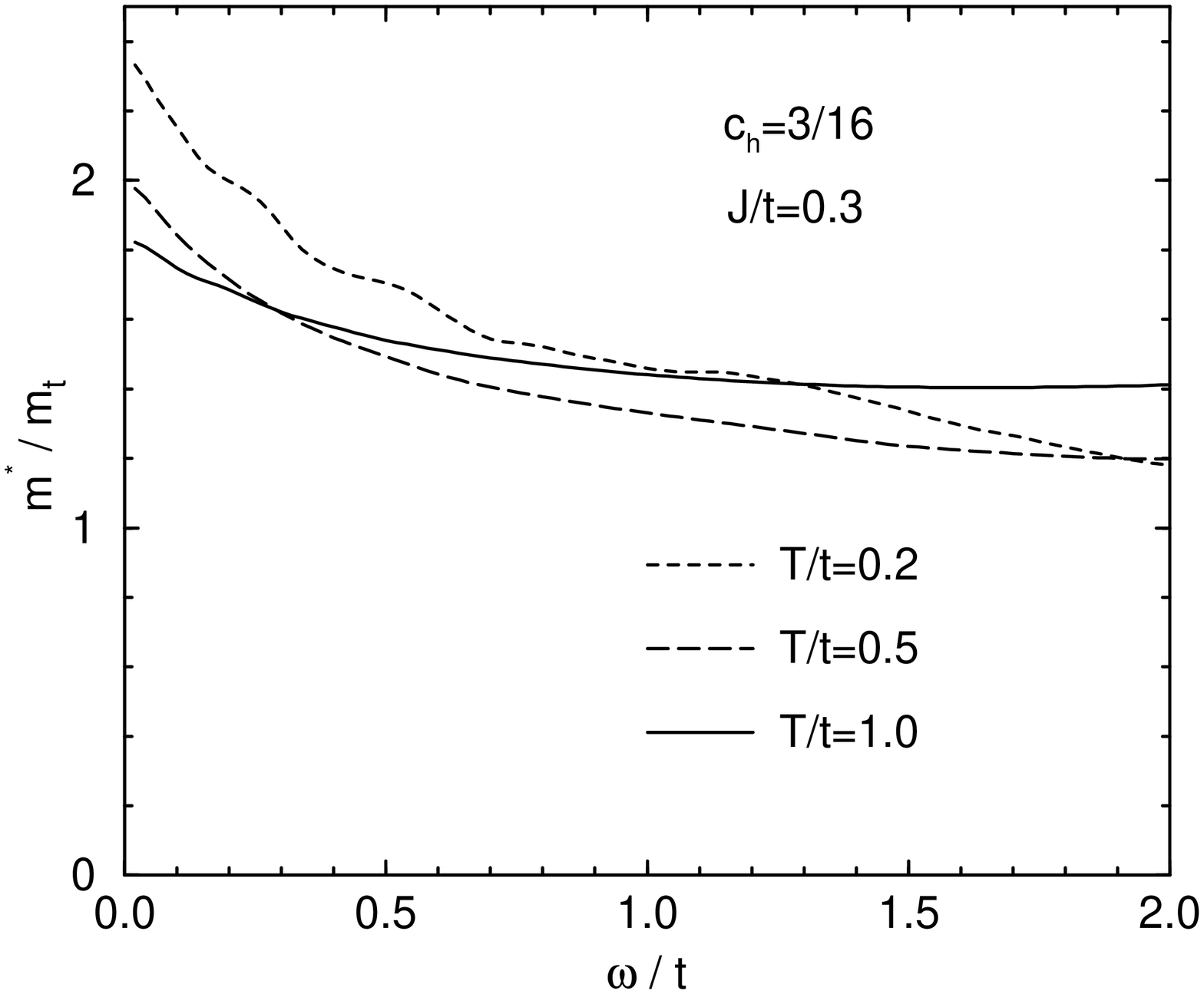,height=7.5cm,angle=-90}}}
\fi
\caption{
(a) Inverse relaxation time $1/\tau$, and (b) mass enhancement $m^*/m_t$
vs. $\omega$, for $c_h=3/16$ and different $T$. The insert in a) shows
the $T$-dependence of $1/\tau(0)$. }
\label{5.5}
\end{figure}

Experimental results and theoretical considerations are often
discussed in terms of the $\omega$-dependent relaxation time $\tau$
and the effective mass $m^*$. These can be uniquely introduced via the
complex $\tilde \sigma(\omega)$ and the corresponding memory function
$M(\omega)$ (G\"otze and W\"olfle 1972),
\begin{equation}
\tilde \sigma(\omega)= {ie_0^2 {\cal S} \over \omega + M(\omega)}, \qquad
\qquad {\cal S} = -\langle H_{kin} \rangle/2N, \label{eo3}
\end{equation}
and
\begin{eqnarray}
{1\over\tau(\omega)} &=&{ M''(\omega)\over 1+ M'(\omega)/ \omega},
\nonumber \\
{m^*(\omega)\over m_t} &=& {2 c_h t\over {\cal S}}
\left(1+{M'(\omega)\over\omega}\right), \label{eo4}
\end{eqnarray}
where $m_t = 1/2ta_0^2$ is the bare band mass. Using
relations (\ref{eo3}),(\ref{eo4}) one can formally rewrite $\tilde
\sigma(\omega)$ in the familiar Drude form
\begin{equation}
\tilde \sigma(\omega) = {i c_h e_0^2 \over m^*(\omega)[\omega +i
/\tau(\omega)]}. \label{eo5}
\end{equation}
Employing the equation (\ref{eo5}) we evaluate both $\tau(\omega)$ and
$m^*(\omega)$ from known $\tilde \sigma(\omega)$. Results for
$c_h=3/16$, corresponding to Fig.~\ref{5.4}, are presented in
Fig.~\ref{5.5}.

It follows from Fig.~\ref{5.5}a, but even more directly from the form
(\ref{eo2}), that in the regime $T\alt J$ and $\omega<t$ we can
describe the behaviour of $1/\tau$ with a linear law
\begin{equation}
\tau^{-1} = 2\pi\lambda (\omega + \xi T), \qquad \lambda
\sim 0.09,~~~\xi \sim 2.7~. \label{eo6}
\end{equation}
This dependence falls within the general framework of the MFL
scenario. It is however not the form (\ref{cp3}) proposed originally
(Varma {\it et al.} 1989, Littlewood and Varma 1990), but rather the
one (\ref{cp3a}) deduced from experiments in cuprates (El Azrak {\it
et al.} 1995, Baraduc {\it et al.} 1995).  It should be however stressed
that the asymptotic form (\ref{eo6}) does not allow for any free
parameter, i.e. constants $\lambda$ and $\xi$ are universal and
independent of any model parameters, whereas within the MFL proposal
$\lambda$ is an adjustable parameter while $\xi = \pi$.

It should be also mentioned that the universal dynamics, as described
by the equation (\ref{eo2}), does not seem to be restricted only to the
particular case of doped AFM, but has a wider applicability, e.g. it
has been recently established also in ladder systems (Tsunetsugu and
Imada 1997).

\subsubsection{Underdoped and overdoped regime} 

\begin{figure}[ht]
\centering
\iffigure
\mbox{
\subfigure[]{
\epsfig{file=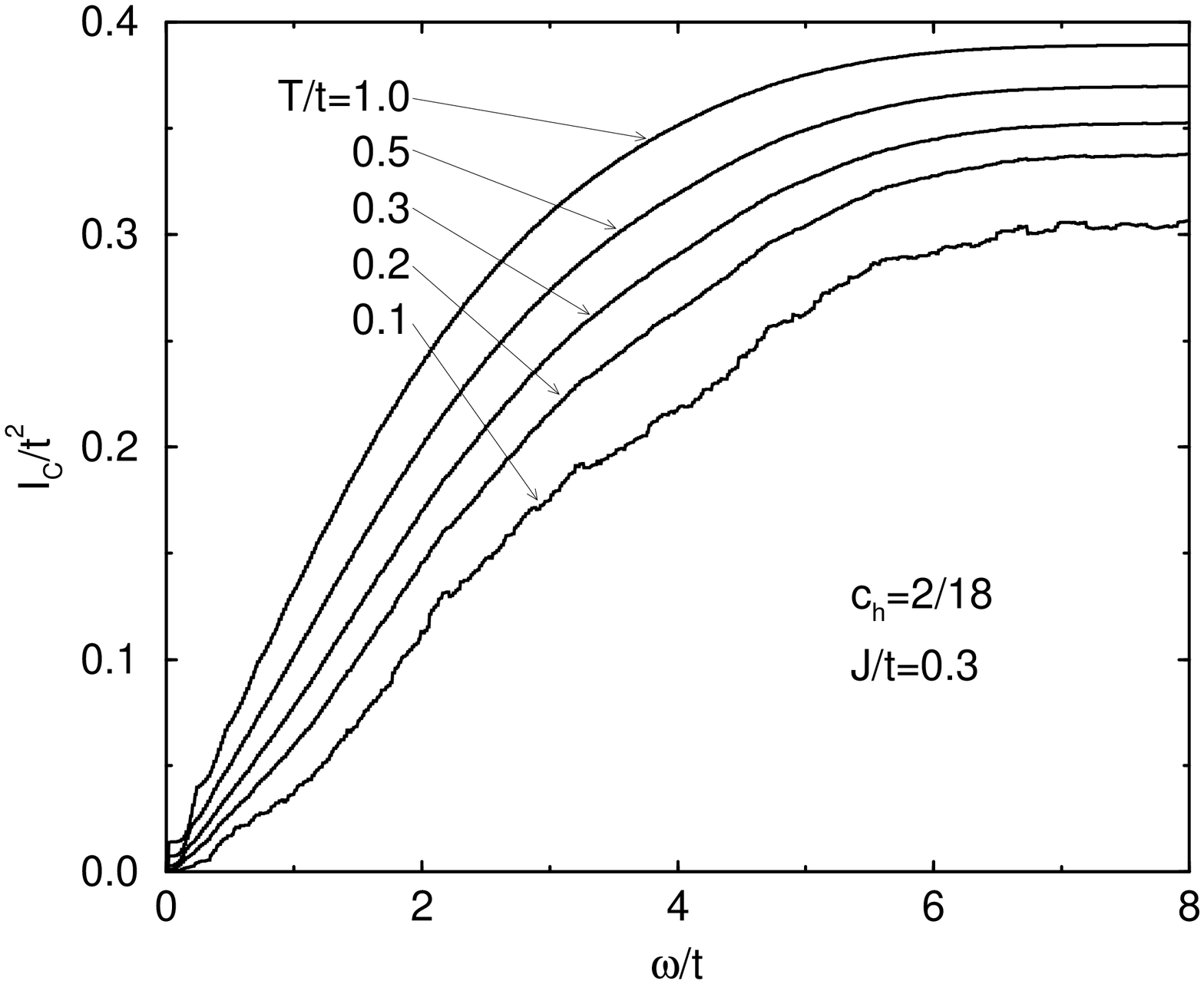,height=8cm,angle=-90}}
\quad
\subfigure[]{
\epsfig{file=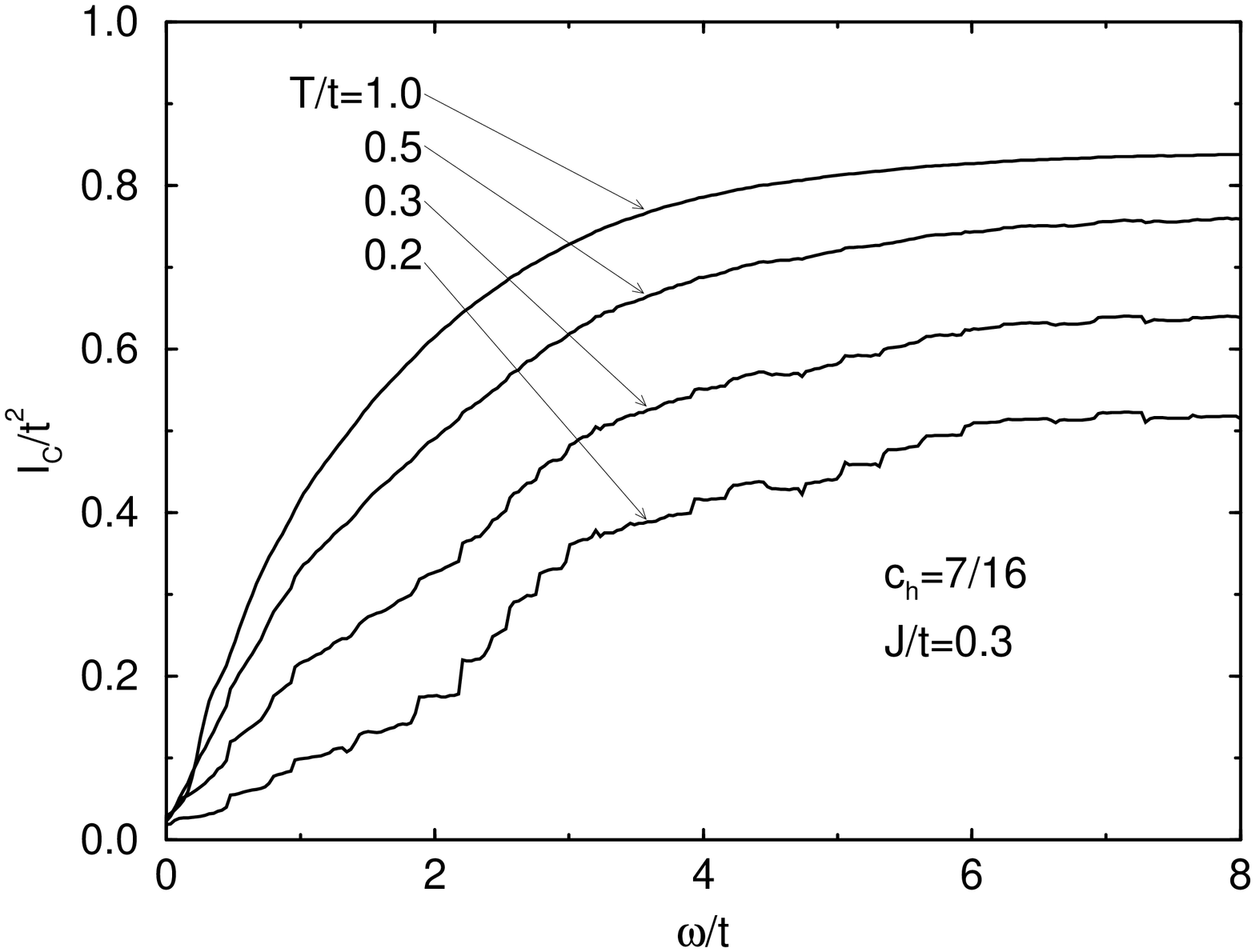,height=8cm,angle=-90}}}
\fi
\caption{
$I_C(\omega)$ for (a) $c_h=2/18$, and (b) $c_h=7/16$, for different $T$.
} \label{5.6}
\end{figure}

Let us turn to the discussion of results at other dopings $c_h$. Results
for $c_h =4/16$ (Jakli\v c and Prelov\v sek 1995c) are in all aspects
very similar to the $c_h=3/16$ case. We show in Figs.~\ref{5.6}
analogous $I_C(\omega)$ for underdoped $c_h = 2/18$,
as well as for overdoped $c_h=7/16$. One feature common to
all $c_h$ considered (including $N_h=1$) is the non-Drude 
behaviour for $\omega>t$. This confirms the belief that the incoherent
motion of holes, dominating the high-$\omega$ response as described
e.g. within the RPA (Rice and Zhang 1989), can remain a valid concept
even at large doping $c_h < 0.5$.

Otherwise it is harder to make conclusions on the most interesting
low-$(\omega,T)$ regime, both for the underdoped as well as for the
overdoped case. For $c_h=2/18$ we note that in contrast to
Fig.~\ref{5.4} the low-$\omega$ slope is not constant at $T<J$ but is
still gradually decreasing with $T$. This results in a modified
variation of the d.c. conductivity $\sigma_0 \propto 1/T^{1-\eta}$
with $\eta >0$, hence also in a sublinear $\rho_0 \propto
T^{1-\eta}$. On the other hand we see already for $T=0.1~t$ a drop of
$I_C(\infty)$ (closely related to the sum rule) which can be
interpreted with a significant persistent (nonscattered) current in a
system with p.b.c. The latter emerges as the coherent part with $D_c>0$
according to the equation (\ref{ec3}), not taken into account in the
presented $I_C(\omega)$. While we cannot say much on the current
relaxation of such a coherent part, it is plausible that an increase
in the coherence should decrease the resistivity $\rho_0$, as found in
underdoped cuprates below the crossover $T^*$. Otherwise results in
Fig.~\ref{5.6}a are close to curves for a single hole in an AFM,
Fig.~\ref{5.2}, with the difference that in the latter case $T^*$
appears even higher.

In the overdoped case in Fig.~\ref{5.6}b $I_C(\infty)$ starts to
deviate from a constant already for $T\alt J$. At $T<J$ a sharp
increase of an unscattered current appears, so we are unable to
speculate on a possible onset of more LFL-like $\rho \propto T^2$.

\subsubsection{Effects of the next-nearest-neighbour hopping}

It is a relevant question whether the anomalous but universal
behaviour of $\sigma(\omega)$ found at the intermediate doping is spoiled
by possible additional terms, e.g. by the introduction of the
n.n.n. hopping $t'$ term (\ref{cm2}), invoked often to obtain a
realistic description of cuprates.  When considering the effect of
$t'$ it is important to realize that at finite doping $t'>0$ tends to
stabilize the AFM ordering while $t'<0$ destabilizes it and tends
toward a Nagaoka-type FM state (Bon\v ca and Prelov\v sek 1989).

\begin{figure}[ht]
\centering
\iffigure
\epsfig{file=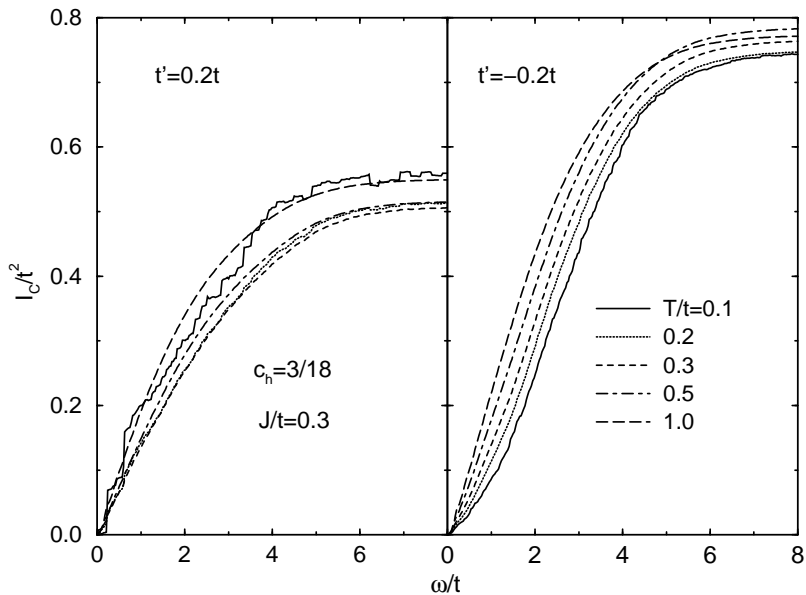,height=11cm,angle=-90}
\fi
\caption{
$I_C(\omega)$ for $c_h=3/18$ and the n.n.n. hopping: a) $t'/t=0.2$, and
b) $t'/t=-0.2$, for different $T$.  } \label{5.7}
\end{figure}

Results for $I_C(\omega)$ at $t' \neq 0$ are presented in
Fig.~\ref{5.7}. We choose rather modest $|t'/t|=0.2$, still quite a
pronounced effect on $I_C(\infty)$ is evident. We should note that in
this case also generalized sum rules, following equations (\ref{ec6})
and (\ref{ec9}), apply. On the other hand, the low-$\omega$ part is
not changed essentially. For $t'/t=0.2$ we note again an universal
behaviour according to the form (\ref{eo2}). Deviations at lowest $T$
are somewhat larger for $t'/t=-0.2$. A possible interpretation of
these results is that $t'\neq 0$ moves the system effectively away
from the starting doping regime, i.e. drives an optimum system either
towards the underdoped or to the overdoped regime.

\subsection{Resistivity}

Finally, we display results for the d.c. resistivities $\rho(T)$, as
extracted from $\sigma(\omega \rightarrow 0)$. The extrapolation
$\omega \rightarrow 0$ is straightforward at higher $T$. It becomes
somewhat more delicate on approaching $T\sim T_{fs}$ due to the
appearance of more pronounced structures and $D_c \ne 0$. Actually,
we can evaluate $\sigma_0$ either from $\sigma(\omega)$ which involves
some smoothing, or from the slope of $I_C(\omega \sim 0)$ which seems
to be more reliable.

\begin{figure}
\centering
\iffigure
\epsfig{file=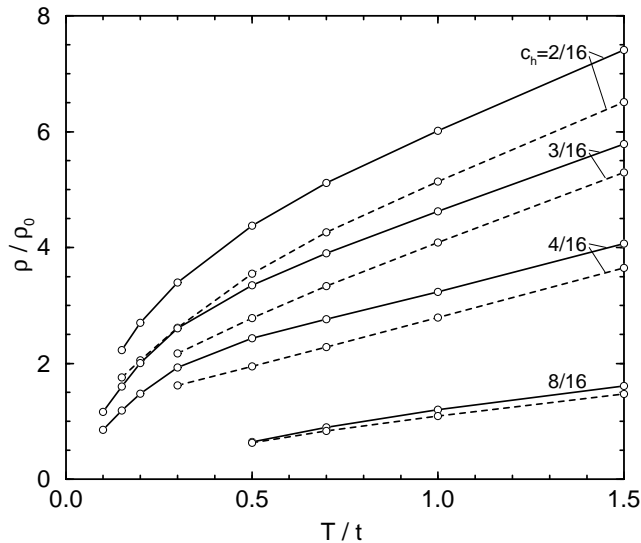,height=10cm,angle=-90}
\fi
\caption{
Resistivity $\rho$ vs. $T$ at different $c_h=N_h/N$ for
$J=~0.3~t$ (full lines) and $J=0$ (dashed lines).
} \label{5.8}
\end{figure}

In Fig.~\ref{5.8} we present results for $\rho(T)$ at various $c_h$
for $J/t=~0.3$, and in comparison also for $J=0$. We first notice that
at $T>t$ calculated $\rho(T) \propto T$.  The slope at $T>t$ is nearly
independent of $J$, the main effect of finite $J>0$ being the upward
shift of $\rho(T)$ curves. The shift also decreases as doping is
increased, since it is plausible that the effect of $J$ nearly
vanishes in the overdoped regime $c_h > 0.3$. The slope at $T>t$ can
be approximately given (for $c_h < 0.25$) as $d\rho/dT \sim \zeta
\rho_0 k_B /c_h t$ with $\zeta \sim 0.4$. This value confirms the
relevance ot the RPA (Brinkman and Rice 1970), which yields for a
square lattice $\zeta \sim 0.72$.

One could expect essential changes in the regime $T\alt J$.  It is
evident from Fig.~\ref{5.8} that at the intermediate doping $0.15< c_h
\alt 0.25$ the curve $\rho(T)$ changes the slope at $T \sim
J$. Nevertheless $\rho(T<J)$ is still linear in $T$, as emerges also
from the observed universal form (\ref{eo2}).  As discussed in
Sec.~5.4.2, the underdoped systems as $c_h=2/16$ show already
deviations from the universality (\ref{eo2}). Two effects should be
mentioned. In the regime where our analysis is fully applicable and
$l_s<L$ we find sublinear $\rho(T) \propto T^{1-\eta}$, which could be
interpreted also as $\rho(T) \sim A+BT$ with a positive
intercept $A>0$. On the other hand, the appearance of $D_c\neq 0$ in
our results seems to be closely related to another scale $T\sim T^*$
where a kink of $\rho(T)$ appears. It should be also noted that the
introduction of the n.n.n. hopping $t' \neq 0$ does not change
qualitative conclusions.

\subsection{Relation to experiments}

In 2D $\sigma$ is naturally expressed in terms of the universal
constant $\rho_0=\hbar/e_0^2$.  The corresponding 3D conductivity of a
stack of 2D conducting sheets with an average distance $\bar{c}$ is
given instead by $\sigma_{3D}= \sigma/\bar c$.  For comparison with
experiments we reduce the 3D measured values to the 2D conductivities
for three different cuprates at intermediate doping, i.e.  LSCO with
$x=0.2$ (Uchida {\it et al.} 1991), Bi$_2$Sr$_2$CaCu$_2$O$_8$ (BISCCO)
(Romero {\it et al.}  1992), and YBCO (Cooper {\it et al.} 1993).
Experimental results are taken at relatively low temperatures
$T<200~K$, i.e. at $T<T_{fs}$ below the sensitivity of our
small-system calculations.  It is an open experimental question
whether spectra reduced in this way are really universal, i.e. whether
they depend only on the hole concentration $c_h$ in conducting CuO$_2$
layers or other details of material structure are still
important. Moreover, effective $c_h$ for presented materials are known
only approximately, except for LSCO where $c_h \sim x$, i.e. it is
estimated that $c_h \sim 0.23$ for BISCCO and YBCO (Batlogg {\it et
al.} 1994).

\begin{figure}[ht]
\centering
\iffigure
\epsfig{file=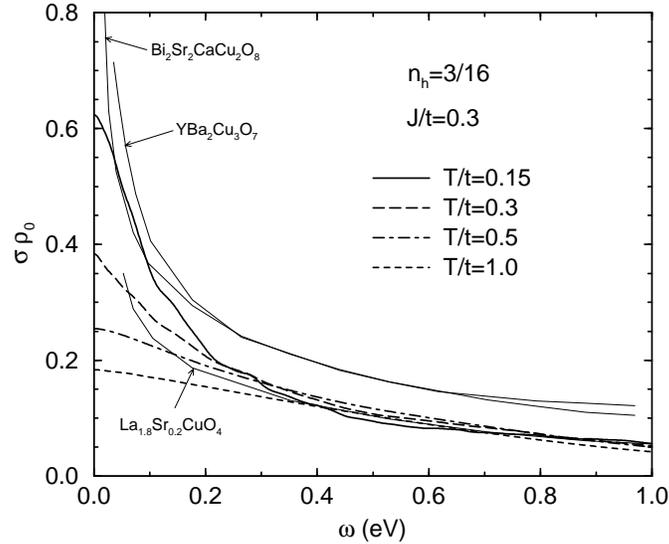,height=10cm,angle=-90}
\fi
\caption{
Sheet conductivities $\sigma(\omega)$ for various $T/t$, in comparison
with measurements in different cuprates. Experimental results refer to
$T<200~K$.  } \label{5.9}
\end{figure}

Taking into account this uncertainty calculated spectra
$\sigma(\omega)$ for $c_h=3/16$ are in a quantitative agreement with
measurements, as seen in Fig.~\ref{5.9}, where the energies are
expressed in eV using $t=0.4$~eV.  We note also that at high $\omega \agt
1~eV$ the fall-off of calculated $\sigma(\omega)$ is faster than that
of measured ones. This could be explained by the emerging contribution
of transitions to excited (charge transfer or upper Hubbard band)
states not taken into account within the $t-J$ model, but clearly
identified experimentally (Uchida {\it et al.} 1991).

It should be noted that also calculated $\tau^{-1}(\omega)$ and
corresponding parameters $\lambda$ and $\xi$, as defined by the
equation (\ref{eo6}) are close to the experimental ones. E.g., an
analysis in a wide frequency range $\omega<1000~$cm$^{-1}$ presented
for YBCO and BISCCO data by El Azrak {\it et al.} (1994) yields
$\lambda \sim 0.11$, while a broader class of optimally doped
materials is consistent with $0.1<\lambda <0.15$ (Baraduc {\it et al.}
1995). A recent investigation (Startseva {\it et al.} 1997) in a
narrower $\omega$ range obtains for an overdoped LSCO ($x=0.22$) similar
values $0.06 <\lambda <0.08$.

Also $m^*/m_t$ in Fig.~\ref{5.5}b qualitatively agrees with
experimental findings (Romero {\it et al.} 1992, El Azrak {\it et al.}
1994, Tanner and Timusk 1992, Cooper {\it et al.} 1993). For a
quantitative comparison we should note that by lowering $T$ to
experimentally investigated values we expect the increase of $m^*$ at
low $\omega$. On the other hand, assuming realistic values $a_0 =
0.38~{\rm nm}$ and $t=0.4~{{\rm eV}}$ we should also take into account
that $m_t \sim 0.6~m_0$.

\begin{figure}
\centering
\iffigure
\epsfig{file=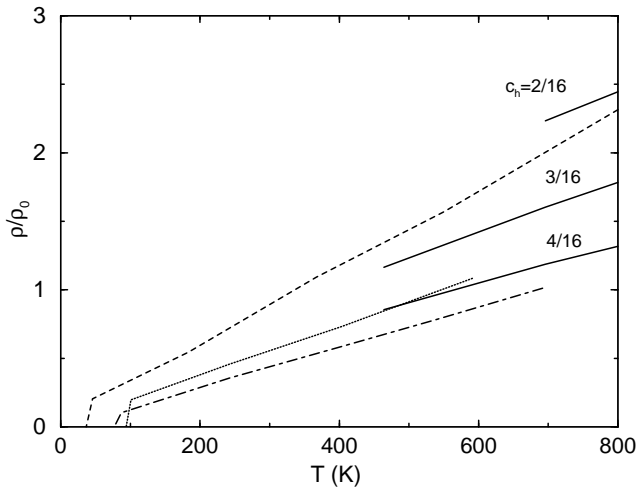,height=10cm,angle=-90}
\fi
\caption{
Sheet resistivities $\rho(T)$ for various dopings (full lines) in
comparison with measurements in LSCO with $x=0.15$ (dotted), BISCCO
(dashed), and YBCO (dash-dotted).  } \label{5.10}
\end{figure}

In Fig.~\ref{5.10} we compare calculated resistivities to the measured
ones. It should be pointed out that there is a restricted $T$ window
where a comparison could be made since $T_{fs} \sim 450 K$, whereby
at $T\sim T_{fs}$ also finite-size effects start to influence our
analysis.  Nevertheless, for the intermediate doping $c_h \sim 0.2$ our
$\rho(T)$ results match quantitatively well experimental ones for
cuprates with comparable hole concentrations, i.e. for BISCCO, YBCO
and LSCO with $x=0.15$ (Takagi {\it et al.} 1992, Batlogg {\it et al.}
1994, Iye 1992).

Turning to underdoped materials, there are certain features which are
reproduced in our results. First, our $\mu(\omega)$ for a single hole,
see Fig.~\ref{5.2}, as well as several $T=0$ numerical studies (Sega
and Prelov\v sek 1990, Dagotto 1994) indicate an existence of a
mid-infrared peak in $\sigma(\omega)$, being a signature of
magnon excitations in a spin background with a longer range AFM
order. This feature has been seen in experiments in LSCO (Uchida {\it
et al.} 1991 ), although its appearance in other materials is
controversial.  We also reproduce the observation that at moderate
doping $c_h>0.1$ the resistivity essentially scales as $\rho \propto
1/c_h$. An unproportional increase of $\rho$ at low doping
$c_h<0.1$, as deduced when comparing values in Fig.~\ref{5.10} with
single-hole mobilities in Fig.~\ref{5.3}, is as well consistent with
experiments, e.g. for LSCO with $x<0.1$ (Takagi {\it et al.}  1992).

\setcounter{equation}{0}\setcounter{figure}{0}
\section{Magnetic Properties}

The static spin response and the spin dynamics in the undoped and doped
AFM state in cuprates have been experimentally studied by the neutron
scattering (see e.g. Shirane 1991), by NMR techniques (see
e.g. Slichter 1994), as well by other methods.

Magnetic excitations in cuprates as measured by the neutron
scattering reveal in insulating materials La$_2$CuO$_4$ (Keimer {\it
et al.}  1992, Hayden {\it et al.} 1996) and YBa$_2$Cu$_3$O$_6$
(Rossat-Mignot {\it et al.} 1991) a remarkable agreement with the
magnons within the planar Heisenberg AFM. On the other hand, a
consistent description of magnetic properties of doped cuprates is
still lacking. Nevertheless it is quite clear that the normal-state
spin dynamics differs qualitatively from the one expected for LFL.  In
doped LSCO NMR and NQR spin-lattice relaxation time $T_1$ on Cu nuclei
is generically nearly $T$ and doping independent in the normal state
($T>T_c$) (Imai {\it et al.} 1993), in contrast to the Korringa
law $1/T_1 \propto T$ in normal metals. Also, in the same regime the
low-$\omega$ dynamical susceptibility in doped systems appears to be
consistent with $\chi''(\omega) \propto \omega/T$ (Shirane 1991,
Keimer {\it et al.} 1992, Sternlieb {\it et al.} 1993).  Experimental
investigations in recent years have been focused on underdoped
materials, which show in the normal phase the emerging spin gap both
in the NMR (Slichter 1994) and in the neutron scattering (Sternlieb
{\it et al.} 1993), but also the appearance of charge stripes related
to AFM domains (Tranquada {\it et al.}  1995). Since the latter
phenomena appear only at lower $T$, hardly accessible in our studies,
we shall comment them only briefly lateron.

Well understood theoretically is so far the isotropic quantum AFM,
with the long-range order at $T=0$ (Manousakis 1991).  For doped
systems several phenomenological explanations (see Sec.~2.2) have been
presented for magnetic properties. NMR and NQR data on the spin
dynamics have been interpreted within the NAFL scenario (Millis {\it
et al.} 1990, Millis and Monien 1992), where the $T$ dependence is
attributed to the variation of the AFM correlation length $\xi(T)$. At
low hole doping the mapping on the quantum critical scaling regime of
the nonlinear sigma model, with the main ingredient $\xi \propto 1/T$,
has been advocated by Sokol and Pines (1993). An alternative scenario
for the low-$(\omega,T)$ behaviour is the MFL scenario (Varma {\it et
al.} 1989, Littlewood and Varma 1991), where $\xi(T)$ is short and not
crucially $T$-dependent.

Most reliable model results for magnetic properties of doped AFM have
been obtained for static quantities by QMC studies of the Hubbard
model and by the g.s. ED of the $t$-$J$ model (see Dagotto 1994), as
well as by means of the HTE analysis (Singh and Glenister 1992a). Much
less conclusive are results for the spin dynamics. While g.s. ED
studies of the spin structure factor $S(\vec q,\omega)$ (Tohyama {\it
et al.}  1995, Eder {\it et al.} 1995) reveal quite an essential
difference between $S(\vec q,\omega)$ and corresponding charge spectra
$N(\vec q,\omega)$, they cannot give reliable conclusions for the
low-$\omega$ behaviour.  There have been only few attempts to address
numerically dynamical properties at $T>0$ (Tohyama {\it et al.} 1993),
prior to the application of the FTLM (Jakli\v c and Prelov\v sek
1995a, b).  It should be reminded that most challenging questions
(related to the normal state) refer to the dynamics at low $\omega$ and to
the d.c. spin response in the strong correlation regime $J<t$ at low
$T<J$.

\subsection{Spin response}

We consider the response of the electronic system to the
time-dependent, spatially modulated magnetic field,
which couples to the spin degrees of freedom via a Zeeman term
\begin{equation}
H' = N M^z_{\vec{q}} B_{\vec{q}},\qquad M^z_{\vec{q}} =
-{g\mu_B\over N}\sum_{i} e^{i\vec{q}\cdot\vec{R_i}} S_{i}^z,  \label{ms1}
\end{equation}
Within the linear response theory the magnetization
response is given by
\begin{equation}
\delta \langle M^z_{\vec{q}}\rangle (\omega)=g^2\mu_B^2
\chi(\vec{q},\omega)B_{\vec{q}}(\omega), \label{ms2}
\end{equation}
where
\begin{eqnarray}
\chi(\vec{q},\omega)&=&i\int_{0}^\infty dt\; 
e^{i\omega t}
\langle[S^z_{\vec{q}}(t),S^z_{-\vec{q}}(0)]\rangle, \nonumber \\
S^z_{\vec{q}}&=& (1/\sqrt{N})\sum_{i}e^{i\vec{q}\cdot\vec{R}_i}
S^z_{i}. \label{ms3}
\end{eqnarray}
$\chi''(\vec{q},\omega)$ can be expressed 
with the dynamical spin structure factor $S(\vec {q},\omega)$
\begin{eqnarray}
\chi''(\vec{q},\omega)&=&\pi(1-e^{-\beta\omega})S(\vec{q},\omega),
\nonumber \\
S(\vec{q},\omega)&=& {1\over \pi}{\rm Re}\int_{0}^\infty dt\; e^{i\omega t}
\langle S^z_{\vec{q}}(t) S^z_{-\vec{q}}(0)\rangle. \label{ms4}
\end{eqnarray}
$S(\vec {q},\omega)$ is the quantity directly evaluated within the
FTLM, following the expression (\ref{fi4}). Consequently the static
structure factor and the static susceptibility can be evaluated,
\begin{eqnarray}
S(\vec{q})&=&\int_{-\infty}^\infty S(\vec{q},\omega)d\omega=
\langle S^z_{\vec{q}}S^z_{-\vec{q}}\rangle, \nonumber \\
\chi(\vec{q})&=&\chi'(\vec{q},0)=\frac{1}{\pi}{\cal P}
\int_{-\infty}^\infty\frac{\chi''(\vec{q},\omega)}{\omega}d\omega.
\label{ms5}
\end{eqnarray}

\subsection{Uniform spin susceptibility and Wilson ratio}

The uniform case $\vec{q}=0$ is particular, since $S_z^{tot}$ is a
conserved quantity. Then one can use for the uniform static
spin-susceptibility $\chi_0$ the expression
\begin{equation}
\chi_0=\chi(\vec{q}=0)=\beta S(\vec{q}=0)=\frac{\langle
(S_z^{tot})^2\rangle}{N T}. \label{mu1}
\end{equation}
The calculation reduces to the one involving only a thermodynamic
average of a conserved operator, and an analysis analogous to the one
used in Sec.~4 applies.  This simplification allows for the
consideration of larger systems, i.e. we calculate $\chi_0(T)$ for a
system of $N=20$ sites in the range $0<c_h<0.3$, while for the undoped
AFM (Heisenberg model) we reach $N=26$ (Jakli\v c and Prelov\v sek
1996).

\begin{figure}
\centering
\iffigure
\epsfig{file=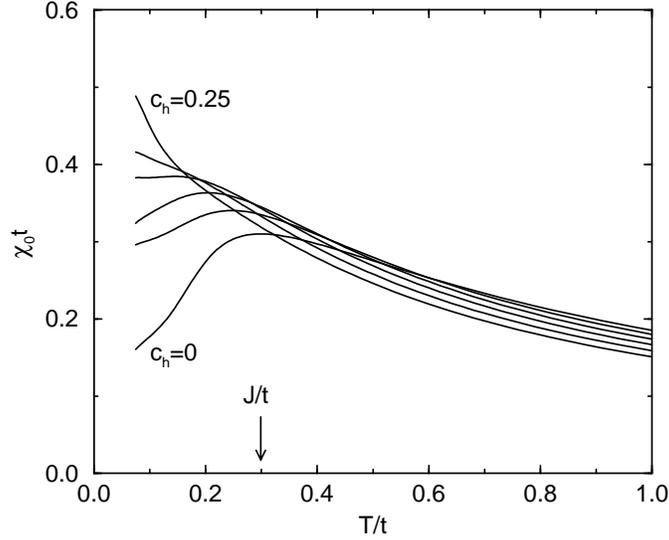,height=10cm,angle=-90}
\fi
\caption{
Uniform susceptibility $\chi_0$ vs. $T$ at several $c_h$ in
steps of 0.05 within a system of $N=20$ sites. $c_h=0$ is obtained
for $N=26$. } \label{6.1}
\end{figure}

Results at several $c_h$ are presented in Fig.~\ref{6.1}. For $c_h=0$
our results agree with the HTE (Singh and Glenister 1992a) down to $T\sim
0.3~J$. In an AFM $\chi_0$ exhibits a maximum at $T=T^*\sim J$, which
reflects the onset of the short range AFM order for $T<T^*$.  Namely,
at $T\to 0$ the longitudinal spin fluctuations in an ordered AFM
gradually freeze out, while transverse ones remain constant, thus
leading to a reduced $\chi_0(T\to 0)$.
   
For a finite doping $\chi_0$ is only weakly diminished for $T>J$, the
reduction being due to a reduced number of spins in the system. On the
other hand, the qualitative change appears for $T<J$.  The maximum
$T^*$ gradually shifts to lower $T$ with doping and finally disappears
at $c_h>0.15$.  In the overdoped regime $c_h \agt 0.15$ we observe a
monotonous Pauli-like $\chi_0$ for $T<0.2~t$, which could signify an
onset of a low-$T$ behaviour consistent with the LFL picture. Still up
to $c_h>0.6$ we do not find the usual LFL behaviour $\chi_0(T)={\rm
const}$, but rather $\chi(T)$ still increases nearly linearly on
lowering $T\to 0$.

Obtained results are quite consistent with experiments on cuprates.
First, $\chi_0$ at $T\to 0$ increases with doping as found in LSCO
(Johnston 1989, Torrance {\it et al.} 1989) and the existence of a
maximum of $\chi_0(T)$ at $T^*(c_h)$ in underdoped cuprates has been
interpreted in terms of a pseudogap scale (Batlogg {\it et al.}
1994). It is evident from our results that such a pseudogap is
inherent within the $t$-$J$ model and is intimately related to the
short range AFM order. It is remarkable that in the model this feature
disappears, i.e. $T^* \to 0$, just at a similar (optimum) doping $c_h
\sim 0.15$ as in cuprates (Batlogg {\it et al.}  1994).  Note also
that the unusual linear increase of $\chi_0(T\to 0)$ in the overdoped
regime, see Fig.~\ref{6.1} for $c_h=0.2$, seems to be well consistent
with experimental results in overdoped LSCO (Loram {\it et al.} 1996).

\begin{figure}
\iffigure
\centering
\epsfig{file=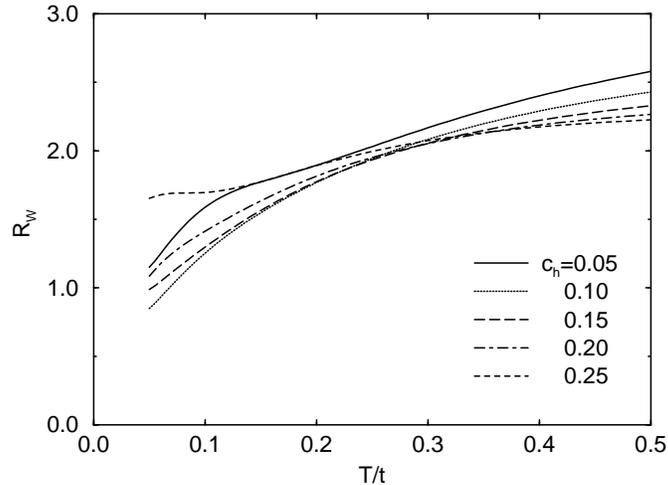,height=10cm,angle=-90}
\fi
\caption{Wilson ratio $R_W$ vs. T at several $c_h$.}
\label{6.2}
\end{figure}

Here we can also comment on the ratio $W=(g^2 \mu_B^2\chi_0)/\gamma$
where $\gamma=C_V/T$. It is used as a test for the concept of nearly
free QP, where one expects $W_0= {1\over 3}(\pi k_B/\mu_B)^2$. The
meaningful measure is thus the Wilson ratio $R_W=W/W_0$ being often
studied in correlated systems, in particular in connection with
heavy-fermion metals. In our case it is convenient to perform the test
on $s$ directly by defining $\tilde \gamma = s/T$ (see also Loram {\it
et al.} 1996). In our notation the dimensionless Wilson ratio can be
expressed as
\begin{equation}
R_W={4 \pi^2 \over 3} {(k_BT/t) (\chi_0 t)\over s/k_B } .\label{mu2}
\end{equation}
We can now easily evaluate $R_W$ by comparing results in
Fig.~\ref{4.3}a with Fig.~\ref{6.1} and results are shown in
Fig.~\ref{6.2}. It is quite striking that in the intermediate regime
$0.1 \alt c_h \alt 0.2$ at low $T\ll J$ we find values, very close to
the free fermion one, i.e. $R_W=1$. In the overdoped regime calculated
values are somewhat larger, but still $R_W<2$. The same seems to be
the case for the underdoped case $c_h=0.05$, where results at lowest
$T$ should be taken with care. With increasing $T$ also $R_W$ is
increasing and seems to reach quite a wide plateau for $T\agt J$ with
$R_W \sim 2$.

This finding is quite puzzling since clearly doped AFM are far from a
simple LFL, and apparently even further from free fermions. Note
however that experimental results also yield $R_W \sim 1$, both for LSCO
and YBCO (Loram {\it et al.} 1996). Moreover, the same experiments
indicate that $s \propto T\chi_0$ in a wide range of $T$ and doping,
so $R_W$ as defined in the equation (\ref{mu2}) is nearly $T$ independent. 

\subsection{Spin structure factor and dynamical susceptibility}

\begin{figure}
\centering
\iffigure
\epsfig{file=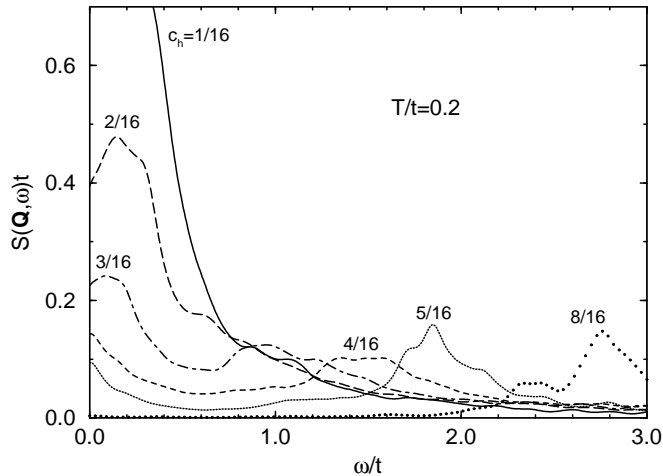,height=10cm,angle=-90}
\fi
\caption{
Spin structure factor $S(\vec Q, \omega)$ at fixed $T=0.2~t <J$ and
different dopings $c_h$.  } \label{6.3}
\end{figure}

Let us first discuss $S(\vec Q, \omega)$ spectra at the AFM wavevector
$\vec Q =(\pi, \pi)$, where the spin response is most
pronounced. In Fig.~\ref{6.3} we present results at fixed $T=0.2~t<J$,
but various dopings $c_h$ (Jakli\v c and Prelov\v sek 1995a).  It
should be noted that $S(\vec Q, \omega)$ is not symmetric around
$\omega=0$, hence a maximum appears in general at $\omega>0$.  The
most interesting feature in Fig.~\ref{6.3} is the qualitative change
of spectra on doping. At low doping $c_h<0.12$ we see that $S(\vec
Q,\omega)$ is dominated by a single central peak, i.e. we are
observing mainly the AFM spin fluctuations of a Heisenberg model. The
main effect of holes is to reduce the AFM correlation length, and
consequently the intensity of the central peak.

In the intermediate regime $0.12<c_h<0.3$ a high-frequency component
with $\omega \agt t$ emerges, coexisting with the remaining
low-$\omega$ fluctuations. It is quite plausible to attribute the
high-$\omega$ dynamics to the free-fermion-like component of the
correlated system, in particular since it appears to be quite
independent of $J$ (provided that $J<t$). Although this observation is
consistent with previous studies (Tohyama {\it et al.} 1993, Putikka
{\it et al.}  1994), the coexistence of spin-fluctuation and
free-fermion timescales at the intermediate doping has been clearly
established only by using the FTLM (Jakli\v c and Prelov\v sek
1995a). Namely, it is harder for other methods, e.g.  within the HTE
method, to resolve coexisting different timescales.  The dual
character is a crucial property, since the free-fermion part
determines to large extent static spin correlations $S(\vec q)$
through the relation (\ref{ms5}), as well as the electron (charge)
density correlations $N(\vec q)$, discussed in Sec.~8.1. The latter
have been in fact interpreted in terms of a quasi FS (Putikka {\it et
al.}  1994).  On the other hand the low-$\omega$ spin dynamics
dominates dynamical and static spin susceptibilities,
i.e. $\chi''(\vec q,\omega)/\omega$ and $\chi(\vec q)$, hence the
low-$\omega$ neutron scattering and NMR processes.  It is also evident
that in the overdoped regime $c_h>0.3$ the low-frequency component is
disappearing and the free-fermion-like fluctuations tend to exhaust
the spectra. $S(\vec Q,\omega)$ is nevertheless quite distinct from the
one for free fermions, even at $c_h \sim 0.5$.

Before we present results for general ${\vec q}\neq \vec Q$, let us
discuss the AFM correlation length $\xi(T)$. One can evaluate $\xi(T)$
from the static real space correlations $S(\vec r)$, corresponding to
$S(\vec q)$, i.e.
\begin{equation}
\xi^2= \frac{1}{4S(\vec{Q})}
\sum_i |\vec{r}_i|^2 \exp(i \vec Q\cdot\vec r_i)S(\vec{r}_i). \label{mf1}
\end{equation}
In the most interesting regime $c_h=0.1-0.3$ we find that $\xi$ is
short, typically $\xi \sim 1$, governed by correlations at $r_i
=1$. It increases less than $30 \% $ between $T=J$ and $T=J/3$. This
finding (Jakli\v c and Prelov\v sek 1995a) is well in agreement with
the HTE studies (Singh and Glenister 1992a), with the QMC
results for the Hubbard model (Furukawa and Imada 1992) as well as
with values $\xi(T=0)$ obtained via the ED within the $t-J$ model
(Dagotto 1994).  Similar values for $\xi$ can be extracted also
considering our results for static $\chi(\vec q)$, although results
are less conclusive at low $T\sim T_{fs}$.

In recent years experiments in cuprates have shown the possibility of
longer range (or even long range) spin correlations with $\vec q \neq
\vec Q$ (Tranquada {\it et al.} 1995, Hayden {\it et al.} 1996). 
This has been related either to an incommensurate spin order or to the
appearance of charge stripes. Such structures are well possible within
the $t$-$J$ model (Prelov\v sek and Zotos 1990, White and Scalapino
1997b) although it is controversial whether they are stable in the
most interesting regime $J\ll t$. Nevertheless we do not expect to
resolve them in our calculations for $J/t=0.3$, since they could
possibly appear only for $T<T_{fs}$ and could be also missed due to
particular clusters which do not prefer expected orderings.
 
\begin{figure}
\centering
\iffigure
\epsfig{file=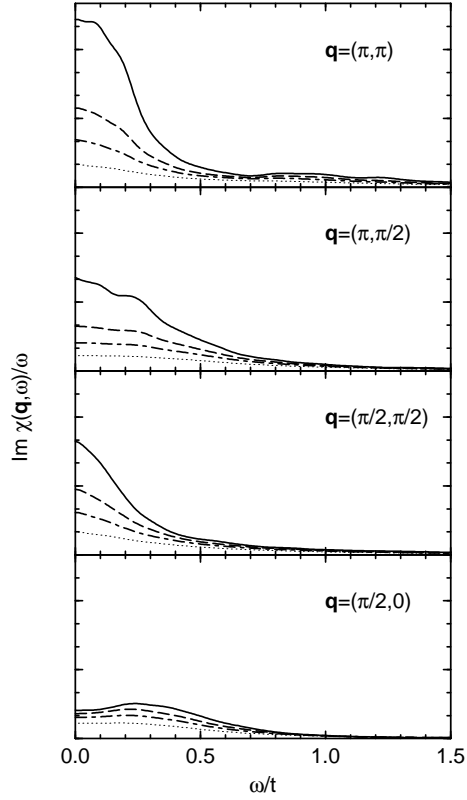,height=11cm}
\fi
\caption{ $\chi''(\vec q,\omega)/\omega$ at $c_h=3/16$ for
nonequivalent $\vec q$ and different $T$: $T/t=0.1$ (full line), $0.2$
(dashed line), $0.3$ (dash-dotted line), and $0.5$ (dotted line).  }
\label{6.4}
\end{figure}

Let us turn to dynamical susceptibilities. Fig.~\ref{6.4} displays
$\chi''(\vec q,\omega)/\omega$ for fixed $c_h=3/16$, but various $T$
and nonequivalent $\vec q$. Note that on a $4\times 4$ lattice $\vec
q= (0,\pi)$ and $\vec q = (\pi/2,\pi/2)$ are equivalent.  In contrast
to Fig.~\ref{6.3}, high-$\omega$ features are now suppressed. Also,
the free-fermion-like component is well separated from the
low-$\omega$ part only for $\vec q \sim \vec Q$, where one expects a
gap in the response of free fermions with a well defined FS.  For
other $\vec q$ the free-fermion contribution persists at larger
$\omega>J$ in the form of a long tail, while in the low-$\omega$
regime it merges with the spin contribution.

The most striking feature of Fig.~\ref{6.4} is the strong $T$
dependence of the low-$\omega$ spectra, whereas the AFM correlation
length $\xi$ has been found to be only weakly $T$ dependent.  This
conclusion on $\chi''(\vec q,\omega)/\omega$ seems to hold within the
correlation regime $|\vec q - \vec Q| < \xi^{- 1}$, where also
$\chi(\vec q) \sim \chi(\vec Q)$.  The relevant volume in the $\vec q$
space clearly increases on doping and exhausts for $c_h=3/16$ already
the substantial part of the Brillouin zone. The variation at low
$\omega$ appears to follow,
\begin{equation}
\chi''(\vec q,\omega)/\omega \propto 1/T,\qquad \omega < T< J,
\label{mf2}
\end{equation}
or equivalently, from the relation (\ref{ms4}) $S(\vec q, \omega)$ is
nearly $T$ and $\omega$ independent in the same regime.  On the other
hand, at the same doping scaling does not hold for $\vec q$ outside
the correlation volume, e.g. for $\vec q=(0,\pi/2)$ in Fig.~6.4, where
$\chi''(\vec q,\omega)/\omega$ is approximately $T$ independent as
expected within the LFL.

Spectra discussed above have as a direct consequence the $T$ variation
of the static $\chi(\vec q)$ for $T<J$. We observe a pronounced $T$
dependence, e.g. $\chi(\vec q) \propto 1/T$ in a wide regime $J/3
< T < t$ for all $q$ within the correlation regime. It should be
however noted that we are quite restricted in the range of $T/J$, so
that more quantitative conclusions on a possible power-law (or
logarithmic) variation with $T$ are not feasible.

\subsection{Local spin dynamics}

\begin{figure}
\centering
\iffigure
\epsfig{file=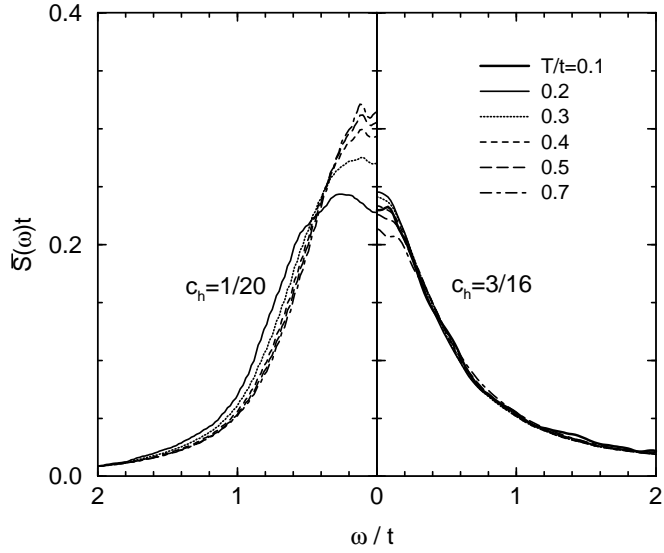,height=10cm,angle=-90}
\fi
\caption{
Local spin correlation function $\bar S(\omega)$ for $c_h=
1/20$ and $c_h=3/16$, and different $T$.
} \label{6.5}
\end{figure}

In order to describe properly the spin correlation function $S(\vec
q,\omega)$ it is very helpful to consider the local spin correlation
function $S_L(\omega)$ and its symmetric part $\bar S(\omega)$
(Jakli\v c and Prelov\v sek 1995b),
\begin{eqnarray}
S_L(\omega)&=&{1\over N} \sum_{\vec q} S(\vec q, \omega), \nonumber \\
\bar S(\omega)&=&S_L(\omega)+S_L(-\omega)= (1+e^{-\beta\omega})  
S_L(\omega). \label{ml1}
\end{eqnarray}
It should be noted that $S_L(\omega)$ and the related susceptibility
$\chi_L(\omega)$ have been directly measured in cuprates by the neutron
scattering (Shirane 1991), while the NMR relaxation on $^{63}$Cu
effectively yields the information on $S_L(\omega \rightarrow 0)$, as
discussed in Sec.~6.5.  An important restriction on $\bar S(\omega)$
is the sum rule following from the equation (\ref{ms5}),
\begin{equation}
\int_0^{\infty}  \bar S(\omega) d\omega = \langle (S_i^z)^2\rangle = 
{1 \over 4} (1-c_h). \label{ml2}
\end{equation}
When we perform the calculation of $\bar S(\omega)$ we omit the
singular $\vec q=0$ term in the summation (\ref{ml1}), so the sum rule
(\ref{ml2}) serves as an useful test.  In Fig.~\ref{6.5} we display
$\bar S(\omega)$ for different dopings $c_h=1/20$ and $3/16$, and
several $T$ in the range $0.1 \le T/t \le 0.7 $ (Jakli\v c and
Prelov\v sek 1995b). It is immediately evident that $\bar S(\omega)$
at the optimum doping $c_h=3/16$ is essentially $T$ independent in a wide
$T$ range, although one crosses the exchange-energy scale $T
\sim J$.

For the underdoped case $c_h=1/20$ as well as for the undoped AFM the
behaviour is analogous for higher $T >T_0 \sim 0.7~J$. Deviations
appear at $T<T_0$, leading at $T\to 0$ to a decrease and to a
flattening of $\bar S(\omega<2J) \sim const$, whereby a weak
enhancement of $\bar S(\omega>2J)$ is then required by the sum
rule. At least this regime of essentially pure Heisenberg model is
well understood theoretically. The behaviour at $T>T_0$ is consistent
with the quantum critical regime within the AFM, whereas for $T< T_0$
the renormalized classical regime applies (Chakravarty {\it et al.}
1989).  In the latter regime we are dealing with longer range
AFM correlations $\xi \gg 1$, hence $\bar S(\omega )$ is essentially
that of an ordered AFM where the simple magnon picture leads in 2D to
$\bar S(\omega<2J) \sim const$.

\begin{figure}
\centering
\iffigure
\epsfig{file=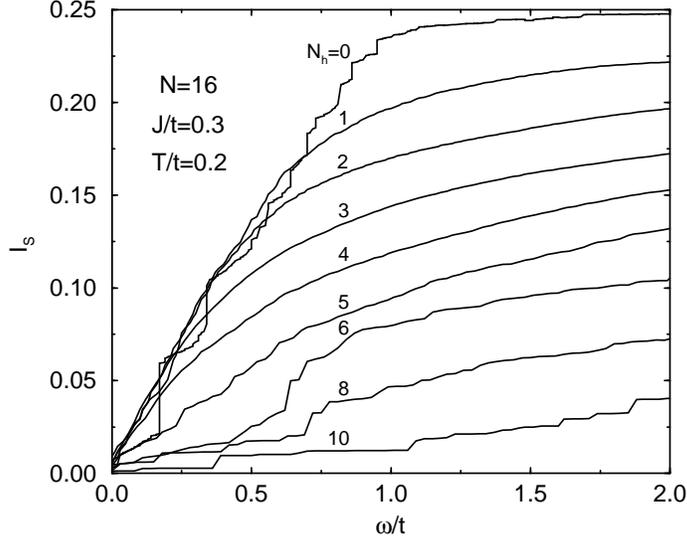,height=10cm,angle=-90}
\fi
\caption{
Integrated spin correlation spectra $I_S(\omega)$ at fixed $T=0.2~t$
and different $c_h=N_h/N$.  } \label{6.6}
\end{figure}

In Fig.~\ref{6.6} we follow the doping dependence of $\bar S(\omega)$
at fixed $T=0.2~t<J$. For convenience we plot again the integrated
spectra, in analogy to the equation (\ref{eo1}),
\begin{equation}
I_S(\omega)= \int_0^{\omega}  \bar S(\omega') d\omega'. \label{ml3}
\end{equation}
For chosen $T$ results are again most reliable at the intermediate
doping.  The most striking message is that the initial slope of
$I_S(\omega)$ and consequently $S_L(\omega \rightarrow 0)$ is nearly
doping independent for $0\le c_h \le 0.25$, as well as $T$ independent
at the intermediate doping.  Only in the overdoped systems with
$c_h>0.25$ the low-$\omega$ behaviour changes qualitatively, where the
low-$\omega$ contribution is strongly suppressed as expected in a
(more) normal FL.  In a pure Heisenberg model the spin dynamics is
nearly exhausted in the range $\omega<3J$ with excitations being
magnons. On the other hand, at the intermediate doping $\bar
S(\omega)$ decreases smoothly up to $\omega \sim 4t$, this being a
consequence of the free-fermion-like component. Still up to $c_h \sim
0.3$ the dominant scale of spin fluctuations remains related to $J$.

Using equations (\ref{ms4}) and (\ref{ml1}) we finally note that the
$T$-independent $\bar S(\omega)$ leads to the universal form
\begin{equation}
\chi''_L(\omega) = {1\over \pi}\tanh\left( {\omega \over 2T} 
\right) \bar S(\omega), \label{ml4}
\end{equation}
where the $T$ dependence is only in the thermodynamic prefactor. It
follows from the equation (\ref{ml4}) that for $T<J$ the only relevant
scale for $\chi''_L(\omega)$ is given by $T$. The same should hold for
the response at fixed $\vec q$. So one can generalize the expression
(\ref{ml4}) to
\begin{equation}
\chi''(\vec q, \omega) \sim {\chi(\vec q) \over \chi_L} \chi''_L(\omega) 
\sim {2\pi \ln^{-1} (\xi q_m) \over |\vec q - \vec Q|^2 +\xi^{-2}} 
\chi''_L(\omega), \label{ml5}
\end{equation}
where the cutoff $q_m\sim \pi$. The scaling is expected to be valid at
least for $|\vec q - \vec Q| < \xi^{-1}$. Since $\xi(T)$ remains
weakly $T$ dependent on decreasing $T$, this introduces an additional
$T$ variation in the expression (\ref{ml5}). Far from the regime $\vec
q \sim \vec Q$, in particular for $q \sim 0$, the response is more
free-fermion like, i.e. $\chi''(\vec q, \omega)$ is $T$ independent,
as seen in Fig.~\ref{6.4} for $\vec q=(\pi/2,0)$.

\subsection{Nuclear magnetic relaxation}

For experimental studies of static and dynamical spin correlations the
NMR and the NQR relaxation are among the most valuable tools. The
hyperfine coupling of electronic spins to $^{63}$Cu and $^{17}$O
nuclear spins $\vec{I}$ within the CuO plane has been established by a
number of authors (Mila and Rice 1989, Shastry 1989, Millis {\it et
al.} 1990, Millis and Monien 1992, Slichter 1994),
\begin{equation}
H_{e-n}=\; \vec{I} \cdot \frac{1}{\sqrt{N}}\sum_{\vec{q}}
{\bf A}(\vec{q})\vec{S}_{\vec{q}}. \label{mn1}
\end{equation}
Form factors are determined by the relative position of
electronic spins (assumed to be centered on Cu sites) and nuclear
spins,
\begin{eqnarray}
^{63}{\bf A}(\vec{q})&=&{\bf A}+2B(\cos q_x a+\cos q_y a),
\nonumber \\
^{17}{\bf A}_\alpha(\vec{q})&=&2C\cos (q_\alpha a/2) \label{mn2},
\end{eqnarray}
where $\alpha=x,y$ refer to Cu-O-Cu axis orientation, and ${\bf A}$,
$B$, $C$ are the hyperfine couplings. ${\bf A}$ is the direct
hyperfine tensor, coupling the electronic and the nuclear spin on the
same Cu site, with components $A_\perp$ and $A_\parallel$,
corresponding to spins oriented perpendicular and parallel to the
CuO$_2$ plane, respectively.

The nuclear spin-lattice relaxation is due to the coupling (\ref{mn1}),
directly related to the low-$\omega$ electronic spin fluctuations.
Let us consider only the magnetic field directed perpendicular to the
CuO$_2$ plane. Then the NMR spin-lattice relaxation rate is given by
(Slichter 1994),
\begin{equation}
{1 \over T_1}= \frac{\zeta}{2N}\sum_{\vec{q}}|A_\perp(\vec{q})|^2 S(\vec
q,\omega_0), \label{mn3}
\end{equation}
where $\omega_0$ is the precession frequency of the nuclear spin in
the magnetic field, and $\zeta=1$. Since $\omega_0 \ll T$, one can
express result as well with the dynamical susceptibility $S(\vec
q,\omega)\approx T\chi''(\vec{q},\omega_0)/\omega_0$.  The NQR
relaxation rate $1/T_1$ is given also by the expression (\ref{mn3}),
but with $\zeta = 4$ (Millis and Monien 1992).  Form factors
(\ref{mn2}) differ essentially for O and Cu sites. Due to AFM
fluctuations spin fluctuations are at maximum around
$\vec{q}=\vec{Q}$, as in the equation (\ref{ml5}).  The weight of this
region enhances $1/T_1$ for $^{63}$Cu, while suppressing it for
$^{17}$O.  This is the most important point explaining the observation
of very different relaxation rates on different nuclei.

In the evaluation of $1/T_1$ within a finite system we again omit
the $\vec{q}=0$ term since $\chi''(\vec q,\omega)/\omega$ is ill
defined for $\omega \rightarrow 0$ due to conserved total $S_z$. A
proper treatment would require a more detailed spin-diffusion
contribution at $q\sim 0$ which however seems to be less important, at
least for the undoped system (Sokol {\it et al.} 1993).  To allow a
direct comparison with experiments we choose hyperfine-coupling
parameters as proposed in the literature (Millis and Monien 1992).

\begin{figure}
\centering
\iffigure
\epsfig{file=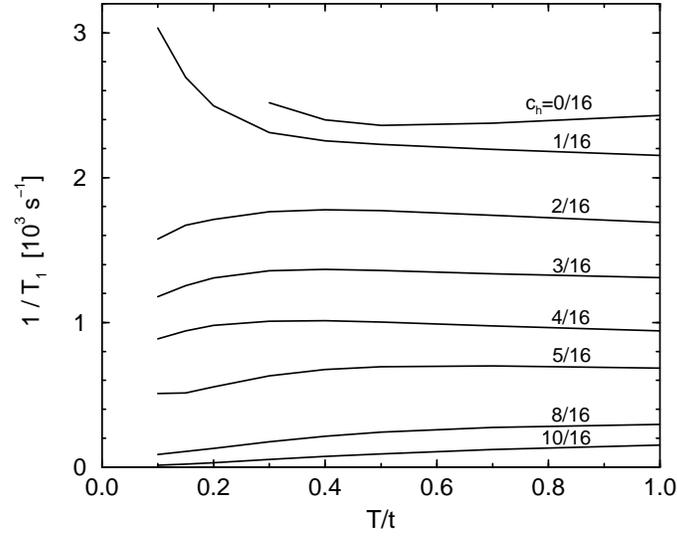,height=10cm,angle=-90}
\fi
\caption{
$^{63}$Cu NQR spin-lattice relaxation rate $1/T_1$ vs.  $T$ for
different dopings $c_h$.  } \label{6.7}
\end{figure}

Results for $^{63}$Cu NQR relaxation rate $1/T_1$ are presented in
Fig.~\ref{6.7} (Jakli\v c and Prelov\v sek 1995a).  For the undoped
case our results agree with Sokol {\it et al.} (1993).  It is
remarkable that $1/T_1$ appears to be nearly $T$ independent for a
broad range of hole concentrations $0.06 <c_h \leq 0.3$.  Only for
overdoped systems with $c_h > 0.5$ the behaviour at $T<t$ approaches
that of a normal LFL with $1/T_1 \propto T$. Since the $^{63}$Cu form
factor is maximum (but slowly varying with $\vec q$) near AFM $\vec
Q$, we get approximately $1/T_1 \propto S_L(\omega_0 \sim
0)$. Previously established universality of $S_L(\omega)$ naturally
explains nearly $T$ independent $1/T_1$, which is moreover also weakly
dependent on $c_h$ for $c_h<0.15$.

Results as shown in Fig.~\ref{6.7} are in agreement, even
quantitatively without any fitting parameters, with remarkable NQR
experiments on doped LSCO (Imai {\it et al.} 1993), which reveal
nearly $T$- and $x$-independent $1/T_1$ at $T>300~{\rm K}$ and
$x<0.15$. In contrast to experiments our results show some variation
of $1/T_1$ with $c_h$ in this doping range. For the optimum doping
lower rates $1/T_1$ are anyhow expected, consistent with the data for
YBCO (Takigawa {\it et al.} 1991, Kitaoka {\it et al.} 1991), where
for $T>T_c$ the rate $1/T_1$ is again only weakly $T$ dependent.

\begin{figure}
\centering
\iffigure
\epsfig{file=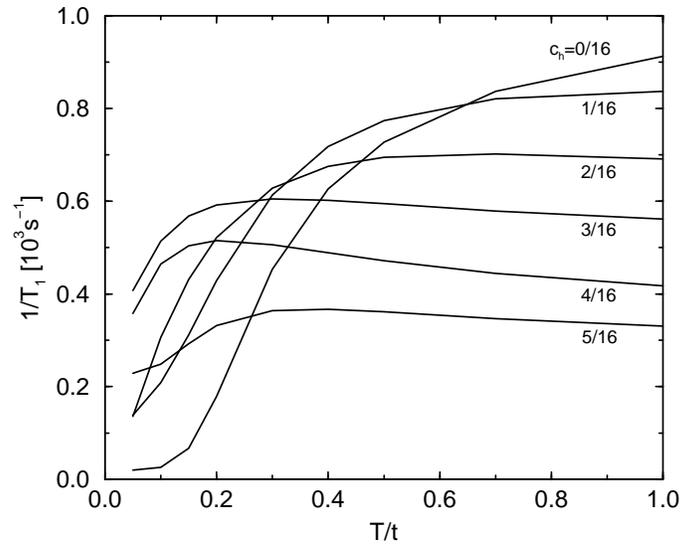,height=10cm,angle=-90}
\fi
\caption{
$^{17}$O NMR spin-lattice relaxation rate $1/T_1$ vs. $T$ for 
different dopings $c_h$.
} \label{6.8}
\end{figure}

Essentially different $T$ dependence of $1/T_1$ on Cu and O nuclei has
been used as an evidence for the importance of strong AFM correlations
and for the non-LFL behaviour in doped cuprates. To evaluate the NMR
relaxation rate $1/T_1$ for $^{17}{\rm O}$ we can use again the
expression (\ref{mn3}) with proper form factors (\ref{mn2}),
projecting out the AFM fluctuations at $\vec q \sim \vec Q$. The
omitted $q \sim 0$ contribution introduces in this case a larger
uncertainty. Nevertheless, for $c_h =1/16$ and $c_h=2/16$ we recover
results very well described with the Korringa behaviour, i.e. $1/T_1
\sim wT $, as seen in Fig.~\ref{6.8} for $T_{fs}<T<J$.  In particular
for $c_h=2/16$ we get $w \sim 0.3$(sK)$^{-1}$, very close to the
actual value $w \sim 0.35$(sK)$^{-1}$ reported for YBCO (Takigawa {\it
et al.} 1991, Kitaoka {\it et al.} 1991).  For $c_h \ge 3/16$
deviations from the Korringa law become more pronounced due to very
short $\xi \sim 1$.

The Gaussian component of the spin-spin relaxation rate $1/T_2$ for 
$^{63}$Cu nuclei can be on the other hand related to static
spin susceptibilities  (Slichter 1994),
\begin{equation}
{1\over T_2} = \sqrt{0.69 \over 8} \left[ {1\over N}
\sum_{\vec q} A_{\parallel}^4(\vec q) \chi^2(\vec q) - \left({1\over
N} \sum_{\vec q} A_{\parallel}^2(\vec q) \chi(\vec q)\right)^2
\right]^{1/2}.\label{mn5}
\end{equation}
The ratio $R=T_1T/T_2$ for $^{63}$Cu nuclei relaxation, which is
approximately $T$ independent in cuprates, has been interpreted as the
evidence for the quantum critical behaviour of the effective spin
system (Sokol and Pines 1993). We find quite analogous $T$ variation
of $R$ (Jakli\v c and Prelov\v sek 1995a) using calculated $\chi(\vec
q,\omega)$ and equations (\ref{mn3}), (\ref{mn5}), together with the
standard parameters (Millis and Monien 1992). The origin of $R(T)\sim
{\rm const}$ is however considerably different from the quantum
critical scenario, since the anomalous $T_2(T)$ dependence can be
related to the $T$ dependence of static $\chi(\vec q)$ which does not
seem to be connected in an evident way with the variation of
$\xi(T)$. Results indicate stronger doping dependence, even at low
doping.  Quantitatively, obtained values are in reasonable agreement
with the experimental ones, e.g. $R \sim 1700~{\rm K}$ at $T=300~{\rm
K}$ for YBCO ($c_h \sim 0.23$), while $R \sim 2400~{\rm K}$ for ${\rm
YBa}_2 {\rm Cu}_3{\rm O}_{6.63}$ (Sokol and Pines 1993).

\subsection{Neutron scattering}

The standard neutron scattering using the thermalized neutrons from the
reactor is restricted to investigations in the energy regime $\omega
<40~$meV.  For cuprates this means only the low-energy part of the
spin dynamics with $\omega<J$.  There have been several measurements,
probing the local $S_L(\omega)$ by measuring $S(\vec q ,\omega)$
integrated over $\vec q \sim \vec Q$ (see Shirane 1991).  To compare
with our results one can simplify the relevant expression (\ref{ml4})
by $\bar S(\omega)\sim \bar S_0$.  Such a form has indeed been used to
describe experiments (Keimer {\it et al.} 1992, Sternlieb {\it et al.}
1993).  In this connection we should take into account that we are not
able to establish in our model calculations (the main reason being too
high $T_{fs}$) the existence of the pseudogap $\omega_g \sim 0.1~J$,
observed in cuprates at low $T\agt T_c$ (Rossat-Mignod {\it et al.}
1991, Sternlieb {\it et al.} 1993) and invoked in the detailed form of
$S_L(\omega)$ used in describing neutron scattering experiments.

\begin{figure}
\centering
\iffigure
\epsfig{file=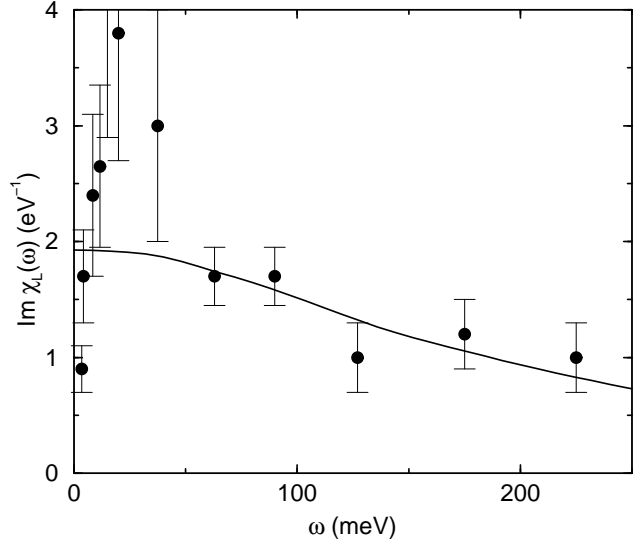,height=10cm,angle=-90}
\fi
\caption{ Local susceptibility $\chi_L^{\prime\prime}(\omega)$
as calculated for $c_h=3/16$ and $T\to 0$, compared with the neutron
scattering result for LSCO with $x=0.14$ (Hayden {\it et al.}
1996). } \label{6.9}
\end{figure}

The introduction of spallation neutron sources has expanded crucially
the accessible energy range of the spin dynamics to $\omega <
1$eV. This gives a possibility to measure the whole relevant range
of the spin dynamics in cuprates, results being available for undoped
La$_2$CuO$_4$ and SC LSCO with $x=0.14$ (Hayden {\it et al.}
1996). Data for the undoped material are well described with the
standard magnon excitations in an ordered AFM, so more challenging are
results for the doped material. Since the authors present the
calibrated local $\chi_L''(\omega)$ up to $\omega \sim 0.25$~eV, we
can compare data quantitatively with our results e.g. for $c_h=3/16$,
as shown in Fig.~\ref{6.9}.  For the latter case measurements were
performed at $T=17~$K$\ll J$, so we present in Fig.~\ref{6.9} the $T
\to 0$ limit of our result following equation (\ref{ml4})
$\chi_L''(\omega>T) \sim \bar S(\omega)/\pi$. We observe that the agreement
is quite satisfactory (note that we use both units in eV assuming
again $t=0.4$~eV) for $\omega>40~$meV. There is however an additional
peak at $\omega \sim 20~$meV, which does not appear in our
analysis. This feature can be interpreted as a low-$T$ phenomenon
related to the onset of longer range incommensurate spin order found
in the same material (Hayden {\it et al.} 1996), clearly beyond the
reach of our method.

\subsection{Orbital diamagnetism}

In comparison with the spin response discussed in previous subsections,
the diamagnetic contribution $\chi_d$ to the d.c. susceptibility has
been much less investigated, both experimentally (Walstedt {\it et
al.}  1992, Miljak {\it et al.} 1993) and theoretically (Rojo {\it et
al.}  1993). A diamagnetic response emerges from the orbital motion of
mobile carriers. For independent electrons it corresponds to the
Landau diamagnetism, which is essentially $T$ independent. For
strongly correlated electrons such a behaviour is far from
evident. Moreover, Rojo {\it et al.} (1993) have shown a relation of
the off-diagonal Hall conductivity in an external homogeneous magnetic
field $B$ (perpendicular to the plane) to the orbital diamagnetic
susceptibility
\begin{equation}
\sigma_{xy} = B{\partial\chi_d \over \partial c_h}
{\partial c_h\over \partial \mu}. \label{mo1}
\end{equation}
Since the Hall effect is known to be anomalous in cuprates, in
particular its $T$ dependence (Ong 1990), one could speculate on similar
anomalies in $\chi_d(T)$, however so far both experimental and
theoretical answers are lacking.

Homogeneous perpendicular field $B$ can be introduced into a tight
binding model with a Peierls construction analogous to the equation
(\ref{ec5}), $t_{ij} \to t_{ij} {\rm exp}(i \theta_{ij})$, where in
the Landau gauge we can write
\begin{equation}
\theta_{ij}=-e_0\vec A(\vec R_i) \cdot \vec{R}_{ij},\qquad 
\vec{A}=B(0,x,0). \label{mo2}
\end{equation}
Our aim is to evaluate the d.c. orbital susceptibility $\chi_d$ via
 the free energy density ${\cal F}$ 
\begin{equation}
\chi_d = -\left. {\partial^2 {\cal F} \over \partial B^2}
\right|_{B=0}. \label{mo3}
\end{equation}
It is nontrivial to incorporate in a small (tilted) lattice the phases
(\ref{mo2}) arising from $B\neq 0$, being at the same time compatible
with the p.b.c. (Veberi\v c {\it et al.} 1998). This can be achieved
only for quantized fields $B=m B_0$ where the smallest field
$B_0=\phi_0/ N$ corresponds to a unit flux $\phi_0$ per 2D
system. Discrete $B$ make the calculation of $\chi_d$ via equation
(\ref{mo3}) less reliable.  It is also a general observation that
properties involving finite $B$ (also the Hall effect) are much more
sensitive to finite-size effects, while at the same time translational
symmetry is lost due to phases (\ref{mo2}) and hence computational
requirements are enhanced. On the other hand, we are evaluating only a
thermodynamic quantity - free energy density ${\cal F}$ allowing for
simplifications discussed in Sec.~4.

\begin{figure}
\centering
\iffigure
\epsfig{file=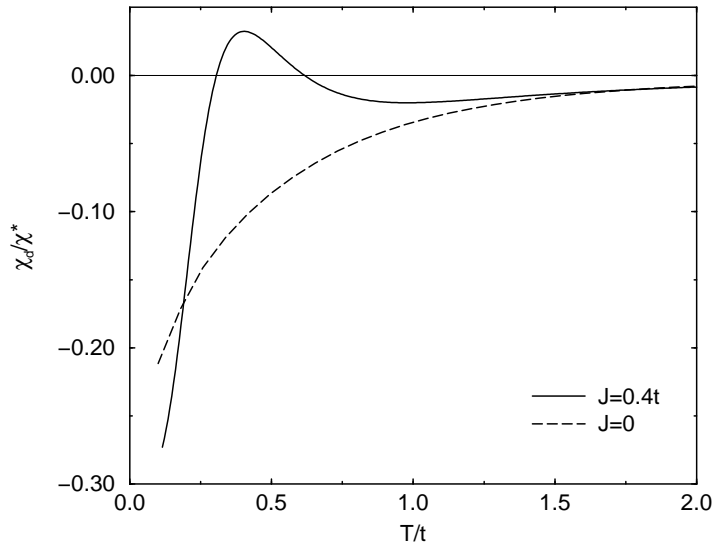,height=9cm,angle=-90}
\fi
\caption{
Diamagnetic susceptibility $\chi_d/\chi^*$ vs. $T$ for a single hole
in an spin background with $J/t=0.4$ (full line), and $J=0$ (dashed
line).}
\label{6.10}
\end{figure}

We consider here only the case of a single hole $N_h=1$, doped into a
magnetic insulator. Nevertheless we expect that results remain
relevant for low $c_h>0$. Namely, by assuming the independence of
charge carriers (spin polarons) $\chi_d$ should scale linearly with
the doping, i.e. $\chi_d \propto c_h$ at least for $c_h \ll 1$.  We
present in Fig.~\ref{6.10} results for $\chi_d(T)/\chi^*$, as obtained
by the analysis of a single hole in an AFM with $J/t=0.4$ on a system
with $N=20$ sites.  Here, $\chi^*=e^2ta_0^4/\hbar^2$ is a
characteristic diamagnetic susceptibility, which can be e.g. compared
with the Pauli susceptibility of tight-binding free fermions (with
constant average density of states) $\chi_P$, $\chi^* \sim 4\chi_P
(m_0/m_t)^2$ (Veberi\v c {\it et al.} 1998). We calculate $\chi_d(T)$
using the equation (\ref{mo3}) and the difference between lowest fields,
$B=0$ and $B=B_0$. For comparison we present also the $J=0$ case,
where finite-size results can be checked with the HTE (Veberi\v c {\it
et al.} 1998).

The $J=0$ case seems easier to study and to understand. The HTE and small
systems show a gradual transition from the high-$T$ regime of
an incoherent hopping to the low-$T$ Nagaoka FM state, accompanied with a
monotonous increase of $|\chi_d|$. At $T\to 0$ $\chi_d$ is expected to
diverge since within the g.s. the dependence on $B$ is nonanalytic,
i.e. $\delta E \propto |B|$ following from a cyclotron motion of a
QP (FM polaron) in a finite $B>0$. Still the asymptotic
behaviour at low $T$ is not simple to establish, since low lying
states (above the Nagaoka state) are not easy to describe.

The behaviour of the AFM polaron with $J>0$ is more involved. At $T\gg
J$ the exchange scale $J$ is not important and results qualitatively
follow that of $J=0$. Quite remarkable is however the vanishing of
the diamagnetic $\chi_d$ (or even the change of its sign) which appears on
lowering $T<t$. This seems to follow from the fact that $J>0$
diminishes and even destroys emerging coherence of a QP. Only at lower
$T<J$ the coherence is reestablished and the standard dynamical
picture of a coherent AFM polaron is dominating the behaviour in the
lowest fields. Reliable results are however quite difficult to obtain
since small systems results (note that we have to follow the variation
with $B$) are quite sensitive to the system shape and boundary conditions
due to a very anisotropic and degenerate QP dispersion.

At low $c_h$ the relevant value of the diamagnetic susceptibility $c_h
\chi_d$ is small compared to the spin contribution (Veberi\v c {\it et
al.} 1998), hence it is not evident that it is directly measurable
(Walstedt {\it et al.}  1992, Miljak {\it et al.}  1993). Nevertheless,
the variation $\chi_d(T)$ is in any case very interesting in
connection with the challenging problem of the Hall effect, related
via the equation (\ref{mo1}), and the anomalous $R_H(T)$. Both
$\chi_d$ as well as the Hall effect emerge from a coupling to orbital
currents, which is the essence of the relation (\ref{mo1}). We note
that at $T \gg J$ the HTE for the Hall constant $R_H(T)$ is analogous
to that of $\chi_d(T)$ discussed here. Crossing the scale $T\sim J$
remains the challenge, whereby it seems that at such $T$ hole-like
$R_H>0$ is even reduced from its high-$T$ value (Shastry {\it et al.}
1993).  Nevertheless Hall effect measurements (Ong 1990) indicate
that $R_H>0$ recovers for $T<T^*$, varying strongly with $T$, and
finally seems to approach the well known quasiclassical result for
$T\to 0$ (Prelov\v sek 1997b).

\setcounter{equation}{0}\setcounter{figure}{0}
\section{Spectral Properties}

One of the most desired quantities, giving valuable information on
electronic properties in interacting systems, is the single-electron
Green's function $G(\vec k,\omega)$ and the corresponding spectral
function $A(\vec k,\omega)$. In theoretical treatments $G(\vec
k,\omega)$ is usually the basis to understand two-particle properties,
such as dynamical electric and magnetic response functions etc. In
strongly correlated systems it appears however that calculations as
well as the measurements of spectral functions are more delicate than
that of most response functions, hence we also treat them towards the
end.

Spectral functions $A({\vec{k}},\omega)$ are directly accessible via
ARPES experiments. There has been in recent years an impressive
advance in the resolution of ARPES as well in the novel information on
cuprates, mostly obtained on convenient layered BISCCO materials (see
review by Shen and Dessau 1995, and references therein).  In the
regime of optimum-doped and overdoped cuprates ARPES experiments
reveal for a wide class of materials a well defined large FS
consistent with the Luttinger theorem and a QP dispersion close to a
tight-binding band (Shen and Dessau 1995, Ding {\it et al.}  1996,
Marshall {\it et al.}  1996).  This seems to imply the validity of the
concept of a metal with an electronic-like FS, although such a
simple picture is in an apparent contradiction with anomalous magnetic
and transport properties.  The LFL interpretation is also spoiled by
the overdamped character of QP peaks. Although a large background
makes fits of particular lineshapes nonunique, the QP inverse lifetime
is found to be of the order of the QP energy, supporting the
possibility of the MFL scenario.

On the other hand, in underdoped materials (Marshall {\it et al.}
1996) well defined QP crossing the FS gradually lose their character
and spectral functions approach a qualitatively different regime of an
undoped AFM (Wells {\it et al.} 1995), where the QP dispersion is that
of a spin polaron in an AFM background.  An excitement has been
induced also by the leading-edge shift (Marshall {\it et al.}  1996),
observed by ARPES in underdoped cuprates in the normal state. It
indicates that a pseudogap consistent with the $d-$wave SC symmetry
persists well above $T_c$. On the other hand recent angle-integrated
PES measurement (Ino {\it et al.}  1997b) indicate another higher
energy pseudogap scale being in a closer correspondence with
$T^*(c_h)$ as deduced from $\chi_0(T)$, $\rho(T)$ and $R_H(T)$
(Batlogg {\it et al.} 1997).

It has been unclear whether above features of spectral functions could
be reproduced within generic models, such as the Hubbard and the
$t$-$J$ model, in particular in the most challenging regime of the
intermediate doping. $A(\vec k,\omega)$ in 2D models have been so far
studied mainly via numerical techniques (Dagotto 1994), in particular
using the ED (Stephan and Horsch 1991, Eder {\it et al.}  1994, Moreo
{\it et al.}  1995) and the QMC (Bulut {\it et al.} 1994, Preuss {\it et
al.} 1995, 1997).  These studies have established a reasonable
consistency of the model QP dispersions with the experimental ones, as
well as the possibility of a large FS, but have not been able to
investigate closer the character of QP, as e.g. manifested in the
corresponding self energies $\Sigma(\vec k,\omega)$ being in the core
of anomalous low-energy properties.

\subsection{Green's function}

When applying the usual definition of the retarded Green's function
(Mahan 1990) to the model of strongly correlated electrons, one should
keep in mind that the $t$-$J$ model (\ref{cm1}) acts within the
restricted fermionic basis.  To avoid the ambiguity it is convenient to
build restrictions directly in the definition of retarded Green's
functions using the projected operators $\tilde{c}^\dagger _{{\vec
k}s}, \tilde{c}_{{\vec k}s}$, being Fourier transforms of
$\tilde{c}^\dagger_{is}, \tilde{c}_{is}$, respectively,
\begin{equation}
G^{R}(\vec{k},\omega)=-i \int_0^\infty dt ~e^{i\omega^+ t}
\langle \{\tilde{c}_{\vec{k}s}(t),\tilde{c}^\dagger_{\vec{k}s}(0)\}_+
\rangle. \label{sg1}
\end{equation}
Note that within the subspace of allowed states the definition is identical
to the usual one. $A(\vec{k},\omega)$ can be represented  as a sum of 
the electron spectral function $A^{+}(\vec{k},\omega)$ 
and the hole spectral function $A^{-}(\vec{k},\omega)$, 
\begin{equation}
A(\vec{k},\omega)=-\frac{1}{\pi}{\rm Im}G^R(\vec{k},\omega)
=A^{+}(\vec{k},\omega)+A^{-}(\vec{k},-\omega), \label{sg2}
\end{equation}
which are expressed in terms of eigenstates as
\begin{eqnarray}
A^{+}(\vec{k},\omega)&=&\Omega^{-1}
\sum_{n,m}|\langle \Psi_m|\tilde c_{\vec{k}s}^\dagger|
\Psi_n\rangle|^2 e^{-\beta
(E_n-\mu N^n_{e})}\delta(\omega+\mu-E_m+E_n), \nonumber \\
A^{-}(\vec{k},-\omega)&=&\Omega^{-1}
\sum_{n,m}|\langle  \Psi_m|\tilde c_{\vec{k}s}| \Psi_n\rangle|^2
e^{-\beta (E_n-\mu N^n_{e})}\delta(-\omega-\mu-E_m+E_n). \label{sg3}
\end{eqnarray}
$A^-$ represents the dynamical response when one electron is taken out
(hole added) from a system. It corresponds (within the sudden
approximation) to PES experiments and is proportional to corresponding
cross sections in ARPES. On the other hand, $A^+$ describes the
dynamics on adding an electron and is related to IPES experiments.  In
the equilibrium electron and hole spectra are related via the Fermi
function $f(\omega)=1/(\exp(\beta\omega)+1)$, e.g.
$A^{-}(\vec{k},-\omega)=f(\omega)A(\vec{k},\omega)$.

The omission of doubly occupied states has important implications, in
particular we note the change in anticommutation relations,
\begin{equation}
\{\tilde{c}^\dagger_{\vec{k}s}, \tilde{c}_{\vec{k}s}\}_+=
{1\over N}\sum_i \{\tilde{c}^\dagger_{is}, \tilde{c}_{is}\}_+= 
{1\over N} \sum_i (1- n_{i-s}). \label{sg4}
\end{equation}
In the paramagnetic phase where $\langle n_{is} \rangle = (1-c_h)/2$,
the equation (\ref{sg4}) leads to a modified sum rule for the spectral
function (usually normalized to unity), i.e.
\begin{equation}
\alpha= \int_{-\infty}^\infty d\omega ~A(\vec{k},\omega)= \frac{1}{2}(1+c_h)<1.
\label{sg5}
\end{equation}
An important consequence of the relation (\ref{sg5}) is an upper bound for
the momentum distribution function,
\begin{equation}
\bar n_{\vec{k}}=\langle c^\dagger_{\vec{k}s}c_{\vec{k}s}\rangle
=\int_{-\infty}^\infty A^{-}(\vec{k},\omega)d\omega < \alpha. \label{sg6}
\end{equation}
The anomalous sum rule (\ref{sg5})  leads to some ambiguity in 
the definition of the self energy $\Sigma(\vec k,\omega)$. 
We keep the usual one,
\begin{equation}
G^R({\vec{k}},\omega)=1 /(\omega-\Sigma({\vec{k}},\omega)), \label{sg7}
\end{equation}
in order to retain the standard definition of QP parameters. An
alternative would be to replace the nominator with $\alpha$ (Prelov\v
sek 1997a). The equation (\ref{sg7}) implies that the self energy does
not vanish for $|\omega| \to \infty$, but shows rather an asymptotic
form $\Sigma({\vec{k}},\omega\to\infty)\to(1-1/\alpha)\omega$.  Note
also that the $t$-$J$ model doesn't allow for a free propagation even
at $J=0$, therefore we do not include any free term
$\epsilon_{\vec{k}}$ in the definition (\ref{sg7}).

It is straightforward to calculate $A(\vec k, \omega)$ using the FTLM
(Jakli\v c and Prelov\v sek 1997), again following the equation
(\ref{fi4}). Since operators $\tilde c^\dagger_{\vec k s},\tilde
c_{\vec k s}$ act between subspaces with different $N_h$, i.e. $N_h\to
N_h \mp 1$, we in fact calculate separately $A^+$ and $A^-$ according
to definitions (\ref{sg3}).  In calculations at low $T$ and fixed
$c_h=N_h/N$ we also replace the grand-canonical average with the
canonical one in the subspace of states with $N_h$ holes.  For a
proper interpretation of results it is important to locate correctly
the chemical potential $\mu(T,c_h)$, discussed in Sec.~4.1.  We
have a nontrivial check of $\mu$ via the one-particle DOS
\begin{equation}
{\cal N}(\varepsilon)={2 \over N} \sum_{\vec{k}}
A({\vec{k}},\varepsilon-\mu), \label{sg8}
\end{equation}  
which relates $\mu$ to $c_h$,
\begin{equation}
\int_{-\infty}^\infty {\cal N}(\omega+\mu)
(e^{\beta\omega}+1)^{-1}d\omega=1-c_h. \label{sg9}
\end{equation}
We find a very good agreement between $\mu$ calculated from ${\cal
N}(\varepsilon)$ and the one determined from thermodynamics in
Sec.~4.1.

The advantage of the FTLM over the g.s. ED is that it gives smooth
$G^R(\vec k,\omega)$ even for quite low $T<J$ provided that
$T>T_{fs}$.  This allows for a meaningful calculation of $\Sigma(\vec
k,\omega)$, which gives a new insight in the properties of QP.  We are
thus able to evaluate also QP parameters defined via $\Sigma(\vec
k,\omega)$ (Mahan 1990), in particular the QP energy $E_{\vec k}$,
the weight $Z_{\vec k}$ consistent with the equation (\ref{cp4}), and
the damping $\Gamma_{\vec k}$,
\begin{eqnarray}
E_{\vec{k}}&=&\Sigma'({\vec{k}},E_{\vec{k}}), \nonumber \\
Z_{\vec{k}}^{-1}&=&1- \partial \Sigma'({\vec{k}},\omega)/
\partial \omega|_{\omega = E_{\vec k}},\label{sg10} \\
\Gamma_{\vec{k}}&=&Z_{\vec{k}}|\Sigma''({\vec{k}},E_{\vec{k}})|.
\nonumber
\end{eqnarray}

\subsection{Single hole in the antiferromagnet}

We begin by considering the spectral properties of a single hole
injected in an AFM. Since in the undoped model particle-number
fluctuations are suppressed, the chemical potential is not a well
defined quantity in this case and we use here unreduced energy
$\varepsilon$ instead of $\omega=\varepsilon-\mu$.

\begin{figure}
\centering
\iffigure
\epsfig{file=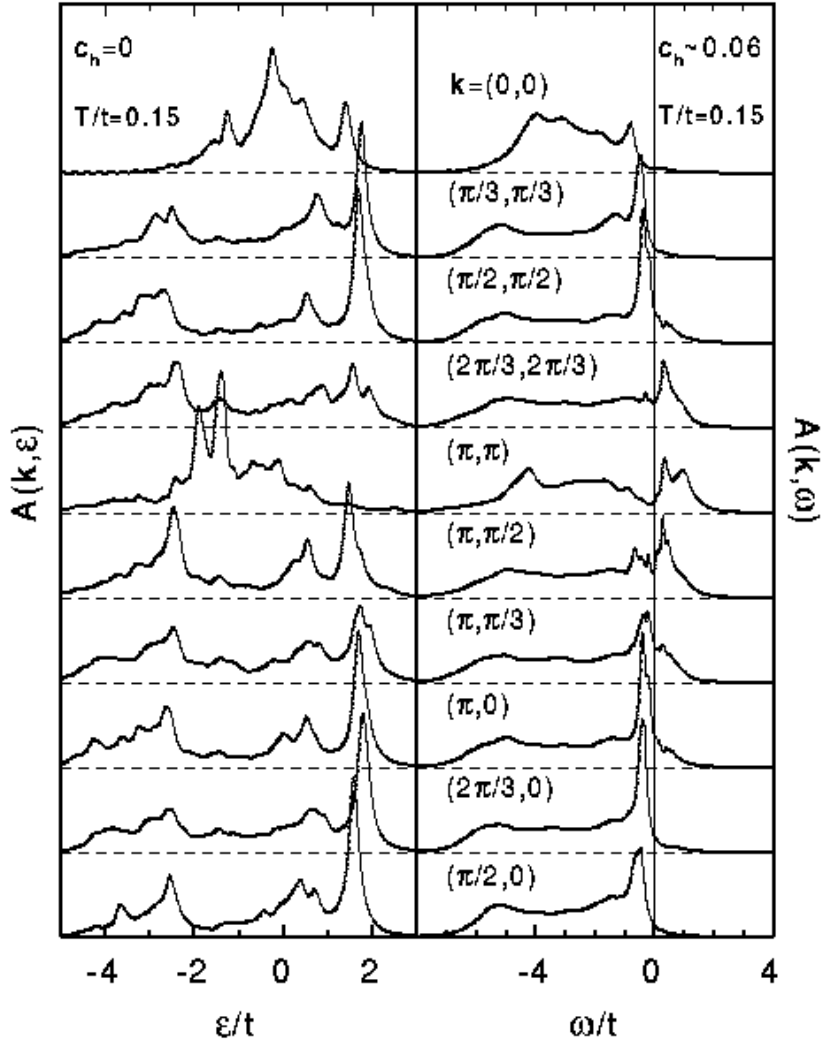,height=16cm}
\fi
\caption{
Spectral functions for the undoped AFM (left), and for a system with
finite doping $c_h=1/N$ (right), both obtained on lattices with
$N=16,18$ sites.  }
\label{7.1}
\end{figure}

On the left side of Fig.~\ref{7.1} the spectral function of a hole in
an AFM is presented.  Although $T=J/2$ is relatively high, implying
only short-range AFM correlations, still a coherent QP peak is
clearly observed at low energies for most available $\vec{k}$.  For
$\vec{k}=(\pi/2,\pi/2)$ and $\vec{k}=(2\pi/3,0)$ peaks coincide with
the g.s. minima for systems with $N=16$ and $N=18$ sites,
respectively.  This indicates that even at $T>0$ the major
contribution to the coherent spectral weight comes from transitions
between the g.s. configurations with $N_h=0$ and $N_h=1$.  The
coherent QP peak shows a dispersion on the energy scale comparable to
$J$, consistent with the self-consistent Born approximation
(Schmitt-Rink {\it et al.}  1988, Kane {\it et al.}  1989), which
yields a bandwidth $W\sim 0.6~t$ for $J/t=0.3$ (Martinez and Horsch
1991). Since no long-range AFM order is expected in small systems, the
propagation of the hole appears to be determined by short-range AFM
correlations. In addition to the QP band at $\varepsilon \alt
\varepsilon_0 \sim 2t$, we observe also high-energy
features at $\varepsilon \ll \varepsilon_0$, which are only weakly
dependent on $\vec{k}$ and can be attributed to the incoherent hole
propagation. High-energy peaks have been observed also in g.s. ED
studies, but the structure was claimed to disappear at large $N$
(Poilblanc {\it et al.} 1993). Our study shows a nontrivial structure
consistent at all considered $N=16-20$, so we would rather attribute
peaks to resonance (excited) states of the AFM spin polaron.

\begin{figure}[ht]
\centering
\iffigure
\epsfig{file=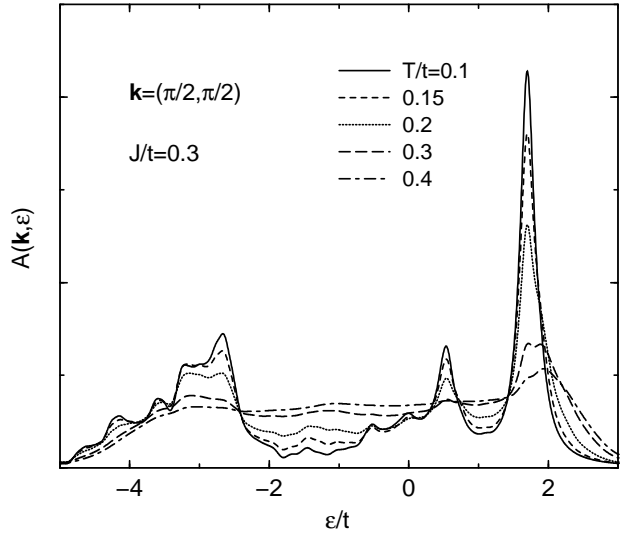,height=10cm,angle=-90}
\fi
\caption{
Spectral function $A(\vec k^*,\varepsilon)$ for the undoped AFM at
several $T$.} \label{7.2}
\end{figure}

In Fig.~\ref{7.2} we present the $T$-dependence of the spectral
function at $\vec{k}^*=(\pi/2,\pi/2)$, which corresponds to the
g.s. momentum of a hole in an AFM.  We realize from Fig.~\ref{7.2}
that there is a significant spectral change going from $T\sim 0$ to
$T>J$. Namely, in the high-$T$ regime the QP peak becomes
progressively broader and merges with the incoherent background, which
at the same time loses the detailed structure (resonance peaks). This
development is plausible since for $T>J$ we are dealing essentially
with the hole propagation in a random spin background well accounted
for within the RPA (Brinkman and Rice 1970).

At $T<J$ the QP energy $E_{\vec k}$ and the corresponding weight
$Z_{\vec k}~$, as determined by equations (\ref{sg10}), show only a
weak $T$ dependence and we get $\tilde Z= Z_{\vec k^*}(T)\sim 0.16$.
We note that this value is obtained for the spectral function
normalized according to the relation (\ref{sg5}) with $\alpha =
1/2$. It is easy to see that in order to make a comparison with the
usual definition of the QP weight of an AFM polaron (Kane {\it et al.}
1989, Martinez and Horsch 1991) we should rather take ${\cal Z}=\tilde
Z/\alpha$ and our value ${\cal Z} \sim 0.32$ is consistent with the
one ${\cal Z} \sim 0.284$, obtained within the self-consistent Born
approximation for $J/t=0.3$.

\subsection{Spectral functions at finite doping}

\begin{figure}
\centering
\iffigure
\epsfig{file=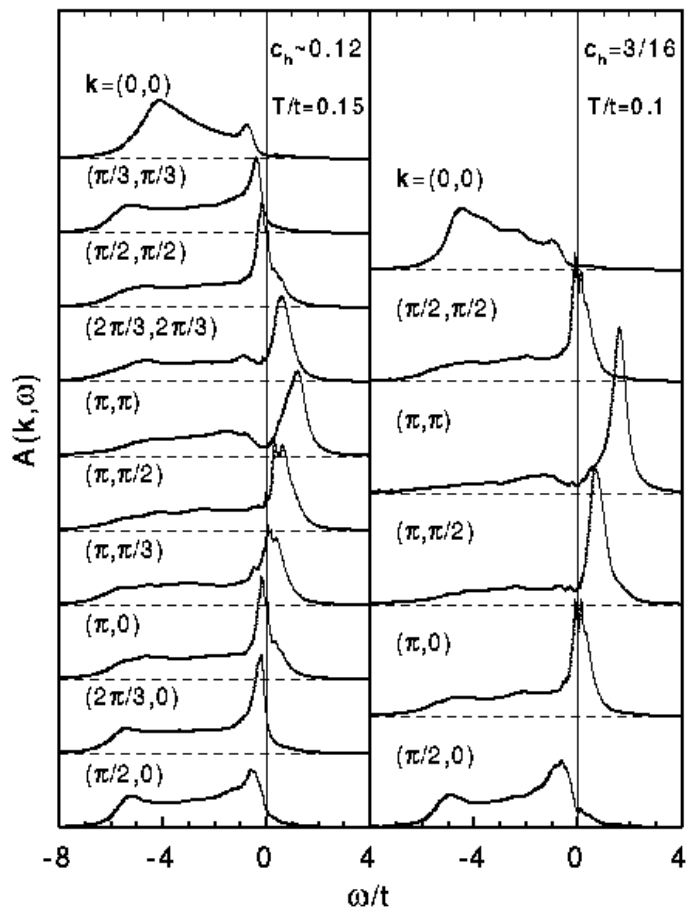,height=14cm}
\fi
\caption{
$A(\vec k,\omega)$ for $N_h=2$ holes on $N=16,18$ sites,
and for $N_h=3$ holes on $N=16$ sites.  } \label{7.3}
\end{figure}

In Figs.~\ref{7.1},~\ref{7.3} we present the gradual development of
spectral functions towards the optimum doping.  For $c_h\sim 0.06$ and
$c_h\sim 0.12$ results obtained on systems with $N=16,18$ sites are
combined for $N_h=1$ and $N_h=2$ holes, respectively.  As distinct to
the undoped case, the spectra are broadened with Lorentzians of variable
width according to
$\delta=\delta_0+(\delta_\infty-\delta_0)\tanh^2(\omega/\Delta)$, with
$\delta_\infty =0.2t$, $\delta_0=0.04t$, and $\Delta=1.0t$.  In this
way sharper (well resolved) $\omega \sim 0$ features remain
unaffected, while some of finite-size structures at higher
$|\omega|>t$ are smoothened out. In any case, $\delta$ is always taken
smaller than the energy widths of main spectral features.

At finite doping we observe in $A({\vec{k}},\omega)$ at all available
$\vec{k}$ a coexistence of sharper features, associated with coherent
QP peaks, with a broad incoherent background, as already established
in ED studies (Stephan and Horsch 1991, Moreo {\it et al.} 1995).
Coherent peaks disperse through $\omega=0$ as ${\vec{k}}$ crosses the
FS. Within the given resolution in the ${\vec{k}}$ space the FS
appears to be large for all $c_h \agt 0.12$, consistent with the
Luttinger theorem (Luttinger 1960). An analogous dispersion is
observed for the underdoped case $c_h\sim 0.06$, although the latter
one requires a more careful interpretation as given lateron.  The
total QP dispersion $W$ is broadening as $c_h$ is increasing,
qualitatively consistent with the slave boson result where $W \propto
c_h t + \chi J$ (Baskaran {\it et al.} 1987, Wang {\it et al.} 1991).

\subsubsection{Intermediate doping}

\begin{figure}
\centering
\iffigure
\epsfig{file=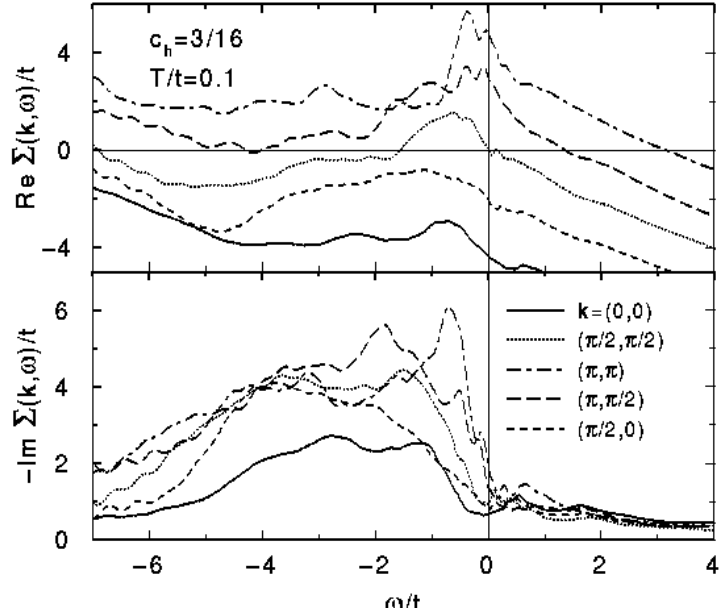,height=13cm,angle=-90}
\fi
\caption{
Self energy $\Sigma(\vec k,\omega)$ for $c_h=3/16$ at various
${\vec{k}}$.  } \label{7.4}
\end{figure}

We first discuss in more detail the regime of intermediate doping
(Jakli\v c and Prelov\v sek 1997), in our lattices $c_h \sim 0.12$ and
$c_h=3/16\sim 0.19$.  In Fig.~\ref{7.4} we show
$\Sigma({\vec{k}},\omega)$ evaluated for $c_h=3/16$ at $T=0.1~t\sim
T_{fs}$. We first notice an asymmetry between the PES ($\omega<0$) and
the IPES ($\omega>0)$ spectra at all
${\vec{k}}$. $\Sigma''({\vec{k}},\omega)$ are small for $\omega>0$, as
compared to $\omega<0$. For ${\vec{k}}$ outside the FS this implies a
modest QP damping, consistent with sharp IPES peaks seen in
$A({\vec{k}},\omega)$ in Fig.~\ref{7.3}, containing the major part of
the spectral weight. $\Sigma'({\vec{k}},\omega)$ shows an analogous
asymmetry, in the region $\omega>0$ resembling moderately renormalized
QP. Due to the definition (\ref{sg7}) the slope of $\Sigma'$
is not zero even at $|\omega|\gg t$.

The behaviour on the PES ($\omega<0$) side is very different.  For all
${\vec{k}}$, $-\Sigma''(\omega)$ are very large at $\omega <-t$,
leading to overdamped QP structures.  We should however distinguish
two cases. For $\vec k$ well outside the FS, large
$|\Sigma''(\omega<0)|>t$ does not invalidate a well defined QP at
$\omega>0$, but rather induces a weak reflection (shadow) of the IPES
peak at $\omega <0$, as well seen in Fig.~\ref{7.3} for ${\vec{k}}
=(\pi,\pi)$.  On the other hand, for ${\vec{k}}$ inside or near the FS
the variation with $\omega$ is more regular, and can be directly
related to the QP damping. Particularly remarkable feature in
Fig.~\ref{7.4} is a linear $\omega$ dependence of $\Sigma''(\omega\alt
0)$ for ${\vec{k}}= (\pi/2,0)$ and ${\vec{k}}=(\pi/2,\pi/2)$.
Meanwhile ${\vec k}=(0,0)$, being further away from the FS, seems to
follow a different more LFL-type behaviour.  Similar general
conclusions follow also from results obtained for the lower doping
$c_h=2/18$.

\begin{figure}
\centering
\iffigure
\epsfig{file=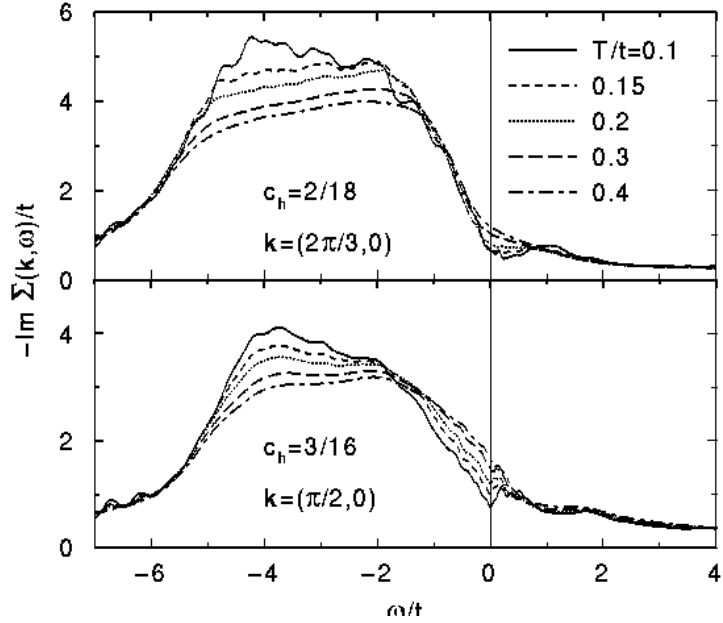,height=13cm,angle=-90}
\fi
\caption{
$\Sigma''({\vec{k}},\omega)$ for $c_h=2/18$ and $c_h=3/16$ at
different $T$ and selected ${\vec{k}}$ inside the FS.  }
\label{7.5}
\end{figure}

To address the latter point in more detail, we show in Fig.~\ref{7.5}
the $T$ variation of $\Sigma''(\vec k,\omega)$ for both dopings at
selected ${\vec{k}}$ inside the FS. For $c_h=3/16$ the linearity of
$\Sigma''(\omega)$ is seen in a broad range $-2t\alt \omega
\alt 0$ at the lowest $T$ shown. Moreover, for this optimum 
doping the $T$ dependence is close to a linear one, taking into
account a small residual (finite-size) damping $\eta_0>0$ at
$\omega=0$.  Data can be well described by
\begin{equation}
-\Sigma''(k\alt k_F,\omega)=\eta_0+\gamma(|\omega|+\xi T),
\qquad \gamma\sim 1.4, \quad \xi\sim 3.5. \label{ss1}
\end{equation}
Such a dependence is consistent with one of the proposed forms
(\ref{cp3a}) within the MFL scenario, as well as with the conductivity
relaxation $1/\tau(\omega)$ in the equation (\ref{eo6}).  In contrast, the
$T$ dependence for $c_h=2/18$ seems somewhat different, and we find
$-\Sigma''\propto \omega$ only in the interval $-t\alt\omega\alt
-T$. This would indicate the consistency with the original MFL form
(\ref{cp3}) (Varma {\it et al.} 1989), however we should be aware that
in the underdoped regime finite-size effects are larger at given $T$.

Here we should comment on the manifestation of the FS in small
fermionic systems. At $T,\omega\sim 0$ we are dealing in the
evaluation of the expression (\ref{sg3}) with the transition between
g.s. of systems with $N_h$ and $N_h'=N_h \pm 1$ holes,
respectively. Since these g.s. states have definite momenta
${\vec{k}}_0$, they induce sharp QP peaks for particular
${\vec{k}}={\vec{k}}'_0- {\vec k}_0$ defining in this way for a small
system the FS, apparently satisfying the Luttinger theorem. At the
same time the corresponding QP damping vanishes,
i.e. $\Sigma''({\vec{k}},\omega \sim 0)\sim 0$.

From $\Sigma({\vec{k}},\omega)$ we can calculate QP parameters, as
defined in equations (\ref{sg10}).  Results for $c_h \sim 0.12$ and
$c_h \sim 0.19$ are presented by Jakli\v c and Prelov\v sek (1997).
We note that parameters are of a limited meaning for $\vec k$ well
inside FS due to a large damping $\Gamma_{\vec{k}}$. Otherwise,
$E_{\vec{k}}$ shows the enhancement of the dispersion with $c_h$,
while $\Gamma_{\vec{k}}<E_{\vec{k}}$ for $|{\vec{k}}|>k_F$.  To
establish the relation with the FL theory one has to evaluate QP
parameters at the FS ${\vec k} \sim {\vec{k}}_F$. Of particular
importance is the renormalization factor $\tilde Z=
Z_{{\vec{k}}_F}(T=0)$. We find that at lowest $T\sim T_{fs}$
$Z_{\vec{k}_F}(T)$ is still decreasing as $T$ is lowered, e.g. a
variation (cca. 20\%) is found within the interval $0.1<T/t<0.3$. This
is not inconsistent with the MFL form $\tilde
Z^{-1}\sim\ln(\omega_c/T)$, although such a conclusion should be taken
with care due to narrow $T$ interval.

Regarding the size of $\tilde Z$ at low but finite $T>0$ we note, that
the value of the momentum distribution function is very close to the
maximum (\ref{sg6}) for all $k<k_F$, i.e. $\bar n_{\vec{k}}\sim
(1+c_h)/2$ (Stephan and Horsch 1991).  Taking the FS volume according
to the Luttinger theorem $V_{FS}/V_{BZ}=(1-c_h)/2$ and assuming that
$\bar n_{\vec{k}}$ falls monotonously with $k$, this implies an
inequality for the discontinuity $\tilde Z$,
\begin{equation}
\tilde Z= \delta \bar n_{\vec{k}_F} < {2c_h \over 1+c_h}.
\label{ss2}
\end{equation}
One should keep in mind that a discontinuity is meaningful at $T
\sim 0$, while at $T>0$ it indicates only a gradual step.  Still we find
a consistent result $\tilde Z= 0.28$ for $c_h=3/16$, while for
$c_h=2/18$ the value $\tilde Z= 0.35$ is somewhat larger, possibly due
to higher $T$ used in calculations.

An analogous argument can be used to explain the electron-hole
asymmetry of $A(\vec{k},\omega)$.  Holes added to the system at
$k<k_F,\omega<0$ move in an extremely correlated system, strongly
coupled to the spin dynamics (Prelov\v sek 1997a), the latter also
exhibiting the anomalous MFL-type behaviour, as given by the
expression (\ref{ml4}). On the other hand, states for $k>k_F$ are not
fully populated, allowing for a moderately damped motion of added
electrons with $\omega>0$.

Let us comment here on the relevance of our results to ARPES spectra
near the optimum doping. The main observation is that for $k<k_F$ we find
the linewidth typically $\Gamma\sim t\sim 0.4~{\rm eV}$ well
compatible with experiments at ${\vec{k}}$ away from the FS and at
intermediate doping (Shen and Dessau 1995, Ding {\it et al.} 1996,
Marshall {\it et al.} 1996). Also the MFL form has been claimed to
describe better the experiments (Olson {\it et al.} 1990), although
this point is still controversial (see Shen and Dessau 1995).

\subsubsection{Overdoped and underdoped regime} 

\begin{figure}
\centering
\iffigure
\epsfig{file=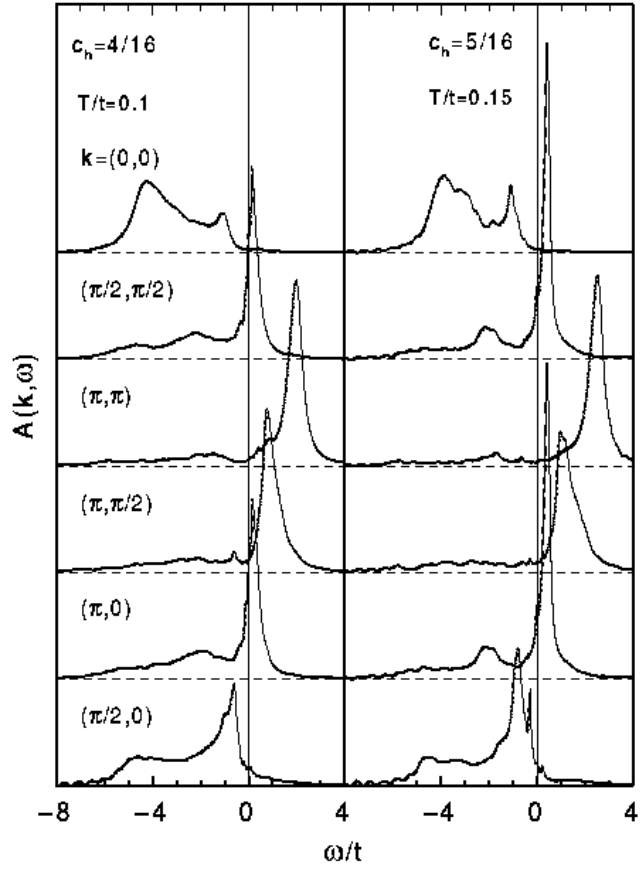,height=14cm}
\fi
\caption{
$A(\vec k,\omega)$ for $c_h=4/16$ and $c_h=5/16$.  }
\label{7.6}
\end{figure}

Let us turn from the intermediate doping first to the overdoped
regime.  Spectra at higher doping $c_h=4/16$ and $c_h=5/16$ are shown
in Fig.~\ref{7.6}. While $A({\vec{k}},\omega)$ at $c_h=4/16$ are still
qualitatively similar to the spectra at intermediate doping, at
$c_h=5/16$ they are already showing a substantially different
behaviour. In the latter case the incoherent background in the PES
spectrum is reduced while QP peaks for $k<k_F$ are sharpened and all
of them are essentially underdamped with $\Gamma_{\vec k}<E_{\vec k}$.
Nevertheless, we still have $\Gamma_{\vec k} \sim E_{\vec k}$ so it is
not evident whether this is already a (more) normal LFL.

The proper analysis of the underdoped regime is even more delicate.
At $c_h \sim 0.06-0.12$, as presented in Figs.~\ref{7.1}, \ref{7.3}, a
'shadow' feature is clearly seen in $A({\vec{k}},\omega)$ for
$k>k_F$. Namely, along with the principal peak at $\omega>0$ a weak
bump in the $\omega<0$ part of the spectrum appears. In $\Sigma'(\vec
k,\omega)$ for $k>k_F$ this effect emerges as a strong oscillation,
most pronounced for ${\vec{k}}=(\pi,\pi)$. It leads even to a double
solution of the equation (\ref{sg10}) for $E_{\vec k}~$, analogous to
the phenomena within the spin-bag scenario (Kampf and Schrieffer
1990). This effect is becoming less visible at larger doping, e.g. at
$c_h=3/16$, so its disappearance is related to the reduction of the
AFM correlation length $\xi \sim 1$.

Let us try to make the contact with recent ARPES measurements on
underdoped BISCCO (Marshall {\it et al.} 1996), which show the opening
of the pseudogap even for $T>T_c$, in particular near the $\vec
k^{**}=(\pi,0)$ point, where the FS seems to disappear. On the other
hand the QP peak disperses through the FS along the $\Gamma$-M
direction where the FS seems to be well defined.  If we look closer at
Fig.~\ref{7.1} for the lowest finite doping $c_h \sim 0.06$, we can
recognize very similar features. We realize that close to $\vec
k^{**}$, the PES part shows spectra very analogous to the undoped AFM
in Fig.~\ref{7.1}, where due to the doubling of the Brillouin zone the
dispersion is expected to reach maximum at $\vec k^{*}=(\pi/2,\pi/2)$
(note however that on a $4\times4$ lattice with n.n. hopping $k^*$ and
$k^{**}$ are equivalent) and folds back for $\vec k$ outside the
reduced AFM zone. E.g., in Fig.~\ref{7.1} for $c_h \sim 0.06$ we can
clearly recognize PES peaks also for $\vec k=(\pi,\pi/3)$ and $\vec
k=(\pi,\pi/2)$, being lower in $\omega$ than the QP peak at $\vec
k^*$. Such a dispersion is also seen in ARPES.  At the same time there
is for both mentioned $k>k^*$ also a visible peak in the IPES part,
which can be interpreted as a QP dispersing through the FS. The PES
and the IPES peak in this case are separated by a gap, which does not
seem to be a finite-size effect, and is qualitatively consistent with
the ARPES feature (Marshall {\it et al.}  1996). It should be also
noted that along the $\Gamma$-M direction the QP dispersion is closer
to a normal metallic one with much less pronounced 'shadows' for
$k>k_F$.

\subsubsection{Influence of next-nearest-neighbour hopping} 

It is evident from spectral functions at the intermediate doping, e.g.
in Figs.~\ref{7.3}, \ref{7.6} that the shape of the FS obtained within
the $t$-$J$ model is closer to a circular one than to the one found in
cuprates via ARPES experiments (Shen and Dessau 1995), where
e.g. $\vec k^{**}$ is inside the FS. It has been well established that
such a FS topology can be reproduced by adding the n.n.n. hopping term
(\ref{cm2}) with $t'<0$ (Tohyama and Maekawa 1994). It is however
still an open question whether the modified hopping and different FS
topology are essential for the anomalous properties of cuprates.

\begin{figure}
\centering
\iffigure
\epsfig{file=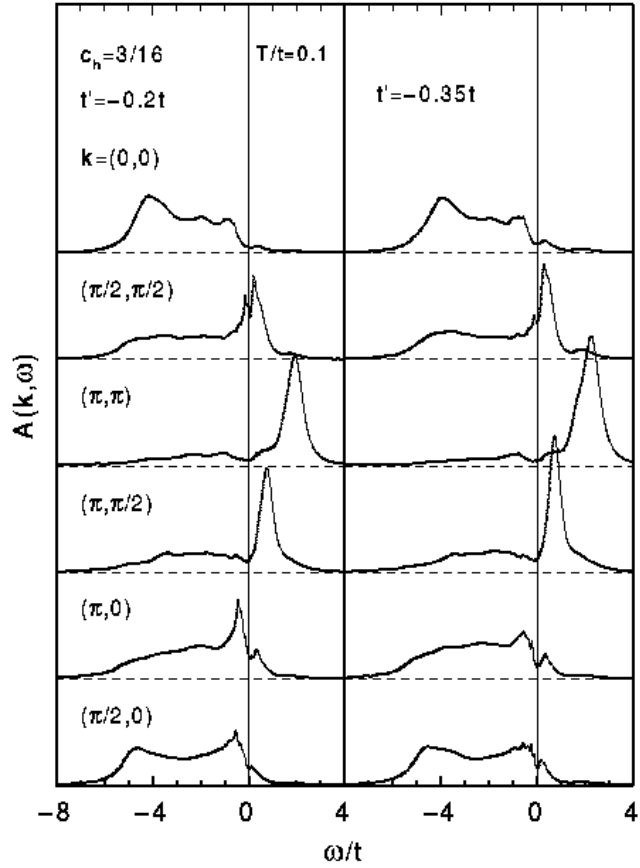,height=14cm}
\fi
\caption{
$A(\vec k,\omega)$ for $c_h=3/16$ with nonzero n.n.n. hopping:
$t'/t=-0.2$ and $t'/t=-0.35$.}
\label{7.9}
\end{figure}

In order to see the qualitative changes introduced by the n.n.n.
hopping, we present in Fig.~\ref{7.9} $A(\vec k,\omega)$ at fixed
doping $c_h=3/16$ for two different $t'/t=-0.2$ and $t'/t=-0.35$. It
is evident that $t'<0$ lifts the degeneracy between of the $\vec k^*$
and $\vec k^{**}$ even in the $4\times 4$ lattice. Also the QP
dispersion changes and the QP peak with $\vec k^*$ is now at
$\omega>0$, i.e.  outside the FS, while the QP with $\vec k^{**}$
moves inside the FS.  The effect is enhanced for larger $|t'/t|$. It
is however important to realize that other QP properties are not
essentially changed, in particular we find that the self energy
$\Sigma''(\vec k,\omega)$ remains qualitatively similar to the case
with $t'=0$.

\subsection{Density of states}

\begin{figure}
\centering
\iffigure
\epsfig{file=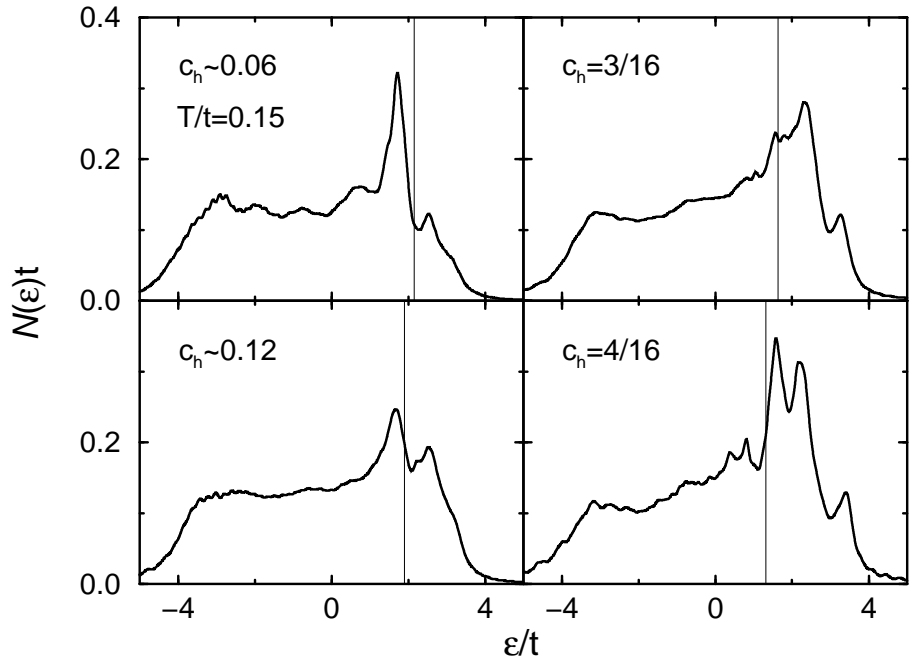,height=13cm,angle=-90}
\fi
\caption{
Single-particle DOS ${\cal N(\varepsilon)}$ for different dopings at
fixed $T/t=0.15$. For $c_h \sim 0.06$ and $c_h \sim 0.12$ we present
joined DOS for $N=16,18$ systems with $N_h=1$ and $N_h=2$,
respectively. The thin vertical lines denote the chemical potential
$\mu(T)$.  } \label{7.7}
\end{figure}

Finally we present in Fig.~\ref{7.7} the doping dependence of the
single-particle DOS ${\cal N}(\varepsilon)$ (Jakli\v c and Prelov\v
sek 1997), as defined in the equation (\ref{sg8}). In a weakly doped
system with $c_h\sim 0.06$, a QP coherent peak (of width $\sim 2J$) is
seen at $\omega =\varepsilon-\mu\alt 0$. Besides that, a broad
background due to well understood incoherent hole motion is dominating
$\omega \ll 0$. At such low doping the electron $\omega>0$ part of the
DOS is weak, with the total intensity $2 c_h$ as compared to $1-c_h$
of the hole part. With increasing $c_h$ the hole incoherent background
does not reduce appreciably in intensity, while the coherent
dispersion near the Fermi energy widens and cannot be well
distinguished in ${\cal N}(\varepsilon)$ from the background. At the
same time, the electron part of the DOS is increasing, both in the
weight and in the width. Note that oscillations which appear for
$\omega>0$ at higher doping $c_h>3/16$ are essentially finite-size
effects. Namely we are dealing with a restricted number of finite-size
$\vec k$ while QP peaks are becoming quite narrow. Such effects are
even more pronounced in overdoped systems with $c_h>0.25$, where the
incoherent part is finally losing its intensity and the coherent QP
are dominating the whole $\varepsilon$ regime.  It should be also
mentioned that the introduction of the n.n.n. hopping $t'\neq 0$, in spite
of a significant change of the FS shape, does not seem to induce an
essential change within the DOS.

Let us comment here the relation with the entropy $s$, Fig.~\ref{4.3}a.
If we assume the low-$T$ form as follows from the LFL theory (Abrikosov
{\it et al.} 1965), we get
\begin{equation}
s ={ \pi^2 {\cal N}(\mu)\over 3\tilde Z}~T. \label{sd1}
\end{equation}
We see from Fig.~\ref{7.7} that ${\cal N}(\mu)$ is only weakly doping
dependent at intermediate $c_h$ and actually also quite close to the
free-fermion value (assuming a tight-binding model with the hopping
parameter $t$). Taking $\tilde Z \sim 0.28$ for $c_h = 3/16$, we get $s
\sim 0.29~k_B$ at $T=0.1~t$, quite consistent with
thermodynamic results in Sec.~4.2.  Here the surprising fact is that
such $s$ represents a large increase over the undoped AFM and appears
due to a very low concentration of mobile holes introduced into an
AFM.

\begin{figure}[ht]
\centering
\iffigure
\mbox{
\subfigure[]{
\epsfig{file=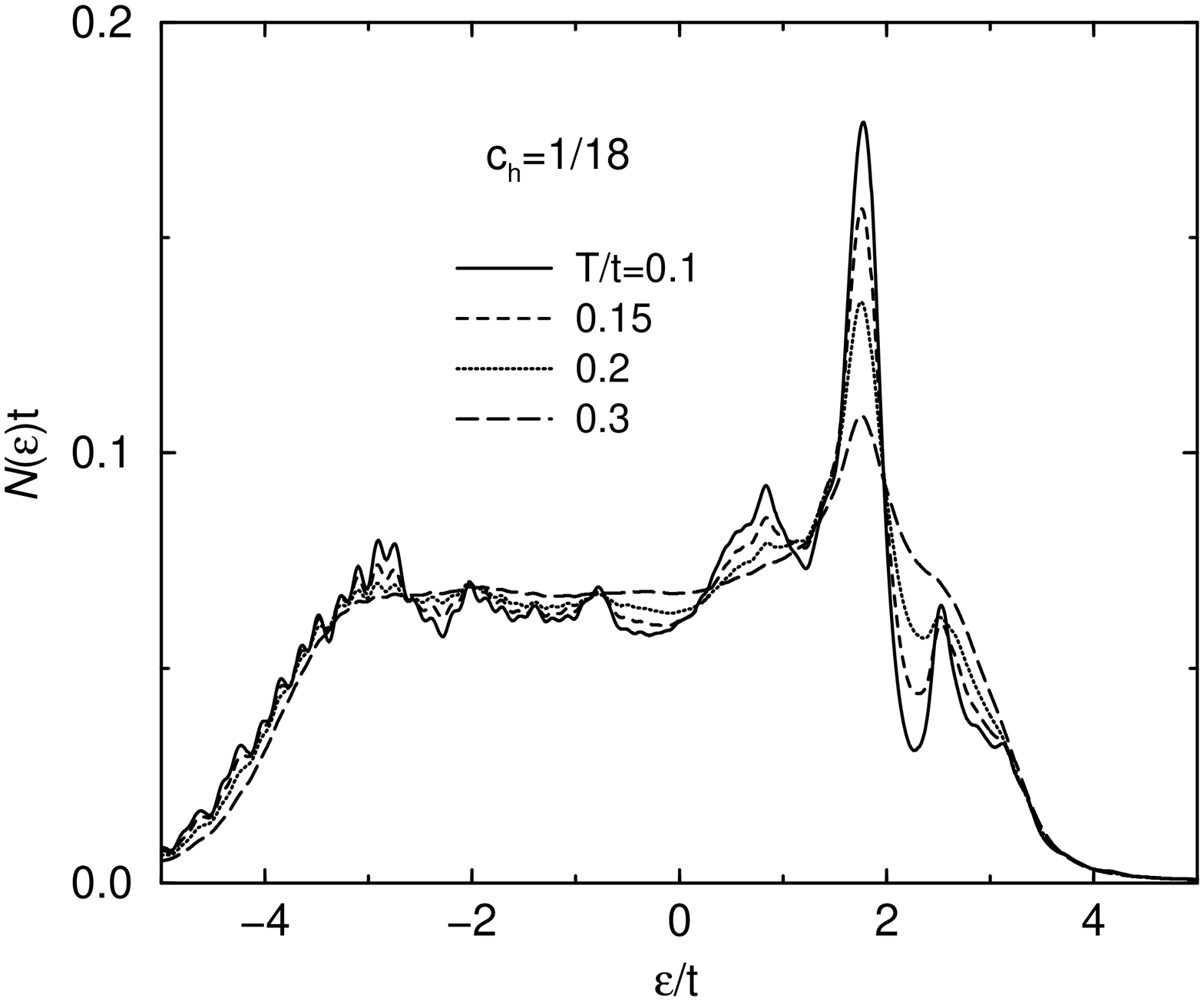,height=8cm,angle=-90}}
\quad
\subfigure[]{
\epsfig{file=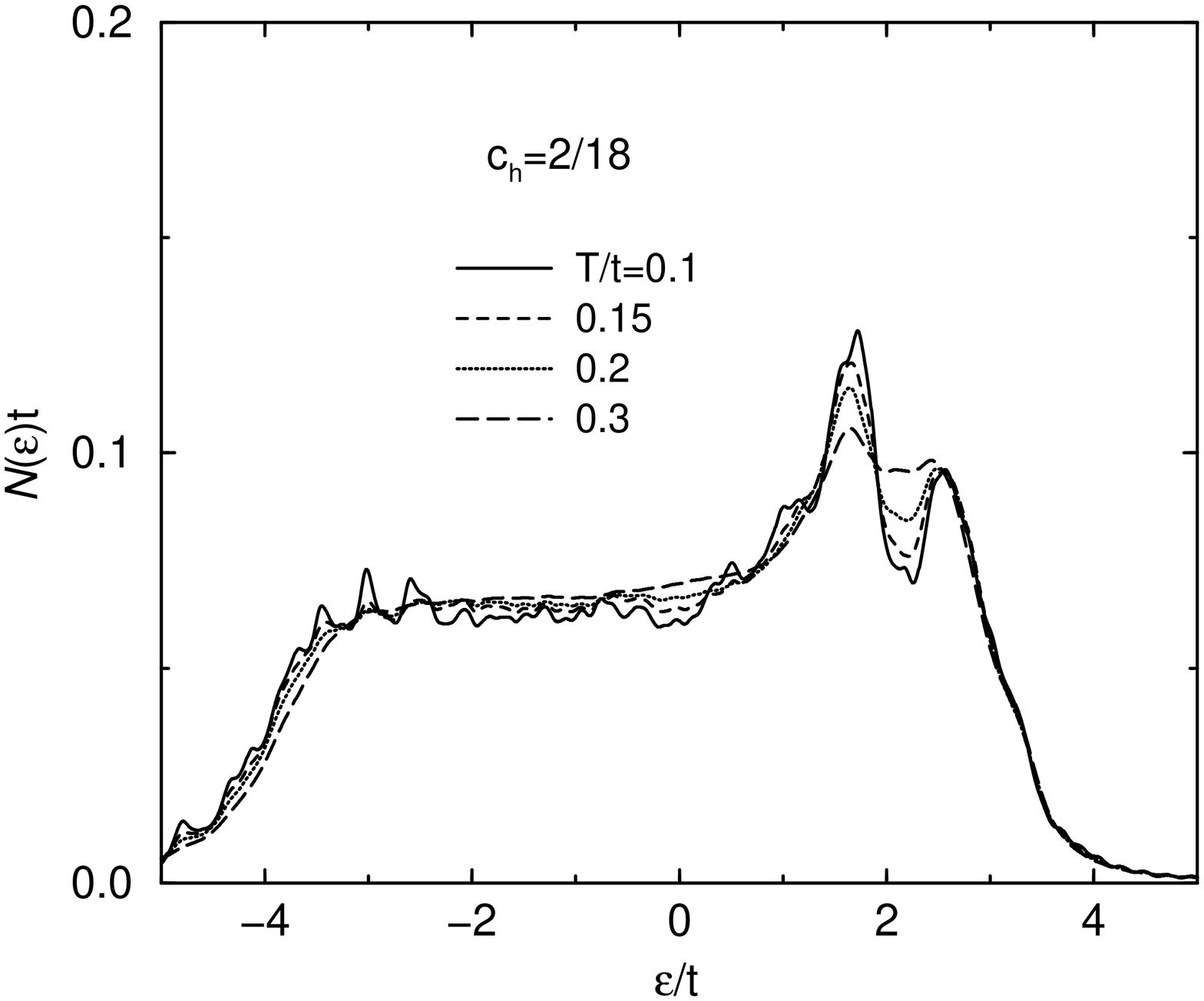,height=8cm,angle=-90}}}
\fi
\caption{
${\cal N(\varepsilon)}$ at various $T$ for dopings:
(a) $c_h=1/20$ and (b) $c_h =2/18$.
} \label{7.8}
\end{figure}

Let us now look closer on ${\cal N}(\varepsilon)$ in the underdoped
regime.  We realize quite clearly that at $c_h=0.12$ a pseudogap
starts to appear in the DOS at $\omega=\varepsilon-\mu \sim 0$ and
becomes even more pronounced at $c_h \sim 0.06$. It is evidently
related to a similar phenomenon in $A(\vec k,\omega)$ in
Fig.~\ref{7.1}. In order to establish the origin of such a gap, we
follow in Fig.~\ref{7.8} the variation of ${\cal N}(\varepsilon)$ with
$T$. We find that the gap structure disappears at $T>J$, so it must be
related to the onset of the short-range AFM order. Also the gap energy
$\Delta\epsilon \alt t$ seems to be determined rather by $J$. It is
plausible that the onset temperature is related to pseudogap scale
$T^*$, discussed before in relation with the maximum in $\chi_0(T)$,
and with the crossover in $\rho(T)$.

Recently the DOS has been measured via the integrated PES and IPES for
LSCO compound in a wide range of doping $0<c_h<0.3$ (Ino {\it et al.}
1997b). More reliable are PES spectra proportional to ${\cal
N}^+(\omega)={\cal N}(\omega+\mu)f(\omega)$, giving information on the
hole part $\omega<0$.  Some features are well consistent with our
results.  In the overdoped regime $c_h>0.2$ ${\cal N}^+(\omega<0)$
appears quite flat. At the same time, there is a substantial and broad
IPES contribution indicating an enhanced DOS at $\omega>0$. On the
other hand, for $c_h<0.17$ a pseudogap starts to emerge gradually at
$\omega \alt 0$. E.g., the inflection point in ${\cal N}^+(\omega)$
moves from $\omega \sim 0$ toward $\omega \sim -0.2~$eV at lowest
dopings. This is quite close to values found in Figs.~\ref{7.8}.  Thus
the gap scale indeed seems to be related to the pseudogap scale $T^*$
discussed in connection with $\chi_0$. Note that the latter is
substantially larger than the leading-edge shift $\sim 30$~meV found
in ARPES (Marshall {\it et al.}  1996). Still it is well possible that
both phenomena are closely related.

\setcounter{equation}{0}\setcounter{figure}{0}
\section{Other properties}

\subsection{Electron density correlations}

It is interesting to study also electron-density dynamics, as
contained within the dynamical charge susceptibility $\chi_c(\vec
q,\omega)$ and the corresponding density correlation function $N(\vec
q,\omega)$, defined by
\begin{eqnarray}
\chi_c''(\vec{q},\omega)&=&e_0^2(1-e^{-\beta\omega})N(\vec{q},\omega),
\nonumber \\
N(\vec{q},\omega)&=& {\rm Re}\int_{0}^\infty dt\; e^{i\omega t}
\langle n_{\vec{q}}(t) n_{-\vec{q}}(0)\rangle, \label{od1} \\
n_{\vec{q}}&=& (1/\sqrt{N})\sum_{i}e^{i\vec{q}\cdot\vec{R}_i}
n_{i}. \nonumber 
\end{eqnarray}
Recently several numerical and analytical studies within the $t$-$J$
model have been devoted to static $N(\vec q)$ (Putikka {\it et al.}
1994) as well as to dynamical $N(\vec q,\omega)$ (Tohyama {\it et al.}
1995, Eder {\it et al.} 1995, Khaliullin and Horsch 1996) at $T=0$,
mainly in connection with the interesting conjecture of the
charge-spin separation (Anderson 1987, Baskaran {\it et al.} 1987) in
layered cuprates.  A parallel study of charge correlations $N(\vec
q,\omega)$ and the spin structure factor $S(\vec q,\omega)$ however
reveals quite an essential difference between an 1D Luttinger liquid
and a 2D doped AFM (Tohyama {\it et al.} 1995), as well as between
charge and spin fluctuations. Typically, $N(\vec q,\omega)$ show a
broad peak at high $\omega \sim 6~t$ at $\vec q \sim \vec Q$ due to
the incoherent motion of holes, while at $q \to 0$ density
fluctuations narrow in a collective acoustic-like mode (Khaliullin and
Horsch 1996). In contrast to $S(\vec q,\omega)$, $N(\vec q,\omega)$
spectra seem to scale linearly with $c_h$ at low doping (Eder {\it et
al.} 1995). It should be also noted that quite generally $N(q\to
0,\omega)$ fluctuations can be related to (anomalous) current
correlations $C(\omega)$ through the particle-number conservation, so
an anomalous behaviour is quite possible for $N(\vec q,\omega)$ as
well.

The investigation of $N(\vec q,\omega)$ at $T>0$ has not been
performed so far. We present here some results, using the FTLM in
analogy with the evaluation of $S(\vec q,\omega)$ in Sec.~6. Our aim
is again twofold. On one hand, FTLM gives smoother spectra at
$T>T_{fs}$, which could be compared to g.s. ED results.  More
interesting is the low-frequency regime $\omega <J$, which cannot be
studied reliably in g.s. calculations, and moreover the $T$ dependence
of spectra in this regime.

\begin{figure}
\centering
\iffigure
\epsfig{file=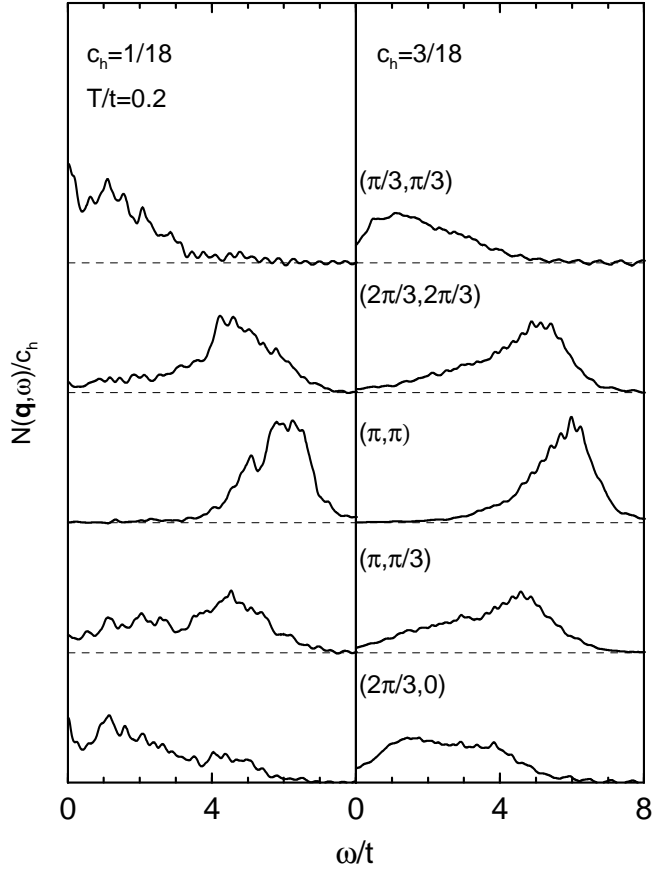,height=14cm,clip=}
\fi
\caption{
Normalized density correlation spectra $N(\vec
q,\omega)/c_h$ at fixed $T/t=0.2$ and $J/t=0.4$ at two dopings:
$c_h=1/18$ and $c_h=3/18$. } \label{5.11}
\end{figure}

In Fig.~\ref{5.11} we present normalized spectra $N(\vec
q,\omega)/c_h$ for nonequivalent $\vec q$ on a lattice of $N=18$ sites
and fixed $T=0.2~t$.  In this case we use $J/t=0.4$. Shown are results
for $c_h=1/18$ and $c_h=3/18$, belonging to regimes of low doping and
optimum doping, respectively.  We notice that normalized spectra are
very similar for both $c_h$, at least for larger $q$, taking into
account also that a more detailed structure for $c_h=1/18$ is partly due
to finite size effects. Our results are in general close to previous
numerical (Tohyama {\it et al.} 1995, Eder {\it et al.} 1995) and
analytical (Khaliullin and Horsch 1996) results. It is characteristic
that at $\vec q \sim \vec Q$ the intensity is nearly exhausted by a
large peak at $\omega \sim 6t$. As expected from conservation laws
spectra are narrowing for $q \to 0$.

\begin{figure}[ht]
\centering
\iffigure
\epsfig{file=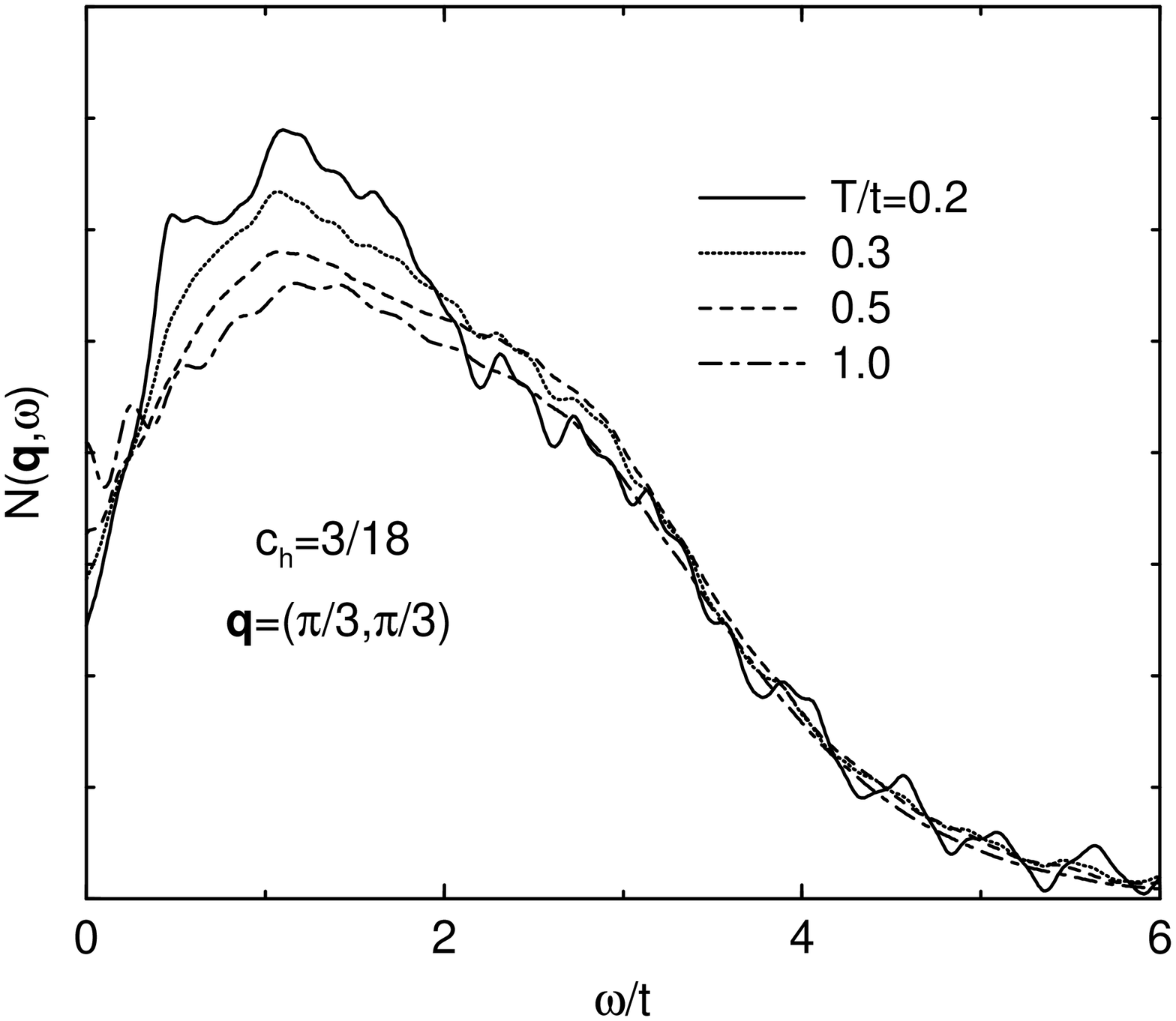,height=10cm,angle=-90}
\fi
\caption{ $N(\vec q,\omega)$ for  $\vec q=(\pi/3,\pi/3)$ and different
$T$ at fixed $c_h=3/18$.  } \label{5.12}
\end{figure}

It is however remarkable that in spite of quite low $T$ in
Fig.~\ref{5.11} $N(\vec q, \omega \to 0)$ does not seem to approach
zero for any $\vec q$, the effect being more visible for smaller
$q$. We follow the evolution of density fluctuations with $T$ at
intermediate doping $c_h=3/16$ for chosen $\vec q=(\pi/3,\pi/3)$ in
Fig.~\ref{5.12}. It is evident that $N(\vec q,\omega)$ is nearly $T$
independent even for $\omega<T$. The behaviour is quite similar to the
one observed for local spin correlations $\bar S(\omega)$ in
Fig.~\ref{6.5} and is consistent with the MFL assumption (\ref{cp2})
for the charge susceptibility $\chi_c''(\omega)$, requiring via the
equation (\ref{od1}) $N(\vec q,\omega \to 0)=const$ independent of
$T$. The MFL behaviour of $N(\vec q, \omega)$ at intermediate doping
is not entirely surprising due to the connection to the anomalous
MFL-type $C(\omega)$ and optical conductivity $\sigma(\omega)$, as
well due to the relation of the electron dynamics to anomalous spin
fluctuations $S(\vec q,\omega)$ (Prelov\v sek 1997a).

$N(\vec q,\omega)$ have been mainly the subject of theoretical
considerations so far. It should be however noted that a relation
could be established with experimentally relevant dielectric function
$\epsilon(\vec q,\omega)$, if the model would incorporate also the
long-range Coulomb repulsion between electrons. Such an additional
interaction could be possibly treated within a framework of an
random-phase-like analysis. It would be desirable since $\epsilon(\vec
q,\omega)$ can be measured also in cuprates via the electron-energy
loss spectroscopy (N\"ucker {\it et al} 1989) which yields directly
Im$[1/\epsilon(\vec q,\omega)]$.

\subsection{Electronic Raman response}

One of very useful probes for the investigation of excitations in an
AFM has been the Raman scattering. Using the latter method, it has
been also established that the reference insulating cuprates
correspond well to a planar Heisenberg AFM, where pronounced Raman
resonance processes at low $T$ are attributed to two-magnon
excitations (Lyons {\it et al.} 1988, Singh {\it et al.} 1989).

A general framework for the theoretical explanation of Raman processes
in correlated systems has been so far given within the Hubbard model,
where the effective Raman operator for resonant and off-resonant
conditions has been derived (Shastry and Shraiman 1990). Still there
are several unresolved questions concerning Raman processes in
cuprates, and more generally in doped AFM.  One puzzling aspect
concerns the observed pronounced $T$ dependence of the two-magnon peak
in undoped cuprates (Knoll {\it et al.} 1990). The latter has been
interpreted as the phonon-induced broadening, but also invoking
higher-order resonant processes (Chubukov and Frenkel 1995). Another
problem is the doping dependence of Raman spectra. Recent experiments,
performed on YBCO materials in the resonant regime (Blumberg {\it et
al.} 1994), show a dramatic increase of the broadening of the
two-magnon peak with doping, so that spectra appear essentially flat
in the normal phase when approaching the optimum doping. At the same
time, the peak position does not move appreciably.

Here we consider only the resonant Raman processes, since most
experiments are performed in the resonant regime.  In contrast to
response functions considered in previous sections, the operator
relevant for the resonant Raman scattering cannot be determined
uniquely within the $t$-$J$ model, since it necessarily involves
higher (resonant) levels. Still we can adopt a view that more complete
(e.g. three-band) model for cuprates can be mapped onto an effective
Hubbard model (Hybertsen {\it et al.} 1990) for resonant transitions
of interest.  Within the Hubbard model near half-filling the
Raman-scattering operator has been derived in the limit $~t/U \ll 1$
(Shastry and Shraiman 1990) yielding the well known form for the
Heisenberg AFM (Parkinson 1969),
\begin{equation}
R= A \sum_{\langle ij\rangle} (\vec \epsilon_{inc}\cdot \vec r_{ij})
(\vec \epsilon_{sc}\cdot \vec r_{ij}) (\vec S_i\cdot \vec S_j - 
{1\over 4} n_i n_j), \label{or1}
\end{equation}
where $\vec \epsilon_{inc}, \vec \epsilon_{sc}$ are the incident and
the scattered electric-field unit vectors, respectively, and the
amplitude factor $A = 4 t^2/(U-\omega_{inc}) \propto J$ incorporates
the resonance at the incident-light frequency $\omega_{inc} \sim U$.
The Raman spectral function is then given by
\begin{equation}
I(\omega) = {1\over \pi N}{\rm Re}\int_0^\infty dt ~e^{i \omega t}
\langle R(t)R(0)\rangle, \label{or2}
\end{equation}
which can be calculated at $T>0$ using the FTLM analogous to other
correlation functions.

Within the $t$-$J$ model with finite doping the Raman intensity
$I(\omega)$, corresponding to the operator (\ref{or1}), has been
evaluated so far at $T=0$ using the ED (Dagotto and Poilblanc 1990),
while for the undoped AFM also $T>0$ has been studied using the full
diagonalization (Bacci and Gagliano 1991). Results obtained via the
FTLM have an advantage over previous ones also for low $T$, since the
Raman intensity is not expected to be singular for $T\to 0$, so much
smoother spectra are obtained by using small but finite $T>T_{fs}$.

We restrict our analysis to the dominant $B_{1g}$ scattering geometry
with $\vec \epsilon_{inc} =(\vec e_x +\vec e_y)/\sqrt{2}$ and $\vec
\epsilon_{sc} = (\vec e_x -\vec e_y)/\sqrt{2}$. The undoped
AFM at $T>T_{fs}\sim J/2$ has been studied by Prelov\v sek and
Jakli\v c (1996), and reveals at low $T$ a two-magnon Raman peak at
$\omega \sim 3.3~J$, consistent with experiments. The width is
however quite narrow and starts to broaden substantially only at higher $T
\sim J$, where a gradual transition to a broad featureless spectrum
occurs. Hence other mechanisms have to be invoked to account for the
observed pronounced $T$-dependent width at lower $T\ll J$ (Knoll {\it
et al.} 1990).

\begin{figure}[ht]
\centering
\iffigure
\epsfig{file=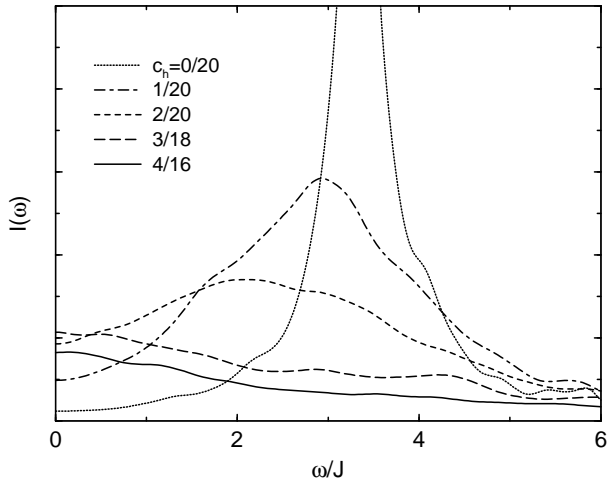,height=8cm,angle=-90}
\fi
\caption{Raman intensity $I(\omega)$ for different dopings $c_h$ at
fixed $T=0.15~t$. } \label{8.1}
\end{figure}

In doped systems $T$ plays a less essential role and typically we
observe only a weak decrease of the Raman intensity in the interval
$T/J = 0.3 - 1.0$ (Prelov\v sek and Jakli\v c 1996). On the other
hand, the dependence on doping is essential, as evident from
Fig.~\ref{8.1}, where we present $I(\omega)$ for various dopings $c_h
\le 0.25$ at lowest $T=0.15~t> T_{fs}$. Already the smallest nonzero
doping $c_h =0.05$ increases dramatically the width of the two-magnon
peak, while spectral features become overdamped on approaching the
optimum doping $c_h \sim 0.15$. It is however remarkable, that the
peak position does not shift appreciably in the underdoped regime $c_h
\leq 0.1$. Only for $c_h > 0.15$ the spectra change to a broad
central-peak form with a maximum at $\omega = 0$.

How can we interpret the above results for the $B_{1g}$ resonant Raman
scattering ? Up to the overdoped situation $c_h \sim 0.3$ the
scattering seems to be dominated by the spin-exchange part
(\ref{or1}), since the latter determines the low-energy fluctuations
even at the optimum doping. Still, there are evident changes with
doping, since the reduced AFM correlation length at larger $c_h$
induces a large broadening of the Raman two-magnon peak and a
reduction of its intensity. In this respect the doped system behaves
quite similarly to an undoped AFM but at elevated $T\sim J$.

A systematic resonant-Raman scattering study has been recently
performed on a sequence of YBCO materials (Blumberg {\it et al.}
1994).  It seems that in the underdoped regime our model results well
reproduce experimental ones, both regarding the shape of the Raman
spectra and their intensity variation with doping. At optimum doping
experiments still reveal a weakly pronounced peak at the same
frequency $\omega \sim 3~J$, while our results on Fig.~\ref{8.1}
already reveal a maximum at $\omega=0$, nevertheless both spectra are
nearly flat.

\subsection{Thermoelectric power}

Among other transport properties the thermoelectric power (TEP) or
the Seebeck coefficient $S$ is one of the most frequently
investigated. Within the cuprate family TEP has been measured for
several materials, for an overview see e.g.  Kaiser and Uher (1991)
and more recent analyses (Tallon {\it et al.} 1995,
Cooper and Loram 1996). Again an universal behaviour has emerged
depending mainly on hole doping. For weakly doped materials $S$ is
large and positive, with some $T$ dependence at higher $T$.  With
increasing doping TEP $S$ decreases rapidly. It also falls off nearly
linearly with $T$, but with a rather small slope. Quite consistently
$S$ is nearly vanishing for the optimum doping and changes sign to a
negative one in overdoped materials. This variation has clearly some
parallel with the doping dependence of the Hall coefficient $R_H$,
also showing a transition from a hole-like behaviour $R_H>0$ at low
doping into an electron-like $R_H<0$ in overdoped samples. It seems
plausible that both anomalous properties are related and emerge from
strong correlations in these materials.

Theoretically TEP in doped AFM or more generally in strongly
correlated metals has not attracted much attention so far (Hildebrand
{\it et al.} 1997). This is not surprising, since it seems to require
the understanding of both electrical and thermal currents.

Within the linear response theory (Mahan 1990) the TEP can be
expressed in terms of particle current $j$ and energy current
$j_E$ correlation functions,
\begin{equation}
S={1\over e_0T}\left[ \mu -{C_{j_Ej}(\omega \to 0) \over
C(\omega \to 0)} \right], \label{ot1}
\end{equation}
where mixed correlation function $C_{j_Ej}(\omega)$ is defined in
analogy with the expression (\ref{ec1}) for $C(\omega)$.

The evaluation of $S(T)$ thus requires the calculation of
$C_{j_Ej}(\omega)$, in addition to the current-current correlation
function $C(\omega)$ discussed extensively in Sec.~5. It is
straightforward to derive the expression for $j_E$, nevertheless the
operator is more involved due to three-site terms. We do not intend to
give here a more complete analysis of $C_{j_Ej}(\omega)$ which will be
presented elsewhere. We mention however only a preliminary
observation, that $C_{j_Ej}(\omega)$ and $C(\omega)$ appear to be
closely related, 
\begin{equation}
C_{j_Ej}(\omega) = \zeta C(\omega), \label{ot2}
\end{equation}
i.e. $\zeta$ is nearly $\omega$ independent for $\omega<t$, but as
well $T$ independent at low $T<J$. Since we are dealing at $c_h>0$
with a metal, although with a strange one, we expect a finite $|S(T\to
0)| < \infty$.  Hence we deduce from the equation (\ref{ot1}) the condition
$\zeta = \mu(T=0)$. Our numerical results for $\zeta$ are
indeed consistent with this assumption.

\begin{figure}[ht]
\centering
\iffigure
\epsfig{file=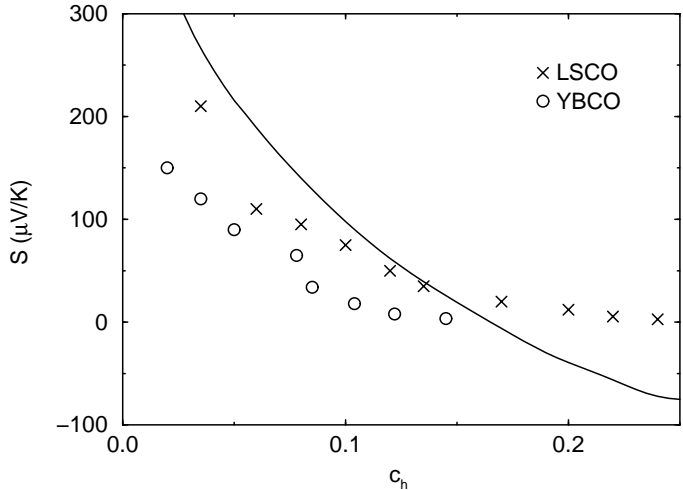,height=10cm,angle=-90}
\fi
\caption{ Thermoelectric power $S$ vs. $c_h$ for $T/t=0.1$.
Experimental result for LSCO and oxygen deficient YBCO are taken from 
Cooper and Loram (1996).  } \label{8.2}
\end{figure}

We thus arrive at the simplified expression,
\begin{equation}		
S \sim {1\over e_0T}[\mu_h(T=0)-\mu_h(T)], \label{ot3}
\end{equation}
which involves only $\mu_h(T)$, studied in Sec.~4.1 and shown in
Fig.~\ref{4.1}. In the low-$T$ regime of interest we claim in a broad
regime of doping $c_h<0.3$ a linear variation (\ref{t2a}) of 
$\mu_h(T)$. This leads directly to $S\sim - \alpha S_0$ where $S_0
k_B/e_0 = 86 \mu V/K$. It is evident from Fig.~\ref{7.3} that $\alpha$
is changing sign from a negative for $c_h<c_h^*$ to a positive one for
$c_h>c_h^*$. In Fig.~\ref{8.2} we plot our result for $S(c_h)$. Note
that due to simplifications involved in the expression (\ref{ot3}) we
do not intend to consider the $T$ variation, hence our results apply
to lowest reachable $T\alt T_{fs} \sim 0.1~t$. For comparison we
present on the same plot also experimental results for LSCO and oxygen
deficient YBCO, taken from Cooper and Loram (1996), whereby data
refer to the normal state at $T \sim 300K$. Qualitative agreement
between theory and experiment is quite reasonable. In LSCO values of
$S$ are decreasing with doping, while $S$ is quite high at low doping
$c_h<0.05$. Towards optimum doping $S$ essentially vanishes. Similar
is the trend for YBCO data, although there are clearly quantitative
differences. The main disagreement between the calculated and
experimental $S$ is in the overdoped regime, where our results
indicate $S<0$ with probably too large values. It is however well
possible that in this regime our analysis is not adequate, first due
to $\mu_h(T)$ approaching more normal LFL $T^2$ dependence, as well as
due to the breakdown of the relation (\ref{ot2}). 

\setcounter{equation}{0}\setcounter{figure}{0}
\section{Discussion}

In the absence of accepted microscopic or phenomenological theories,
which would at least qualitatively describe normal-state properties of
cuprates in the whole doping regime, it is also hard to summarize our
numerical results in a compact manner.

It is first important to understand which phenomena determine the
optimum doping in cuprates and in microscopic models. In cuprates the
optimum is defined by $T_c(c_h^{opt})=max$, but at least very close in
doping, i.e., $c^*_h \sim c_h^{opt}$ are also materials where the
low-$T$ electronic entropy $s$ and the uniform susceptibility $\chi_0$
are maximum (Loram {\it et al.} 1996), where the pseudogap scale
disappears, i.e. $T^* \to T_c$ (Batlogg 1997), etc.

Within the $t$-$J$ model we can associate the optimum so far only with
respect to normal-state properties, e.g. with $s(c_h^*)=max$. Such a
maximum has to exist at a nontrivial $c_h^*>0$ due to the relation
(\ref{t5}) with $\mu_h(T)$, where the latter must change the sign from a
weakly doped AFM (semiconductor-like regime) at $c_h \agt 0$ to a
regime of nearly free 2D fermions at $c_e \sim 0$. It is plausible
that within the prototype model (\ref{cm1}) $c_h^*$ is determined by
the interplay of the spin exchange energy $E_J=\langle H_J \rangle$
and the (hole) kinetic energy $E_t=\langle H_t \rangle$. Since at low
doping we get $|E_t| \propto t c_h$ and $|E_J| \propto J$ we can get a
rough estimate $c_h^* \propto J/t \ll 1$. From such an argument it
seems plausible that at $c_h \sim c_h^*$ the AFM correlation length
becomes very short $\xi \sim 1$, while AFM correlations are
essentially irrelevant in the overdoped regime $c_h > c_h^*$.

\subsection{Universality at intermediate doping}

It is evident from our results that normal-state properties are
universal at the optimum doping $c_h \sim c_h^*$. Since the maximum
$s(c_h)$ is broad within the $t$-$J$ model also this regime is quite
extensive, i.e. $0.15<c_h<0.25$ for chosen $J/t = 0.3$. The main
feature of this regime obtained in our study is the MFL-type dynamics
in several response functions: the local spin susceptibility
$\chi_L''(\omega)$, the optical conductivity $\sigma(\omega)$, density
correlations $N(\vec q,\omega)$, and spectral functions $A(\vec
k,\omega)$. It is characteristic that the dynamics in the low-$\omega$
($\omega<J$) regime seems to be universal, i.e. it is determined
solely by $T$ itself without any additional parameters, as concluded
from equations (\ref{eo2}) for $\sigma(\omega)$ and (\ref{ml4}) for
$\chi_L''(\omega)$. It should be however reminded that this
universality could be restricted to a certain $T$ range, e.g.  we can
show its validity only for $T_{fs}<T<J$.

It seems that the origin of the universality is in the large
degeneracy of the spin subsystem, where the frustration in induced by
the hole doping. A plausible argument is that holes tend to induce FM
correlations to minimize kinetic energy, while spins prefer an AFM
ordering. In any case, only spins have enough degrees of freedom to
explain large entropy $s \sim 0.4~k_B$ even at $T<J$.

Let us attempt an explanation for the universality of $\bar S(\omega)$
at $c_h\sim c_h^*$. From the explicit representation in terms of
eigenstates,
\begin{equation}
\bar S(\omega) = (1+ e^{-\beta\omega}) 
\sum_{n,m} {e^{-\beta E_n}\over Z} |\langle \Psi_n| S_i^z|\Psi_m \rangle|^2 
\delta (\omega - E_m+E_n), \label{d1}
\end{equation}
we can first discuss conditions that the response is $T$ independent
for $\omega > T$.  Since in this case the prefactor is constant, a
naive requirement appears to be that low lying states $E_n -E_0 < T$
have similar matrix elements to excited states with $E_m-E_n > T$.
Nevertheless the validity of this observation clearly depends on the
character and on the density of low-lying many-body states, and
e.g. does not apply at low $T$ to a system with a unique g.s.  As
realized from the entropy results in Sec.~4.2 the degeneracy and the 
density of many-body states is clearly very large at the intermediate
doping. To explain the $T$ independence of $\bar S(\omega)$ even for
$\omega<T$, we recall the sum rule (\ref{ml2}) and assume that
there is no characteristic scale $\omega_c< T$ which could introduce
an additional low-$\omega$ structure in $\bar S(\omega)$.  A natural
candidate for such a scale in an AFM is the (gap) frequency $\omega_c
\sim c /\xi$ where $c$ is the spin wave velocity and $\xi$ the AFM
correlation length.  Here originates the essential difference between
an undoped and the optimally doped AFM. In a pure AFM within the
renormalized classical regime $\xi$ is exponentially large for $T\ll
J$ (Chakravarty {\it et al.}  1989) and consequently $\omega_c \ll T$.
On the other hand, at the intermediate doping $\xi$ appears to be
short and not strongly $T$ dependent, i.e. $\xi \propto 1/\sqrt{c_h}$.
The latter situation excludes $\omega_c<T<J$, hence leading to an
universal $\bar S(\omega)$.

Turning to the discussion of $\sigma(\omega)$, as given by the equation
(\ref{eo2}), we realize that it is not meaningful to associate
$1/\tau(0) \propto T$ with a current relaxation rate. Namely the
current-current correlation function $C(\omega)$ shows a broad uniform
spectrum indicating on a very fast current decay with $1/\tilde \tau
\sim 2t$. Thus it seems that a general condition for the validity of
the specific MFL form (\ref{eo2}) is a fast current decay with $1/\tilde
\tau \gg T$. This could happen due to holes moving in a disordered
spin subsystem, which leads to entirely incoherent collisions among
holes and the appropriate mean free path $l_s$ is only few unit
cells. Hence it seems that we are dealing with a novel phenomenon of
quantum diffusion which follows an universal form (\ref{eo2}).

In an analogous way also the relaxation of single electron states near
FS is affected, as manifested in the MFL form of the self energy
$\Sigma(\vec k,\omega)$, equation (\ref{ss1}). Anyhow it is plausible
that the QP damping $\Sigma''(k\sim k_F,\omega)$ and the current
relaxation $1/\tau(\omega)$, as given by the equation (\ref{eo6}), are
closely related. There are also phenomenological arguments (Littlewood
and Varma 1991) which indicate that the MFL-type dynamics of a boson
(spin) subsystem implies a MFL behaviour of the electron
propagator. Recently such a relation has been derived by one of the
authors (Prelov\v sek 1997a), employing a decoupling approximation for
the projected $G^R(\vec k,\omega)$.

On the other hand it should be stressed that within the same regime
certain electronic properties appear quite normal, i.e. close to the
LFL behaviour. First, spectral functions indicate on a rather well
defined large FS following the Luttinger theorem. This seems to
indicate that correlated electronic state evolves continuously from a
noninteracting one, whereby the FS can be traced to momenta of
lowest (g.s.)  many-body states. It is much harder to explain quite
LFL behaviour of the entropy $s \propto T$ at low $T$, as seen in
Fig.~\ref{4.3}a, and moreover a free-fermion-like Wilson ratio $W \sim
W_0$, as found in Sec.~6.2.

It should be noted again that such an universality could be restricted
to a certain $T$ window. Namely it is well possible that at low $T\sim
T_{coh}$ some coherence or ordering could appear. Note however that
e.g. the onset of a spin ordering would invalidate our arguments on
the MFL-type universality in the low-$T$ phase. This is relevant for
cuprates showing stripe structures (Tranquada {\it et al.} 1995) or
longer range incommensurate ordering (Hayden {\it et al.} 1996). Even
in this case the dynamics at higher-$\omega$ is possibly not affected
by the ordering, as seen in the analysis of the neutron scattering
experiments in Sec.~6.6.

At the intermediate doping our numerical results do not give an
indication for a coherence temperature $T_{coh}$, where the breakdown
of the universality is expected. This is not surprising in view of the
experimental facts for cuprates where at optimum doping $T_{coh} \sim
T_c$ and our $T_{fs}>T_c$. Nevertheless we note that certain
restrictions follow already from the MFL-type dynamics. E.g., the
particular MFL form (\ref{eo2}) for $\sigma(\omega)$ leads at $T\to 0$
to a divergent integral on the lower bound $\omega \to 0$ and hence
violates the optical sum rule (\ref{ec9}). Therefore such a form
cannot apply to arbitrary low $T$ giving an indication for the
existence of a lower crossover $T_{coh}$.

\subsection{Energy scales in underdoped antiferromagnet}

The underdoped regime $c_h<c_h^*$ is less convenient for the FTLM
approach. One reason is that $T_{fs}$ is increasing towards the undoped
AFM. At the same time the low-energy (pseudogap) scale is increasing
and its effects are for certain quantities hardly resolvable from
finite-size effects, hence they increase the uncertainty of results.

Still there are clear indications for the pseudogap scale $T^*$ within
the $t$-$J$ model. Most evident is the maximum in $\chi_0(T)$, as
presented in Sec.~6.2. Quite a similar behaviour can be followed also
for $s(T)/T$, as deduced e.g. from Fig.~\ref{4.3}a. This is related to
nearly constant Wilson ratio $R_W(T) \sim 1$, as defined in the equation
(\ref{mu2}). Note that such a behaviour has been found also
experimentally by Loram {\it et al.} (1996), extending to low-$T$
regime not reachable in our calculations. It is quite evident from
$\chi(T)$ that the pseudogap scale $T^*$ is related to the onset of
short range AFM correlations.

It seems that the same phenomenon governs the pseudogap appearing in
the DOS $\cal N(\varepsilon)$ in the underdoped regime, as discussed
in Sec.~7.4. While the pseudogap is clearly visible for lowest $c_h$,
e.g. for $c_h\sim 0.06$, it is becomes more shallow towards the
optimum doping $c_h \sim c_h^*$. This is consistent with PES
measurements by Ino {\it et al.} (1997b). Experimentally one of the
indications for $T^*$ is also the change of the slope in
$\rho(T)$. Here our results are less conclusive and appear to be
finite-size limited for $T<T^*$.  A possible interpretation is that
the QP scattering in the presence of longer range AFM correlations
becomes a different one, in particular the mean free path $l_s$
increases due to the opening of the spin pseudogap. In our small
systems this possibly leads to $l_s(T<T^*)>L$ and consequently to a
finite-size $D_c(T<T^*)>0$ which cannot be interpreted uniquely. Note
that so far there is neither a satisfactory theory nor a numerical
calculation of the mobility $\mu_0(T < J)$, even for a single hole in an
AFM which could serve as a guide for the limit $c_h \to 0$.

Experiments indicate the existence of a lower crossover scale $T_{sg}$
(Batlogg 1997), see Fig.~~\ref{2.1}, identified as a spin gap in the
NMR relaxation and appearing also in the ARPES, $\sigma_c(\omega)$
etc. In our calculations it would be easiest to resolve the existence
of a spin gap in $\chi_0(T)$ and in $\bar S(\omega)$, still we do not
find an indication for that down to $T\sim 0.1~t$.

\subsection{Conclusions and open questions}

One of the main conclusions of this work is that the prototype $t$-$J$
model, in spite of its apparent simplicity, accounts surprisingly well
for a variety of normal state properties of cuprates. This confirms
the belief that most unusual properties are dominated by strong
correlations and by the interplay of antiferromagnetism and itinerant
character of electrons, both effects being inherent within the model.
The agreement is a qualitative but as well a quantitative one, hence
the parameters of the model appear to be really microscopic ones in
the whole range of doping and not just some effective parameters
changing with doping.

There are evidently many open questions regarding the interpretation
and the understanding of our results, and related experimental facts
on cuprates:

\noindent a) The universal MFL-type dynamics in the intermediate regime, which
seems to be well founded by our results on $\sigma(\omega)$,
$\chi_L''(\omega)$ and self energies, as well as by experimental
facts, lacks a proper theoretical explanation, although some relations
have been proposed already (Littlewood and Varma 1991, Prelov\v sek
1997a).

\noindent b) There are more fundamental problems in the underdoped
regime. One of them is the evolution of the FS in the limit $c_h \to
0$. A number of authors investigated the evidence for the small FS -
hole pockets (Eder and Becker 1991, Eder {\it et al.} 1994) and its
possible consequences (Trugman 1990) at low doping, also stimulated by
recent ARPES measurements (Marshall {\it et al.}  1996) of underdoped
materials. In a strict sense at $T\to 0$ the transformation from a
large to a small FS seems to require a phase transition with a
qualitative change in a number of properties. There is no evidence for
that neither in experiments nor in our small-system analysis. On the
other hand our spectral functions at lowest $c_h>0$ reveal a folding
of QP dispersion analogous to a single hole in an AFM. Still it is
possible that such a feature coexists with a normal QP dispersion
crossing of a large FS, however with a weight $\tilde Z \to 0$,
i.e. gradually vanishing with doping as predicted by the equation
(\ref{ss2}). Such a scenario is not in contradiction with ARPES
results. It has been in fact recently deduced from the integrated PES
experiments (Ino {\it et al.}  1997b), still it would be hard to
establish it beyond doubt.

\noindent c) Related is the question of the thermodynamics at very low
doping. There is a distinction whether at low $T$ QP behave as a
degenerate Fermi gas with finite QP weight $\tilde Z >0 $ even for
$c_h \to 0$, or as a nondegenerate system of spin polarons with
$\tilde Z \to 0$.  In the first case one could expect for $c_h,T \to
0$ a finite electron compressibility $\kappa < \infty$, as discussed
in Sec.~4.1, while the alternative would induce $\kappa \to
\infty$. Our results are not conclusive in this respect, there is
however also a numerical evidence for the latter scenario within the
Hubbard model (Furukawa and Imada 1992, Assaad and Imada 1996).

There are several quantities which we did not discuss in this review.
One of most challenging is the Hall effect, which is known to be
anomalous. Unfortunately the inclusion of a real magnetic field $B>0$
increases computational efforts (see Sec.~6.7), while at the same time
also the physics of the Hall conductivity is less local. The anomalous
properties of the orbital diamagnetism $\chi_d(T)$ and its relation to
the Hall effect, equation (\ref{mo1}), nevertheless confirm the anomalous
behaviour of $R_H(T)$. Also perpendicular $\sigma_c(\omega)$ is known
to be a challenge for theoreticians. Still it appears that the
interplane transport is essentially incoherent for $T>T_c$, so it
could be possibly related to the knowledge of planar spectral
functions $A(\vec k,\omega)$.

The major open question within the $t$-$J$ (or Hubbard) model is
still the existence of additional low energy scales and of related
low-$T$ transitions. There are certain indications that such scales
must exist in such models, it remains however a subject of future
studies to find whether such transitions lead to an ordered spin and
charge structure or to desired superconductivity.

\vskip 0.5 truecm
{\bf Acknowledgments}

The work is supported by the Ministry of Science and Technology of
Slovenia.  One of the authors (J.J.) wants to thank for the support
and the hospitality of the Max-Planck Institut f\"ur Physik Komplexer
Systeme, Dresden, where part of the work has been completed.

\newpage

\section{References}

\begin{list}{}{\setlength{\itemindent}{-1cm}\setlength{\itemsep}{0pt}
\setlength{\parsep}{0pt}}
\item $^*$ Present address: Cadence Design Systems, D-85540 Haar,
Germany

\item

\item ABRIKOSOV, A. A., GOR'KOV, L. P., and DZYALOSHINSKII, I. E., 1965,
{\it Quantum Field Theoretical Methods in Statistical Physics}
(Pergamon, Oxford), p. 169.

\item ANDERSON, P. W., 1987, {\it Science} {\bf 235}, 1196.

\item ANDERSON, P. W., and ZOU, Z., 1988, {\it Phys. Rev. Lett.} {\bf 60},
132.

\item ASSAAD, F. F., and IMADA, M., 1996, {\it Phys. Rev. Lett.} {\bf
76}, 3176.

\item BACCI, S., and GAGLIANO, E., 1991, {\it Phys. Rev.} B
{\bf 43},  6224.

\item BANG, Y., and KOTLIAR, G., 1993, {\it Phys. Rev.} B {\bf 48},  9898.

\item BARADUC, C., EL AZRAK, A., and BONTEMPS, N., 1995,
{\it J. of Supercond.}, {\bf 8}, 1.

\item BARNES, T., and RIERA, J., {\it Phys. Rev.} B {\bf 50},
6817 (1994).

\item BASKARAN, G., ZOU, Z., and ANDERSON, P. W., 1987, {\it Solid State
Commun.} {\bf 63}, 973.

\item BATLOGG, B., {\it et al.}, 1994, {\it Physica} C
{\bf 235-240}, 130.

\item BATLOGG, B., 1997, {\it Physica} C {\bf 235-240}, 130.

\item BEDNORZ, J. G., and M\"ULLER, K. A., 1986, {\it Z. Phys.} B
{\bf 64}, 189.

\item BIRGENEAU, R. J., {\it et al.}, 1988, {\it Phys. Rev.} B
{\bf 38}, 6614.

\item BLUMBERG, G., {\it et al.}, 1994, {\it Phys. Rev.} B {\bf 49}, 13295.

\item BON\v CA, J., and PRELOV\v SEK, P., 1989, {\it
Solid State Commun.} {\bf 71}, 755.

\item BRINKMAN, W., and RICE, T. M., 1970, {\it Phys. Rev.} B {\bf 2},
1324.

\item BULUT, N., SCALAPINO, D. J., and WHITE, S. R., 1994, {\it
Phys. Rev.} B {\bf 50}, 7215.

\item CASTELLA, H., ZOTOS, X., and PRELOV\v SEK, P., 1995,
{\it Phys. Rev. Lett.} {\bf 74}, 972.

\item CASTELLANI, C., DI CASTRO, C., and GRILLI, M., 1995,
{\it Phys. Rev. Lett.} {\bf 75}, 4650.

\item CHAKRAVARTY, S., HALPERIN, B. I., and
NELSON, D. R., 1989, {\it Phys. Rev.} B {\bf 39}, 2344.

\item CHUBUKOV, A. V., and FRENKEL, D. M., 1995,
{\it Phys. Rev. Lett.} {\bf 74}, 3057.

\item COOPER, S. L., {\it et al}., 1993, {\it Phys. Rev.} B {\bf 47}, 8233.

\item COOPER, J. R., and LORAM, J. W., 1996, {\it J. Phys. I France}
{\bf 6} 2237.

\item DAGOTTO, E., and D. POILBLANC, D., 1990, {\it Phys. Rev.} B
{\bf 42}, 7940.

\item DAGOTTO,  E., 1994, {\it Rev. Mod. Phys.} {\bf 66}, 763.

\item DING, H., {\it et al.}, 1996, {\it Phys. Rev. Lett.} {\bf 76}, 1533.

\item DUFFY, D., and MOREO, A., 1997, {\it Phys. Rev.} B 
{\bf 55}, 12918. 

\item EDER, R., and BECKER, K. W., 1991, {\it Phys. Rev.} B 
{\bf 44}, 6982.

\item EDER, R., OHTA, Y., and SHIMOZATO, T., 1994, {\it Phys. Rev.} B 
{\bf 50}, 3350.

\item EDER, R., OHTA, Y., and MAEKAWA, S., 1995, {\it Phys. Rev. Lett.} 
{\bf 74}, 5124.

\item EL AZRAK, A., {\it et al}., 1994, {\it Phys. Rev.} B {\bf 49}, 9846.

\item EMERY, V. J., 1987, {\it Phys. Rev. Lett.} {\bf 58}, 2794.

\item EMERY, V. J., KIVELSON, S. A., and LIN H. Q., 1990, {\it Phys. Rev.
Lett.} {\bf 64}, 475.

\item FLEURY, P., and LOUDON, R., 1968, {\it Phys. Rev.} {\bf 166}, 514.

\item FULDE, P., 1991, {\it Electron Correlations in Molecules
and Solids}, Springer Series in Solid-State Sciences Vol. 100
(Springer-Verlag, Berlin).

\item FURUKAWA, N., and IMADA, M., 1992, {\it J. Phys. Soc. Jpn.} {\bf
61}, 3331.

\item GEORGES, A., KOTLIAR, G., KRAUTH, W., and ROZENBERG, M. J.,
1996, {\it Rev. Mod. Phys.} {\bf 68}, 13. 

\item GOMEZ-SANTOS, G., JOANNOPOULOS, J. D., and NEGELE, J. W., 1989,
{\it Phys. Rev.} B {\bf 39}, 4435.

\item G\"OTZE, W., and W\"OLFLE, P.,  1972, {\it Phys. Rev.} B
{\bf 6}, 1226.

\item HALDANE, F. D. M., 1981, {\it J. Phys.} C {\bf 14}, 2585.

\item HAYDEN, S. M., {\it et al.}, 1996, {\it Phys. Rev. Lett.} {\bf 76}, 1344.

\item  HAYDOCK, R., HEINE, V., and KELLY, M. J., 1972,
{J. Phys.} C {\bf 5}, 2845.

\item HELLBERG, C. S., and MANOUSAKIS, E., 1997, {\it Phys. Rev.
Lett.} {\bf 78}, 4609. 

\item HILDEBRAND, G., HAGENAARS, T. J., GRABOWSKI, S., SCHMALIAN, J.,
and HANKE, W., 1997, {\it Phys. Rev.} B {\bf 56}, R4317.

\item HIRSCH, J. E., 1985, {\it Phys. Rev.} B, {\bf 31}, 4403. 

\item HUBBARD, J., 1963, {\it Proc. Roy. Soc.} A {\bf 277}, 237.  

\item HWANG, H. Y., {\it et al.}, 1994, {\it Phys. Rev. Lett.} {\bf
72}, 2636.

\item HYBERTSEN, M. S., STECHEL, E. B., SCHL\"UTER, M., and
JENNISON, D. R., 1990, {\it Phys. Rev.} B {\bf 41}, 11068.

\item IMAI, T., SLICHTER, C. P., YOSHIMURA, K., and KOSUGE K., 1993,
{\it Phys. Rev. Lett.} {\bf 70}, 1002.

\item IMADA, M., and TAKAHASHI, M., 1986, {\it J. Phys. Soc. Jpn.}
{\bf 55}, 3354.

\item INO, A., 1997a, {\it et al.}, {\it Phys. Rev. Lett.}
{\bf 79}, 2101. 

\item INO, A., 1997b, {\it et al.}, preprint. 

\item IYE, Y., 1992,  in {\it Physical Properties of High Temperature
Superconductors III}, edited by D. M. Ginsberg (World Scientific, 
Singapore), p.285.

\item JAKLI\v C, J.,  and PRELOV\v SEK, P., 1994a, {\it Phys. Rev.} B  
{\bf 49}, 5065.

\item JAKLI\v C, J., and PRELOV\v SEK, P., 1994b, {\it Phys. Rev.} B
{\bf 50}, 7129.
 
\item JAKLI\v C, J., and PRELOV\v SEK, P., 1995a, {\it Phys. Rev. Lett.}
{\bf 74}, 3411.

\item JAKLI\v C, J., and PRELOV\v SEK, P., 1995b, {\it Phys. Rev. Lett.}
{\bf 75}, 1340.

\item JAKLI\v C, J., and PRELOV\v SEK, P., 1995c, {\it Phys. Rev.}
B {\bf 52}, 6903.

\item JAKLI\v C, J., and PRELOV\v SEK, P., 1996, {\it
Phys. Rev. Lett.}  {\bf 77}, 892.

\item JAKLI\v C, J., and PRELOV\v SEK, P., 1997, {\it Phys. Rev.} B
{\bf 55}, R7307.

\item JARRELL, M., GUBERNATIS, J. E., SILVER, R. N., and SIVIA, D. S., 1991,
{\it Phys. Rev.} B {\bf 43}, 1206.

\item JOHNSTON, D. C., SINHA, S. K., JACOBSON, A. J.,
and NEWSAM, J. M., 1988, {\it Physica} C {\bf 153-155},  572. 

\item JOHNSTON, D. C., 1989, {\it Phys. Rev. Lett.} {\bf 62}, 957.

\item KAISER, A. B., and UHER, C., 1991,
in {\it Studies of High Temperature Superconductors}, Vol.7, edited by
A. V. Narlikar (Nova Science Publishers, New York), p. 353.

\item KAMPF, A. P., and SCHRIEFFER, J. R., 1990, {\it Phys. Rev.} B
{\bf 41}, 6399.

\item KANE, C. L., LEE, P. A., and READ, N., 1989, {\it Phys. Rev.} B
{\bf 39}, 6880.

\item KEIMER, B., {\it et al.}, 1992, {\it Phys. Rev.} B {\bf 46},
14034.

\item KHALIULLIN, G., and HORSCH, P., 1996, {\it Phys. Rev.} B {\bf
54}, R9600.

\item KITAOKA,  Y., {\it et al.}, 1991, {\it Physica} C
{\bf 185 - 189}, 98.

\item KNOLL, P., THOMSEN, C., CARDONA, M., and MURUGARAJ, P.,
1990, {\it Phys. Rev.} B {\bf 42}, 4842.

\item KOHN,  W., 1964, {\it Phys. Rev.} {\bf 133}, A171.
 
\item KOHNO, M., 1997, {\it Phys. Rev.} B {\bf 55}, 1435. 

\item LANCZOS, C., 1950, {\it J. Res. Nat. Bur. Stand.} {\bf 45}, 255.

\item LITTLEWOOD, P. B., and VARMA, C. M., 1991,
{\it J. Appl. Phys.} {\bf 69}, 4947.

\item LORAM, J. W., MIRZA, K. A., COOPER, J. R., and  LIANG, W. Y., 
1993, {\it Phys. Rev. Lett.} {\bf 71}, 1740. 

\item LORAM, J. W., MIRZA, K. A., COOPER, J. R., ATHANASSOPOULOU,
N. A., and LIANG, W. Y., 1996, {\it Proc. of 10$^{th}$ Anniversary HTS
Workshop, Houston}, (World Scientific), p.341.

\item LUTTINGER, J. M., 1960, {\it Phys. Rev.} {\bf 119}, 1153.

\item LYONS, K. B., FLEURY, P. A., REMEIKA, J. P., COOPER, A. S.,
and NEGRAN, T. J., 1988, {\it Phys. Rev.} B {\bf 37}, 2353.

\item MAHAN, G. D., 1990, {\it Many-Particle Physics} (Plenum, New York).

\item MAKIVI\'{C}, M.,  and JARRELL, M., 1992, {\it Phys. Rev. Lett.}
{\bf 68}, 1770.

\item MALDAGUE, P. F., 1977, {\it Phys. Rev.} B {\bf 16}, 2437.

\item MANDRUS, D., FORRO, L., KENDZIORA, C., and MIHALY, L., 1991, 
{\it Phys. Rev.} B {\bf 44}, 2418.

\item MANOUSAKIS, E., 1991, {\it Rev. Mod. Phys.} {\bf 63}, 1.

\item MARSHALL, D. S., {\it et al.}, 1996, {\it Phys. Rev. Lett.}
{\bf 76}, 4841.

\item MARTINEZ, G., and HORSCH, P., 1991, {\it Phys. Rev.} B {\bf 44}, 317.

\item METZNER, W., SCHMIT, P., and VOLLHARDT, D., 1992, {\it Phys. Rev.} 
B {\bf 45}, 2237.

\item MILA, F., and RICE, T. M., 1989,  {\it Physica} C, {\bf 157}, 561.

\item MILJAK, M., ZLATI\'C, V., KOS, I., THOMPSON, J. D., CANFIELD,
P. C., and FISK, Z., 1993, {\it Sol. St. Commun.}, {\bf 85}, 519.

\item MILLIS, A. J.,  MONIEN, H., and PINES, D., 1990, {\it Phys. Rev.} B 
{\bf 42}, 167.

\item MILLIS,  A. J., and MONIEN, H., 1992, {\it Phys. Rev.} B
{\bf 45}, 3059.

\item MONTOUX, P., and PINES, D., 1994, {\it Phys. Rev.} B {\bf 50},
16015. 

\item MOREO, A., HAAS, S., SANDVIK, A. W., and DAGOTTO, E., 1995,
{\it Phys. Rev.} B {\bf 51}, 12045.

\item MORIYA, T., TAKAHASHI, Y., and UEDA, K., 1990,
{\it J. Phys. Soc. Jpn.} {\bf 59}, 2905.

\item MOTT, N. F., and DAVIS, E. A., 1979, {\it Electronic Processes
in Noncrystalline Materials} (Clarendon, Oxford).

\item NAGAOKA, Y., 1966, {\it Phys. Rev.} {\bf 147}, 392.

\item NAGAOSA, N., and LEE, P. A., 1990, {\it Phys. Rev. Lett.}
{\bf 64}, 2450. 

\item NAZARENKO, A., VOS, K. J. E., HAAS, S., DAGOTTO, E., and
GOODING, R. J.,  1995, {\it Phys. Rev.} B {\bf 51}, 8676.

\item N\"UCKER, N., {\it et al.}, 1989, {\it Phys. Rev.} B {\bf 39},
12379.

\item OITMAA, J., and BETTS, D. D., 1978, {\it Can. J. Phys.}
{\bf 56}, 897.

\item OHATA, N., and KUBO, R., 1970, {\it J. Phys. Soc. Jpn.}
{\bf 28}, 1402.

\item OLSON, C. G., {\it et al.}, 1990, {\it Phys. Rev.} B {\bf 42},
381.

\item ONG, N. P., 1990, in {\it Physical Properties of High Temperature
Superconductors}, edited by D. M. Ginsberg (World Scientific,
Singapore), Vol. 2, p.~459. 

\item PANG, H., AKHLAGHPOUR, H., and JARRELL, M., 1996,
{\it Phys. Rev.} B {\bf 53}, 5086.

\item PARLETT, B. N., 1980, {\it The Symmetric Eigenvalue Problem}
(Prentice-Hall, Englewood Cliffs).

\item PARKINSON, J. B., 1969, {\it J. Phys.} C {\bf 2}, 2012.

\item POILBLANC, D., 1991, {\it Phys. Rev.} B {\bf 44}, 9562.

\item POILBLANC, D., ZIMAN, T., SCHULZ, H. J., and DAGOTTO, E., 1993,
{\it Phys. Rev.} B {\bf 47}, 14267.

\item PRELOV\v SEK, P., and ZOTOS, X., 1990, {\it Phys. Rev.} B
{\bf 47}, 5984.

\item PRELOV\v SEK, P., and JAKLI\v C, J., 1996, {\it Phys. Rev.} B
{\bf 53}, 15095.

\item PRELOV\v SEK, P., 1997a, {\it Z. Phys.} B {\bf 103}, 363.

\item PRELOV\v SEK, P., 1997b, {\it Phys. Rev.} B {\bf 55}, 9219.

\item PREUSS, R., HANKE, W., and von der LINDEN, W., 1995,
{\it Phys. Rev.  Lett.} {\bf 75}, 1344.

\item PREUSS, R., HANKE, W., Gr\"ober, C.,  and EVERTZ, H. G., 1997,
{\it Phys. Rev.  Lett.} {\bf 79}, 1122.

\item PRUSCHKE, T., JARRELL, M., and FREERICKS, J. K., 1995, {\it
Adv. Phys.} {\bf 44}, 187.

\item PUCHKOV, A. V., {\it et al.}, 1996, {\it Phys. Rev.  Lett.}
{\bf 77}, 3212.

\item PUTIKKA, W. O., LUCHINI, M. U., and RICE, T. M., 1992,
{\it Phys. Rev. Lett.}  {\bf 68}, 538.

\item PUTIKKA, W. O., GLENISTER, R. L., SINGH, R. R. P., and
TSUNETSUGU, H., 1994, {\it Phys. Rev. Lett.} {\bf 73}, 1994.

\item RAM\v SAK, A., and PRELOV\v SEK, P., 1989, {\it Phys. Rev.} B
{\bf 40}, 2239.

\item RICE, T. M., and ZHANG, F. C., 1989, {\it Phys. Rev.} B
{\bf 39}, 815.

\item RICE, T. M., 1995, in {\it Proceedings of the Les Houches Summer
School, Session LVI}, edited by B. Doucot and J. Zinn-Justin
(Elsevier, Amsterdam), p.19.

\item ROJO, A. G., KOTLIAR, G., and CANRIGHT, G. S., 1993, {\it
Phys. Rev.} B {\bf 14}, 9140.

\item ROMERO, D. B., {\it et al}., 1992, {\it Solid State Commun.}
{\bf 82}, 183.

\item ROSSAT - MIGNOT, J., {\it et al.}, 1991, {\it Physica} C,
{\bf 185 - 189}, 89.

\item SCALAPINO, D. J., WHITE, S. R., and ZHANG, S., 1993,
{\it Phys. Rev.} B {\bf 47}, 7995.

\item SCHMITT-RINK, S., VARMA, C. M., and RUCKENSTEIN, A. E., 1988,
 {\it Phys. Rev. Lett.} {\bf 60}, 2793.

\item SEGA, I., and PRELOV\v SEK, P., 1990, {\it Phys. Rev.} B {\bf
42}, 892.

\item SHASTRY, B. S., 1989, {\it Phys. Rev. Lett.} {\bf 63}, 1288.

\item SHASTRY, B. S., and SHRAIMAN, B. I., 1990,
{\it Phys. Rev. Lett.} {\bf 65}, 1068.

\item SHASTRY, B. S., and SUTHERLAND, B., 1990, {\it Phys. Rev. Lett.}
{\bf 65}, 243.

\item SHASTRY, B. S., SHRAIMAN, B. I., and SINGH, R. R. P., 1993,
{\it Phys. Rev. Lett.} {\bf 70}, 2004.

\item SHEN, Z.-X., and DESSAU, D. S., 1995, {\it Physics Reports}
{\bf 253}, 1.

\item SHIRANE, G., 1991, {\it Physica} C {\bf 185 - 189}, 80.

\item SILVER, R. N., and R\"ODER, H., 1994, {\it Int. J. Mod. Phys.} C
{\bf 5}, 735.

\item SINGH, R. R. P., FLEURY, P. A., LYONS, and SULEWSKI, P. E.,
1989, {\it Phys. Rev. Lett.} {\bf 62}, 2736.

\item SINGH, R. R. P., and GLENISTER, R. L., 1992a, {\it Phys. Rev.} B
{\bf 46}, 11871.

\item SINGH, R. R. P., and GLENISTER, R. L., 1992b, {\it Phys. Rev.} B
{\bf 46}, 14313.

\item SCHLESINGER, Z., {\it et al.}, 1990, {\it Phys. Rev. Lett.}
{\bf 65}, 801.

\item SLICHTER, C. P., 1994, {\it Proceedings of the Los Alamos
Symposium on Strongly Correlated Electronic Materials - 1993}, editors
K. S. Bedell, Z. Wang, D. E. Meltzer, A. V. Balatsky, and
E. Abrahams, (Addison-Wesley), p. 427.

\item SOKOL, A., GAGLIANO, E., and BACCI, S., 1993, {\it Phys. Rev.} B
{\bf 47}, 14646.

\item SOKOL, A., and PINES, D.,  1993, {\it Phys. Rev. Lett.} {\bf
71}, 2813.

\item SUZUKI, M., 1993, editor, {\it Quantum Monte Carlo Methods in
Condensed Matter Physics} (World Scientific, Singapore).

\item STARTSEVA, T., {\it et al.}, 1997, preprint cond-mat/9706145.

\item STEPHAN, W.,  and HORSCH, P., 1991, {\it Phys. Rev. Lett.}
{\bf 66}, 2258.

\item STERNLIEB, B. J., {\it et al.}, 1993, {\it Phys.  Rev.} B
{\bf 47},  5320.
 
\item TAKAGI, H., {\it et al.}, 1992, {\it Phys. Rev. Lett.}
{\bf 69}, 2975.

\item TAKIGAWA, M., {\it et al.}, 1991, {\it Phys. Rev.} B
{\bf 43}, 247.

\item TALLON, J. L., COOPER, J. R., de SILVA, P. S. I. P. N.,
WILLIAMS, G. V. M., and LORAM, J. W., 1995, {\it Phys. Rev. Lett.}
{\bf 75}, 4114.

\item TANNER, D. B., and TIMUSK, T., 1992, in {\it Physical Properties of
High Temperature Superconductors III}, edited by D. M. Ginsberg (World
Scientific, Singapore), p.363.

\item TOHYAMA, T., OKUDA, H., and MAEKAWA, S., 1993, {\it Physica} C
{\bf 215}, 382.

\item TOHYAMA, T., and MAEKAWA, S., 1994, {\it Phys. Rev.} B
{\bf 49}, 3596.

\item TOHYAMA, T., HORSCH, P., and MAEKAWA, S., 1995, {\it Phys. Rev.
Lett.}  {\bf 74}, 980.

\item TORRANCE, J. B., {\it et al.}, 1989, {\it Phys. Rev.} B
{\bf 40}, 8872.

\item TRANQUADA,  J. M., {\it et al.}, 1995, {\it Nature} {\bf 375}, 561.

\item TRUGMAN, S. A., 1990, {\it Phys. Rev. Lett.} {\bf 65}, 500.

\item TSUJI, M., 1958, {\it J. Phys. Soc. Jpn.} {\bf 13}, 979.

\item TSUNETSUGU, H., and IMADA., M., 1997, {\it J. Phys. Soc. Jpn.},
{\bf 66}, 1876.

\item UCHIDA, S., {\it et al}., 1991, {\it Phys. Rev.} B {\bf 43}, 7942.

\item UCHIDA, S., 1997, {\it Physica C} {\bf 282-287}, p.12.

\item VARMA, C. M., LITTLEWOOD, P. B.,  SCHMITT-RINK, S.,
ABRAHAMS, E., and RUCKENSTEIN, A. E., 1989, {\it Phys. Rev. Lett.}
{\bf 63}, 1996.

\item VEBERI\v C, D., PRELOV\v SEK, P., and SEGA, I., 1998,
{\it Phys. Rev. B}, to appear. 

\item VOLLHARDT, D., 1997, {\it Phys. Rev. Lett.} {\bf 78}, 1307.   

\item Von der LINDEN, W., 1992, {\it Phys. Rep.} {\bf 220}, 53.

\item WALSTEDT, R. E., BELL, R. F., SCHNEEMEYER, L. F., WASZCZAK,
J. V., and ESPINOSA, G. P., 1992, {\it Phys. Rev.} B {\bf 45}, 8074.   

\item WANG, Z., BANG, Y., and KOTLIAR, G., 1991, {\it Phys. Rev. Lett.}
{\bf 67}, 2733.

\item WELLS, B. O., 1995,  {\it et al}., {\it Phys. Rev. Lett.}
{\bf 74}, 964.

\item WHITE,  S. R., 1992, {\it Phys. Rev. Lett.} {\bf 69}, 2863.

\item WHITE,  S. R., and SCALAPINO, D., 1997a, {\it Phys. Rev.} B
{\bf 55}, R14701.

\item WHITE,  S. R., and SCALAPINO, D., 1997b, preprint cond-mat/9705128.

\item ZAANEN, J., and OLE\'{S}, A. M., 1988, {\it Phys. Rev.} B
{\bf 37}, 9423.

\item ZHANG, F. C., and RICE, T. M., 1988, {\it Phys. Rev.} B
{\bf 37}, 3759.

\item ZEYHER, R., 1991, {\it Phys. Rev.} B {\bf 44}, 10404.

\item ZOTOS, X., PRELOV\v SEK, P., and SEGA, I., 1990,
{\it Phys. Rev.} B {\bf 42}, 8445.

 \item ZOTOS, X., and PRELOV\v SEK,
P., 1996, {\it Phys. Rev.} B {\bf 53}, 983.

\end{list}
\end{document}